\documentclass[a4paper,11pt]{article}
\pdfoutput=1 % if your are submitting a pdflatex (i.e. if you have
\usepackage{jheppub} % for details on the use of the package, please
\usepackage[T1]{fontenc} % if needed
\usepackage{xfrac}
\usepackage{caption}
\usepackage[dvipsnames]{xcolor}
\usepackage{relsize}
\usepackage{graphicx}% Include figure files
\usepackage{dcolumn}% Align table columns on decimal point
\usepackage{bm}% bold math
\usepackage{epstopdf}
\usepackage{mathrsfs}
\usepackage{amsmath,amssymb,amsfonts,latexsym}
\usepackage{amsmath,bbold}
\usepackage{color}
\usepackage{slashed}
\usepackage{nicefrac}
\usepackage[colorlinks=true,linktocpage=true,linkcolor=blue,citecolor=blue]{hyperref}
\usepackage[utf8]{inputenc}
\usepackage[T1]{fontenc}
\usepackage{accents}
\usepackage[normalem]{ulem}

\def\bs{\boldsymbol} 

\def\bdel{\bs\partial}

\long\def\comment#1{ }

\def\0{{\boldsymbol 0}}
\def\q{{\bm q}}

\def\k{{\boldsymbol k}}

\def\x{{\boldsymbol x}}
\def\y{{\boldsymbol y}}
\def\p{{\boldsymbol p}}

\def\r{{\boldsymbol r}}
\def\z{{\boldsymbol z}}
\def\u{{\boldsymbol u}}

\def\and{\qquad\text{and}\qquad}

\def\omegacbar{\bar\omega_c}

\def\tform{{t_\text{f}}}

\def\tsplit{t_\text{split}}

\def\med{\text{med}}

\def\min{\text{min}}
\def\max{{\text{max}}}
\def\reg{{\text{reg}}}

\def\Kc{{\cal K}}
\def\Ic{{\cal I}}
\def\Cc{{\cal C}}
\def\momega{{\omega}}

\def\mfp{\lambda}
\def\omegaBH{\omega_\text{\tiny BH}}

\def\xBH{x_\text{\tiny BH}}
\def\ho{\text{\tiny HO}}
\def\ioe{\text{\tiny IOE}}

\newcommand{\beq}{\begin{eqnarray}}
\newcommand{\eeq}{\end{eqnarray}}
\newcommand{\be}{\begin{eqnarray*}}
\newcommand{\ee}{\end{eqnarray*}}
\newcommand{\bal}{\begin{align}}
\newcommand{\eal}{\end{align}}
\newcommand{\rmd}{{\rm d}}
\newcommand{\dd}{{\rm d}}
\newcommand{\rme}{{\rm e}}

\def\rmR{{\rm Re}}
\def\rmI{{\rm Im}}
\def\abar{{\rm \bar\alpha}}

\newcommand{\nn}{\nonumber\\ }

\def\qhat{{\hat{q}}}
\def\cK{{\cal K}}
\def\bs{\boldsymbol} 
\title{A unified picture of medium-induced radiation}

\author{Johannes Hamre Isaksen,}
\author{Adam Takacs}
\author{and Konrad Tywoniuk}

\affiliation{Department of Physics and Technology, University of Bergen, 5007 Bergen, Norway}

\emailAdd{johannes.isaksen@uib.no}
\emailAdd{adam.takacs@uib.no}
\emailAdd{konrad.tywoniuk@uib.no}

\abstract{
We revisit the picture of jets propagating in the quark-gluon plasma. In addition to vacuum radiation, partons scatter on the medium constituents resulting in induced emissions. Analytical approaches to including these interactions have traditionally dealt separately with multiple, soft, or rare, hard scatterings. A full description has so far only been available using numerical methods. We achieve full analytical control of the relevant scales and map out the dominant physical processes in the full phase space. To this aim, we extend existing expansion schemes for the medium-induced spectrum to the Bethe--Heitler regime. This covers the whole phase space from early to late times, and from hard splittings to emissions below the thermal scale. Based on the separation of scales, a space-time picture naturally emerges: at early times, induced emissions  
start to build from rare scatterings with the medium. At a later stage, induced emissions due to multiple soft scatterings result in a turbulent cascade that rapidly degrades energy down to, and including, the Bethe--Heitler regime. We quantify the impact of such an improved picture, compared to the current state-of-the-art factorization that includes only soft scatterings, by both analytical and numerical methods for the medium-induced energy distribution function. 
Our work serves to improve our understanding of jet quenching from small to large systems and for future upgrades of Monte Carlo generators.
}

\begin{document} 
    \maketitle
\flushbottom

%%%%%%%%%%%%%%%%%%%%%%%%%%%%%%%%%%%%%%%%%%%%%%%%%%%%
\section{Introduction}
\label{sec:intro}
%%%%%%%%%%%%%%%%%%%%%%%%%%%%%%%%%%%%%%%%%%%%%%%%%%%%

Short-lived droplets of hot and dense nuclear matter, called the quark-gluon plasma (QGP), are produced in relativistic heavy-ion collisions at RHIC and LHC. Embedded in the same high-energy collisions, hard QCD processes are also present, resulting in the production of collimated sprays of energetic particles that are commonly referred to as jets~\cite{Ellis:318585,Dokshitzer:1991wu}. Jets are well-understood, perturbative objects within perturbative QCD and they are described up to high precision in proton-proton collisions~\cite{Larkoski:2017jix,Marzani:2019hun,Dasgupta:2020fwr}. During their propagation, however, jet particles can interact with the surrounding nuclear matter. The modification of jet features, therefore, reflects the properties of the QGP created in heavy-ion collisions~\cite{Gyulassy:1990ye,Gyulassy:2003mc,Peigne:2008wu,Mehtar-Tani:2013pia,Ghiglieri:2015zma,Andrews:2018jcm}. Currently, a vigorous experimental program dedicated to quantifying jet modifications is ongoing at both RHIC and LHC, focusing on a broad set of observables which includes measurements of the modification of the jet spectrum, jet substructure observables, and jet correlations~\cite{Armesto:2015ioy,Connors:2017ptx,Kogler:2018hem,Cunqueiro:2021wls,Apolinario:2022vzg} (for a selection of predictions see Refs.~\cite{Chien:2015hda,Mehtar-Tani:2016aco,KunnawalkamElayavalli:2017hxo,Caucal:2019uvr,Casalderrey-Solana:2019ubu,Ringer:2019rfk,Caucal:2020xad,Caucal:2021cfb,Mehtar-Tani:2021fud,Takacs:2021bpv,Attems:2022ubu}).

High-energy jets are particularly suitable probes of the QGP because their energy scale $Q_{\rm jet}$ is much larger than the typical momentum scale of the medium $Q_{\rm med}$. If this is the case, the impact of medium modifications should therefore not affect the internal structure of the jet, which would still rely on perturbative QCD~\cite{Casalderrey-Solana:2012evi,Caucal:2018dla,Vaidya:2020lih}. A key ingredient when considering jet modifications is the radiation induced by scatterings with the deconfined medium constituents. Such emissions typically appear at scales comparable to $Q_\med$. Emission at scales much higher than $Q_{\rm med}$, on the other hand, are unaffected by the medium, resulting in a factorized picture between vacuum and medium processes~\cite{Mehtar-Tani:2017web,Caucal:2018dla}. Medium-induced emissions redistribute the original jet parton energy to multiple, soft particles over large angles, including out of the jet cone. This leads to a net jet energy loss which, in turn, is manifested as a suppression of the jet spectrum (for an updated discussion of jet quenching see Refs.~\cite{Mehtar-Tani:2021fud,Takacs:2021bpv}, and for applications to substructure see Refs.~\cite{Mehtar-Tani:2016aco,Caucal:2019uvr,Caucal:2021cfb}). Consequently, medium-induced emissions are a crucial component of jet energy loss and thus of phenomenological studies of jet observables in heavy-ion collisions.

The medium-induced emission spectrum was formulated a long time ago~\cite{Baier:1996kr,Zakharov:1997uu,Arnold:2002ja}. Previous solutions were limited to either (i) expanding in the number of scatterings (referred to as the opacity expansion)~\cite{Gyulassy:1999zd,Wiedemann:2000za,Djordjevic:2003zk,Sievert:2018imd}, or (ii) considering multiple soft scatterings (called the harmonic oscillator approximation)~\cite{Baier:1996kr,Zakharov:1997uu,Salgado:2003gb,Armesto:2003jh}. Meanwhile, several works focused on the underlying scales that separate the limiting cases~\cite{Baier:1996kr,Arnold:2008iy,Arnold:2009mr,Blaizot:2012fh,Kurkela:2014tla,Dominguez:2019ges,Andres:2020kfg}. The full problem has also been tackled by numerical techniques~\cite{Zakharov:2004vm,CaronHuot:2010bp} (or more recently in Refs.~\cite{Feal:2018sml,Ke:2018jem,Andres:2020vxs,Schlichting:2020lef}). Not long ago, analytical techniques were developed that provided a unified description of the multiple, soft and rare, hard scatterings in a dense medium ~\cite{Mehtar-Tani:2019tvy,Mehtar-Tani:2019ygg,Barata:2020sav,Barata:2020rdn,Barata:2021wuf}, which better match the full numerical solutions. The main challenge, common to both the numeric and analytic approaches, resides in dealing with multiple interactions with the underlying medium.

In this paper, we revisit the different analytic approaches to resumming multiple interactions for calculating the medium-induced emission spectrum. These include the opacity expansion (OE) and the improved opacity expansion (IOE), which includes harmonic oscillator approximation as the leading term. Moreover, we rigorously derive the resummed opacity expansion (ROE) for the first time, which extends the description of the spectrum to low energy emissions in the Bethe--Heitler regime. We provide a novel unified picture of these resummation schemes by identifying their relevant emergent scales and demonstrating their respective regions of validity. For example, we show that the single scattering approximation, contained in the leading order of OE, is valid even for a big medium, where one would expect more than one scattering if the emitted energy is high enough. We show that the full phase space of medium-induced emissions, spanning from the maximal jet energy to the thermal scale, is covered by a union of these expansions, see also Ref.~\cite{Andres:2020kfg}. Each of the expansions is also associated with the corresponding physical scattering processes, and thus we reinterpret the frequently used terms such as GLV emissions, coherent scatterings, and Bethe--Heitler region in a unified framework. Our framework goes beyond previous attempts to describe all regimes of medium-induced bremsstrahlung by presenting a resummation framework that can be systematically improved and that is valid both in the dilute and dense regimes.

As a next step, we identify the regimes where not only multiple interactions are important, but also multiple emissions \cite{Arnold:2002zm,Blaizot:2013vha}. These conditions are met for sufficiently soft emissions in a large medium. The previously established hierarchy of emergent scales plays a crucial role in mapping out early, rare, and relatively hard emissions and a successive cascade of soft splittings. In this context, hard medium-induced splittings can be thought of as extra sources, in addition to the parent parton, for the full cascade. This description is realized analytically in a novel scheme that combines a fixed order expansion of rare emissions with an all-order resummation of soft splittings. The resulting energy distribution links the asymptotic early and late time behaviors for which analytical solutions exist. Finally, we resum multiple induced emissions numerically, using the previously obtained precise determination of the in-medium splitting rates, to calculate the energy distribution function. We highlight the interplay of rare hard scatterings, coherent soft splittings, and Bethe--Heitler emissions in a finite medium, providing a state-of-the-art resummation. 

Our reorganized picture helps not only with the physical understanding of induced emissions, but provides a fast and efficient way to calculate the medium-induced spectrum, which is a key ingredient for estimating jet energy loss. It also serves to inform Monte Carlo algorithms simulating full jet evolution inside the medium about how to implement multiple medium-induced emissions and how to combine them with vacuum-like emissions, e.g. see in Ref.~\cite{Caucal:2018dla}.

The paper is structured as follows. In Sec.~\ref{sec:heuristic}, as an introduction, we discuss the structure of the induced spectrum in the various regimes using heuristic arguments, and we show how the radiation in the different regimes is related to single soft, multiple soft and single hard scatterings with the medium, see Fig.~\ref{fig:PhaseSpace_PhysicalProcesses}. The spectrum is calculated in detail in Sec.~\ref{sec:spectrum}. We revisit the opacity expansion and the improved opacity expansion schemes, and put on a firm footing a novel resummation scheme, dubbed resummed opacity expansion, which is valid for emissions below the Bethe--Heitler scale. Improving on previous discussions, we provide formulas for the spectrum at arbitrary order and calculate it exactly up to second order in all the expansions, allowing us for the first time to establish regions where they converge. Finally, in Sec.~\ref{sec:multiple} we consider the problem of multiple emissions. We analyze induced particles coming from the full phase space and confirm the importance of considering multiple emissions, especially in the soft sector. In order to facilitate an improved analytical understanding of the problem, we finally suggest a resummation scheme of multiple emissions by iterating in rare, hard emissions and including an arbitrary number of soft splittings. This is compared to the full numerical results. We conclude with an outlook in Sec.~\ref{sec:conclusions}. The appendix contains lots of useful formulas, including the rate of emissions and finite-$z$ corrections that are important for phenomenology. The code we have developed to calculate the kernels and solve for the energy distribution is provided at \url{https://github.com/adam-takacs/kernels.git}.

\begin{figure}
    \centering
    \includegraphics[width=0.3\textwidth]{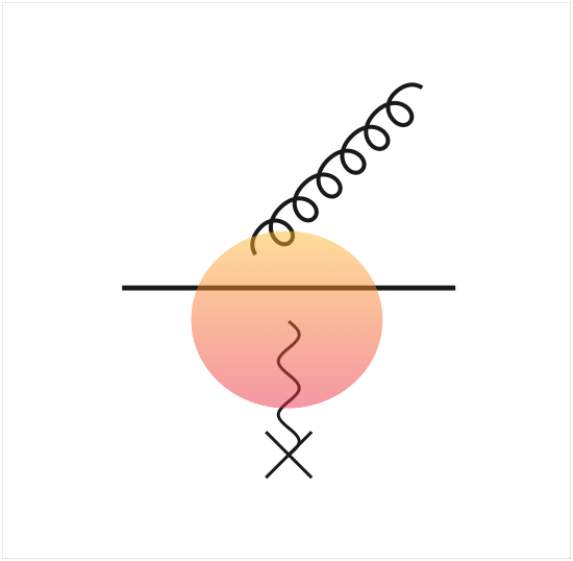}
    \hspace*{-1em}
    \includegraphics[width=0.3\textwidth]{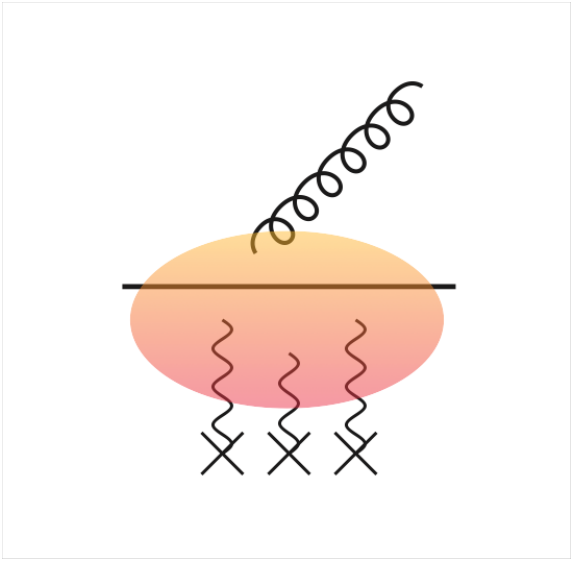}
    \hspace*{-1em}
    \includegraphics[width=0.3\textwidth]{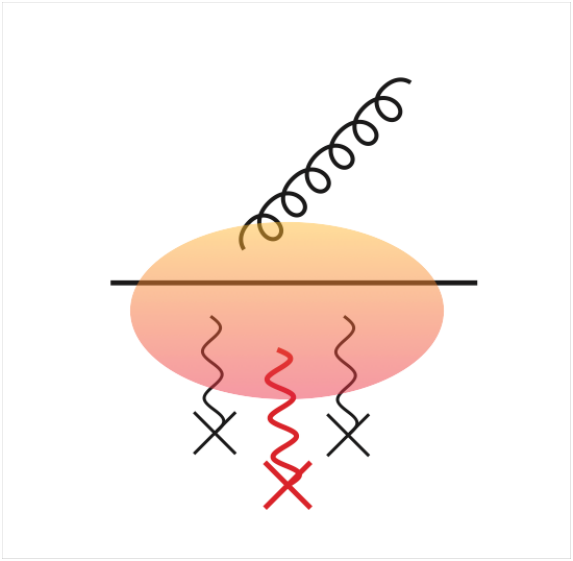}
    \caption{Three regimes of the radiative spectrum in dense media, $L\gg \mfp$: the Bethe--Heitler regime $\omega < \omegaBH$ (leftmost), the BDMPS-Z regime $\omegaBH < \omega < \omega_c$ (middle), and the hard GLV regime $\omega_c < \omega$ (rightmost figure). The size of the blob represents the typical formation time of the emission, which is $\tform < \mfp$ in the leftmost panel and $\tform > \mfp$ in the center and rightmost panels.     Black, wavy lines represent soft interactions with the medium, where the momentum transfer is of the order of the medium scale $|\q |\lesssim \mu$, while the red, wavy line represents a hard scattering event, with $|\q| \gg \mu$.}
    \label{fig:regimes-illustration}
\end{figure}

%%%%%%%%%%%%%%%%%%%%%%%%%%%%%%%%%%%%%%%%%%%%%%%%%%%%
\section{Heuristic discussion of the medium-induced spectrum}
\label{sec:heuristic}
%%%%%%%%%%%%%%%%%%%%%%%%%%%%%%%%%%%%%%%%%%%%%%%%%%%%

The spectrum of gluon emissions induced by scatterings on a deconfined medium plays a central role in the phenomenology of jet quenching. However, a full understanding of all regimes have so far been lacking analytically and was previously only achievable through numerical methods. Here, we present a unified view of all relevant medium scales and their related regimes. Similar heuristic discussions have previously been presented in, e.g., \cite{Baier:1996kr,Peigne:2008wu,Mehtar-Tani:2019tvy,Andres:2020kfg}. In Sec.~\ref{sec:spectrum} we will provide rigorous derivations of the findings argued for here.

We formulate the spectrum of induced emissions by focusing on the relevant length scales:
\begin{itemize}
    \item The \textbf{size} of the medium $L$ (or the length of propagation $t <L$).
    \item The \textbf{mean free path} of the medium $\lambda\sim\frac{1}{n\sigma_{\rm tot}}$, which combines the density $n$ and the scattering strength $\sigma_{\rm tot}\sim\int\rmd\sigma$, and it describes the distance between typical scatterings.
    \item The \textbf{formation time} of an emission $t_{\rm f}=\frac{2\omega}{\bm k^2}$, where $\omega$ is the energy and $\bm k$ is the transverse momentum of the emission.
\end{itemize}
In addition, the relation between the in-medium screening scale $\mu$ and the range of available transverse momenta $|\k|$ is also important. 

The \textit{opacity} $\chi \equiv L/\mfp$ characterizes the denseness of the medium. If the opacity is small ($L\ll \mfp$), the medium is ``dilute'', or weakly interacting, while it is ``dense'', or strongly interacting if $L\gg \mfp$.\footnote{From this perspective, we have ``fixed'' $L$ and vary $\mfp$. Naturally, we could also have identified these two regimes as a ``large'' and ``small'' media, where we ``fix'' the mean free path and vary $L$ instead.} The dilute medium barely consists of scattering centers, however, in the dense medium one should account for an arbitrary number of interactions. 

A big part of this paper will be about calculating the spectrum of medium-induced emissions. For reference, the vacuum spectrum reads,
\begin{equation}
    \omega\frac{\rmd I}{\rmd \omega} \sim \alpha_s \int \frac{\rmd \k^2}{\k^2} \,,
\end{equation}
where currently we did not specify the limits of the transverse momentum integral.
This contains the well-known soft ($\omega \to 0$) and collinear ($\k^2\to 0$) divergences. In contrast, the collinear divergence in the medium spectrum is removed by the need to exchange transverse momentum with the medium. 

Motivated by this, we introduce a heuristic model that captures some of the features of the medium-induced spectrum, given by
\begin{equation}
\label{eq:spectrum_ill}
     \omega\frac{\rmd I}{\rmd \omega}\sim \alpha_s L\int\rmd^2\bm k \,\sigma(\bm k) \sim \alpha_s^3 n L \int \frac{\rmd \bm k^2}{(\bm k^2+\mu^2)^2} \,,
\end{equation}
where $n$ is the medium density and $\mu$ a screening mass. The behavior at high-$\k$, $\sigma\sim \bm k^{-4}$, reproduces the expected Coulomb tail. The factor $L$ arises since the emission can take place anywhere along the medium length. For a more accurate description of medium-induced emissions, we refer the reader to Sec.~\ref{sec:spectrum}.

In our effective description, we focus on the hierarchy among the introduced scales and show the separation of different scattering regions. Firstly, in the $t_{\rm f}\gg L$ limit, the formation of the emission extends beyond the medium, where one naturally should expect vacuum physics to dominate.\footnote{For such soft emissions, medium effects can influence the color coherence properties leading to a modification of the phase space \cite{Mehtar-Tani:2010ebp,Mehtar-Tani:2012mfa}.} We will hence not consider this possibility here. The remaining cases are listed below:

%%%%%%%%%%%%%%%%%%%%%%%%%%%%%%%%%%%%%%%%%
\paragraph{\textbf{Dilute media} ($t_{\rm f} \leq L  \ll \lambda$):}
%%%%%%%%%%%%%%%%%%%%%%%%%%%%%%%%%%%%%%%%%
In case of a low medium opacity, we expect that roughly one scattering occurs. This process will typically transfer a momentum of order of the Debye mass to the emitted gluon, or $\langle\k^2\rangle \sim \mu^2$, leading to $\tform = 2 \omega /\mu^2$. The formation of the gluon has to take place inside the medium, giving rise to the characteristic energy scale in the dilute regime, namely
\begin{equation}
    \label{eq:baromegac}
    \bar \omega_c =\frac{1}{2}\mu^2 L \,.
\end{equation}
This separates two regimes of emissions that are sourced via different scattering processes: on the one side soft gluons with $\omega < \bar \omega_c$, generated via a soft scattering with the medium $\langle\k^2\rangle \lesssim \mu^2$. Hard gluons with $\omega > \bar \omega_c$ can also be generated, however those demand a large momentum exchange with the medium, $\langle\k^2\rangle > \mu^2$, which is comparatively rare. Let us now consider how the spectrum behaves in these two distinct regimes.

According to our discussion above, the soft production should be dominated by soft transverse momentum exchanges with the medium. Hence, we expect that the spectrum of emitted gluons goes as
\begin{equation}
    \label{eq:dilute-soft}
   \left. \omega\frac{\rmd I}{\rmd\omega} \right|_{\omega < \bar \omega_c} \sim \alpha_s L\, \int_0^{\infty}\rmd \bm k^2 \frac{\alpha_s^2 n}{(\bm k^2+\mu^2)^2} \sim \alpha_s \frac{L}{\mfp}  \,,
\end{equation}
where the integral is dominated by $\k^2 \lesssim \mu^2$. This integral gives the proportionality with the inverse mean free path  i.e. $n/\mu^2 \sim 1/\mfp$, resulting in an overall factor of medium opacity $L/\mfp$. This parametric estimate misses an important logarithmic factor $\sim  \ln \bar \omega_c/\omega$, see a further discussion in Sec.~\ref{sec:oe}, which signals that the simplifications pertaining to the ``soft'' regime break down at $\omega \approx \omegacbar$.

\begin{figure}
    \centering
    \includegraphics[width=0.45\textwidth]{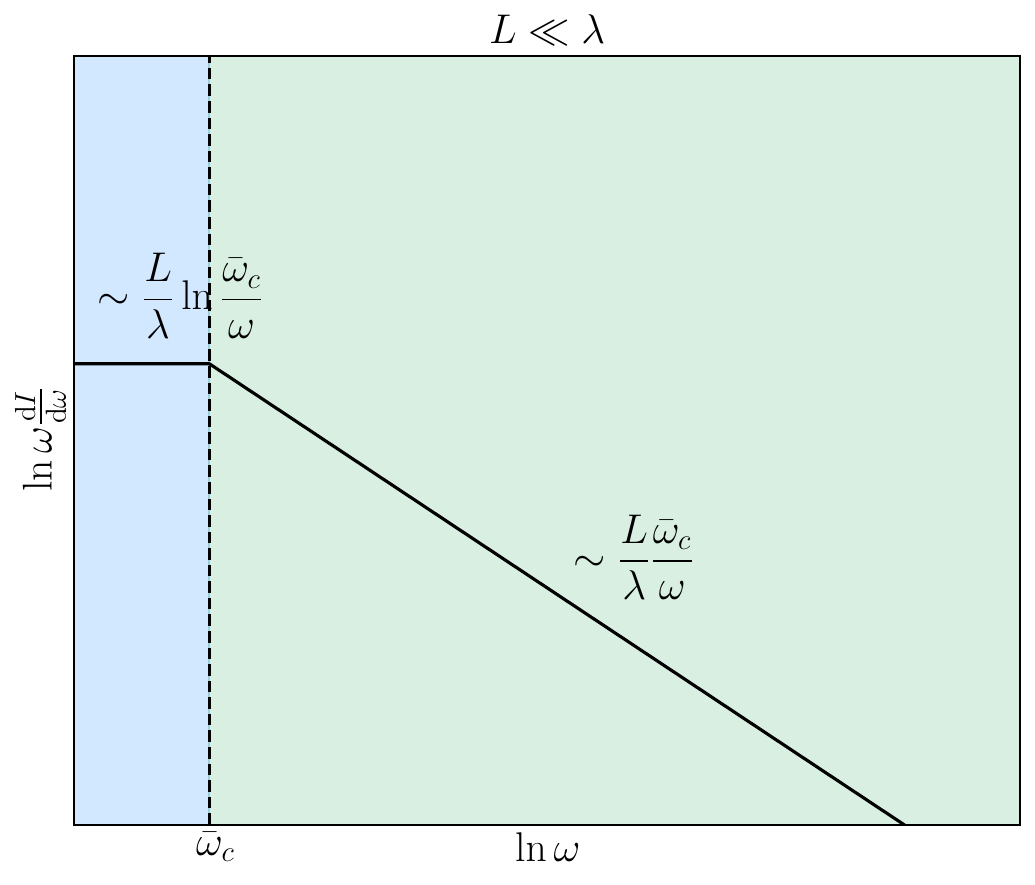}%
    \hspace{1cm}
    \includegraphics[width=0.45\textwidth]{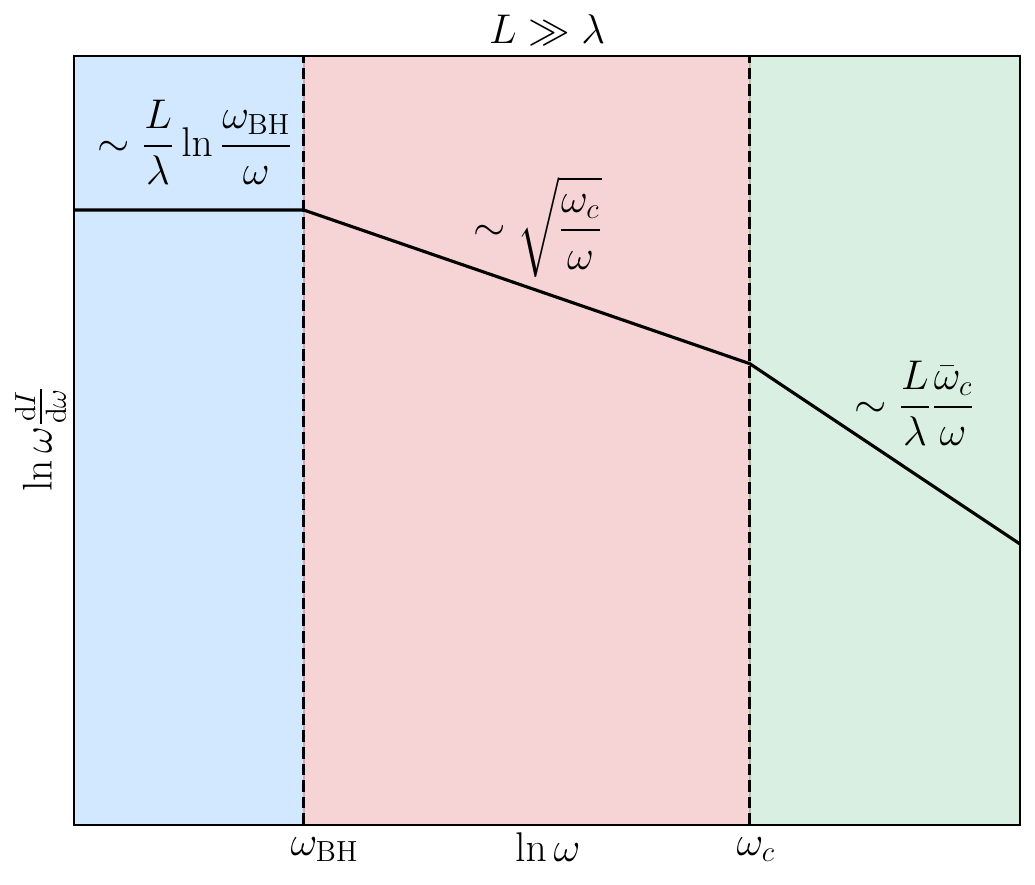}
    \caption{A sketch of the spectrum of medium-induced gluons in a dilute medium $L \ll \mfp$ (left) and in a dense medium $L \gg \mfp$ (right).}
    \label{fig:spectra-sketch}
\end{figure}

For hard emissions, $\omega > \bar \omega_c$, we instead get that
\begin{equation}
    \label{eq:dilute-hard}
   \left. \omega\frac{\rmd I}{\rmd\omega} \right|_{\omega > \bar \omega_c} \sim \alpha_s^3 n L\, \int_{\omega/L}^{\infty} \frac{\rmd \k^2}{\k^4} \sim \alpha_s \frac{L}{\mfp} \frac{\bar \omega_c}{\omega}  \,,
\end{equation}
where we used $\tform=2\omega/\bm k^2<L$, and neglected the screening mass $\mu^2$ in this parametric regime, since $\langle\bm k^2\rangle\gg\mu^2$.
Compared to the soft regime from Eq.~\eqref{eq:dilute-soft} it is suppressed by an additional power of $\bar \omega_c/\omega\ll1$. The complete spectrum in the dilute regime is sketched in Fig.~\ref{fig:spectra-sketch} (left).

%%%%%%%%%%%%%%%%%%%%%%%%%%%%%%%%%%%%%%%%%
\paragraph{\textbf{Dense media with long formation time} ($\lambda \ll t_{\rm f}  \ll L$):}
%%%%%%%%%%%%%%%%%%%%%%%%%%%%%%%%%%%%%%%%%
In a dense medium we should expect that typically many scatterings occur during the emission process, which is illustrated in the middle of Fig.~\ref{fig:regimes-illustration}. This demands a more sophisticated model than what we suggested in Eq.~\eqref{eq:spectrum_ill}. Nevertheless, we can approximate the total transferred transverse momentum by $\langle\k^2\rangle \sim \hat q t$, which resembles a random walk for $t$ time in two dimensions, with $\hat q$ playing the role of a diffusion constant.\footnote{Arbitrary dense medium, would result in overlapping scatterings that description if beyond the scope of this paper. Multiple independent scatterings require well separated scattering centers ($1/\mu\ll\lambda$) see in Ref.~\cite{Baier:1996kr}.} This constant determines the typical transverse momentum accumulated per unit length, or $\hat q \sim n \sim \mu^2/\mfp$. In this case the formation time becomes
\begin{equation}
    \tform = \sqrt{\frac{2 \omega}{\hat q}} \,.
\end{equation}
This is often called the coherence length, since during the formation time, interference effects between multiple scattering with the medium are active and the gluon feels only one effective scattering center. The accumulated transverse momentum during the splitting process is in this case $\langle \k^2 \rangle = \sqrt{2 \omega \hat q}$, which is the celebrated Landau-Pomeranchuk-Migdal (LPM) effect.

Again, comparing the formation time to the medium length, leads to the characteristic energy scale in the dense regime, namely
\begin{equation}
    \label{eq:omegac}
    \omega_c = \frac{1}{2} \hat q L^2 \,,
\end{equation}
and thus $\omega<\omega_c$ for multiple soft scatterings. The maximal possible momentum accumulated via multiple soft scatterings is denoted $ \langle \k^2 \rangle \sim Q_s^2 = \hat qL$. The other limiting scale of the multiple scattering regime arises when considering the minimal formation time in this hierarchy, i.e. $\tform > \lambda$, giving rise the scale $\omega>\omegaBH$, (see later in Eq.~\eqref{eq:omegaBH}). In this case, the accumulated transverse momentum squared reduces to a single soft scattering $\langle \k^2 \rangle \sim \hat q \lambda \sim \mu^2$.

In the multiple soft scattering regime, characterized by $\omegaBH \ll \omega \ll \omega_c$, the mean free path has to be replaced by the formation time in Eq.~\eqref{eq:dilute-soft}, leading to
\begin{equation}
    \label{eq:dense-lpm}
   \left. \omega \frac{\rmd I}{\rmd \omega} \right|_{\omegaBH \ll \omega \ll \omega_c}\sim \alpha_s \frac{L}{\tform} \sim \alpha_s \sqrt{\frac{\hat q L^2}{\omega}} \,.
\end{equation}
This is also often referred to as the BDMPS-Z spectrum in the soft limit.

For hard gluon emissions, $\omega \gg \omega_c$, we also have to demand that $\langle \k^2 \rangle \gg Q_s$. In other words, only a hard scattering can provide sufficient transverse momentum to fulfill all the conditions. The relevant contribution is therefore captured by Eq.~\eqref{eq:dilute-hard} and, remarkably, the spectrum in this limit is identical to the hard tail in the dilute regime, namely
\begin{equation}
    \left. \omega \frac{\rmd I}{\rmd \omega} \right|_{\omega \gg \omega_c} \sim \alpha_s \frac{L}{\mfp}\frac{\bar \omega_c}{\omega} \,.
\end{equation}
This demonstrates that, even in a dense medium, hard emissions mostly are driven by single, rare hard scattering events. An illustration of this can be seen on the right in Fig.~\ref{fig:regimes-illustration}.

%%%%%%%%%%%%%%%%%%%%%%%%%%%%%%%%%%%%%%%%%
\paragraph{\textbf{Dense media with short formation time} ($t_{\rm f} \ll \lambda  \ll L$):}
%%%%%%%%%%%%%%%%%%%%%%%%%%%%%%%%%%%%%%%%%
The picture we just described should hold as long as there indeed is time for multiple scatterings during the emission process, namely that $\tform > \lambda$. However, when the formation time is short the parton will only have time to scatter once before it splits. This is illustrated on the left in Fig.~\ref{fig:regimes-illustration}. The transverse scale is typically soft $\langle \bm k^2 \rangle  \sim \mu^2$, and thus $\tform = \frac{2\omega}{\mu^2}$. This regime is characterized by $\tform \ll \lambda$, or equivalently as a condition on the energy $\omega \ll \omegaBH$, where we have defined the scale
\begin{equation}
    \label{eq:omegaBH}
    \omegaBH =\frac{1}{2}\mu^2 \lambda \,.
\end{equation}
Note that $\bar\omega_c(\lambda) \equiv \omegaBH$. In this case the spectrum becomes
\begin{equation}
    \label{eq:dense-bh}
    \left. \omega \frac{\rmd I}{\rmd \omega} \right|_{\omega < \omegaBH} \sim \alpha_s \frac{L}{\mfp} \,,
\end{equation}
which is similar to the result in Eq.~\eqref{eq:dilute-soft} and it is sometimes referred to as Bethe--Heitler region because of the QED analogue. The $t_{\rm f}\ll\lambda$ condition is satisfactory but not necessary for BH emissions. There are BH emissions with $t_{\rm f}>\lambda$ but only with one real scattering. We show this more rigorously later in Sec.~\ref{sec:spectrum}. Our heuristic analysis fails to capture the additional logarithmic term which in this case comes with $ \ln \omegaBH/\omega$, see Sec.~\ref{sec:roe}.
A sketch of the spectrum in a dense medium can be found in Fig.~\ref{fig:spectra-sketch} (right).

\begin{figure}
    \centering
    \includegraphics[width=0.65\textwidth]{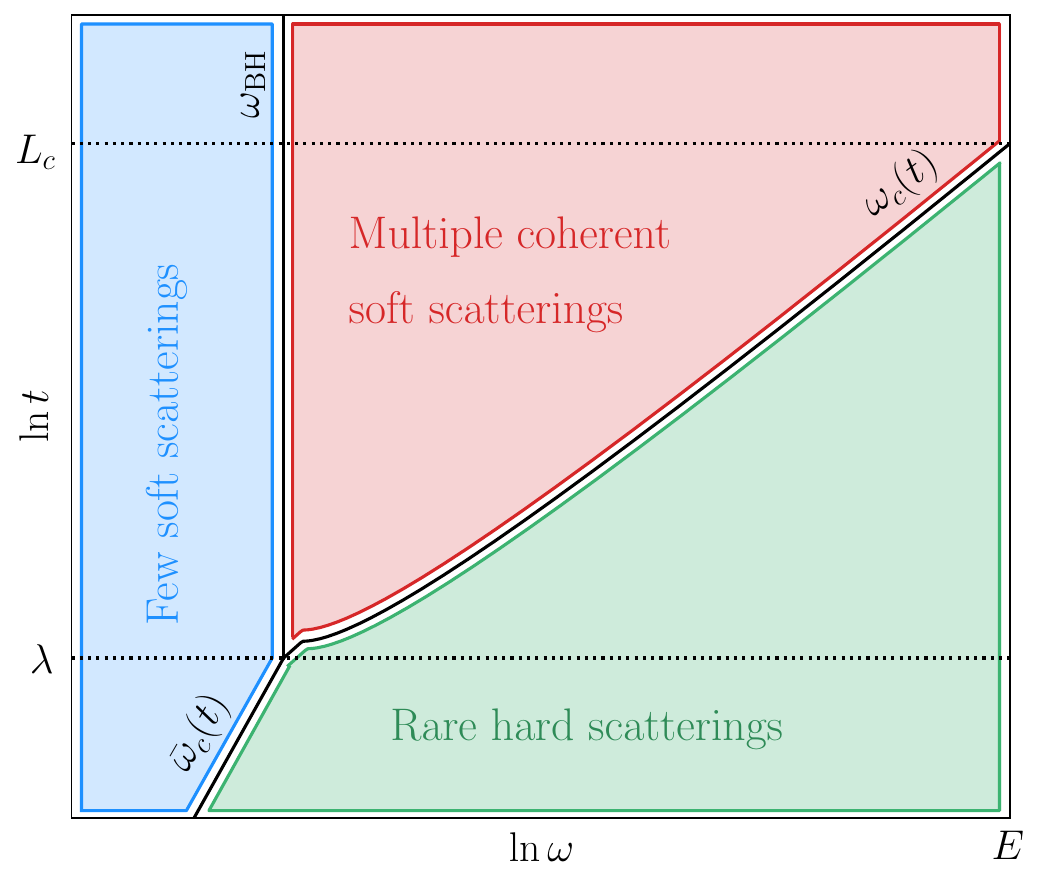}
    \caption{The phase space for medium-induced emissions designated to the leading scattering processes. The main length scales of the problem are $\lambda$ and the critical length $L_c$, corresponding to the characteristic energies $\omegaBH$ and $E$, see text for further details.}
    \label{fig:PhaseSpace_PhysicalProcesses}
\end{figure}

%%%%%%%%%%%%%%%%%%%%%%%%%%%%%%%%%%%%%%%%%%%%%%%%%%%%
\paragraph{Summary:}
%%%%%%%%%%%%%%%%%%%%%%%%%%%%%%%%%%%%%%%%%%%%%%%%%%%%
Bringing together the heuristic arguments of this section, we show a sketch of the emission spectrum for dilute and dense media in Fig.~\ref{fig:spectra-sketch}. The full emission phase space is divided by three lines corresponding to the emergent scales: $\bar \omega_c(t) = \frac12\mu^2 t$ in the dilute regime ($L < \mfp$), and $\omegaBH = \frac12\mu^2 \lambda$ and $\omega_c(t) = \frac12\hat q t^2$ in the dense regime ($L > \mfp$) and they are shown in Fig.~\ref{fig:PhaseSpace_PhysicalProcesses}. The $\omega_c(t)$ line is not completely straight because $\hat q$ in general is $\omega$ dependent (see in Sec.~\ref{sec:ioe}). When the medium size is of the order of the mean free path they all collapse to the same value, i.e. $\omegacbar(\mfp) = \omega_c(\mfp) = \omegaBH$. Typically, we adopt a notation where the scales written without the $t$-argument denote their respective values at $L$, e.g. $\omega_c \equiv \omega_c(L)$. 

These scales delineate three distinct regimes of scattering processes and thus induced emissions that were discussed in the preceding paragraphs. The areas between these scales are governed by few soft, multiple soft and rare hard interactions with the medium, as depicted with colors in Fig.~\ref{fig:PhaseSpace_PhysicalProcesses} and discussed above. We also show two length scales: the mean free path $\lambda$ and the critical medium length $L_c$. The mean free path marks the time where multiple scatterings appear. The critical medium length indicates where rare, hard scatterings will no longer have an effect, that is where $\omega_c(t)=E$, leading to $L_c = \sqrt{2E/\hat q}$.\footnote{When considering a finite splitting fraction $z$ the exact definition turns out to be $L_c = \sqrt{E/(2 \hat q)}$.}

As the parton moves through the medium, at each instant $t$ the emission rate $\omega \rmd I/(\rmd \omega \rmd t)$ is different since the phase space for available emissions changes. The emission spectrum $\omega \rmd I/\rmd \omega$, evaluated at a given $t$ (typically $t=L$), includes the accumulated range of processes that occurred up to that time. The main goal of this work is to consider multiple medium-induced emissions in this system of dynamically evolving scales presented in Fig.~\ref{fig:PhaseSpace_PhysicalProcesses}. In the next section, Sec.~\ref{sec:spectrum}, we will formally derive the results we only have argued for in this section. Finally, in Sec.~\ref{sec:multiple} we will tackle the issue of multiple emissions in this scheme.

%%%%%%%%%%%%%%%%%%%%%%%%%%%%%%%%%%%%%%%%%%%%%%%%%%%%
%%%%%%%%%%%%%%%%%%%%%%%%%%%%%%%%%%%%%%%%%%%%%%%%%%%%
\section{Spectrum of medium-induced emissions}
\label{sec:spectrum}
%%%%%%%%%%%%%%%%%%%%%%%%%%%%%%%%%%%%%%%%%%%%%%%%%%%%
%%%%%%%%%%%%%%%%%%%%%%%%%%%%%%%%%%%%%%%%%%%%%%%%%%%%

In this section, we derive the all order emission spectrum induced by elastic scatterings on a deconfined medium. Of the three expansions we present here, the opacity expansion and the improved opacity expansion have been discussed in depth in previous works, see Refs.~\cite{Gyulassy:1999zd,Wiedemann:2000za} and \cite{Mehtar-Tani:2019tvy,Mehtar-Tani:2019ygg,Barata:2020sav,Barata:2020rdn,Barata:2021wuf} respectively. The resummed opacity expansion has been argued for before, see \cite{Wiedemann:2000za,Andres:2020kfg,Schlichting:2021idr}, but is here derived rigorously for the first time. This paper strives to be a comprehensive reference for all of the expansions, and hence they are all presented in detail. Furthermore, we extend previous calculations to all orders and present results to order $N = 2$ for the opacity expansion and $N_r = 2$ for the resummed opacity expansion, and extract the relevant limits analytically. This provides valuable insight into the underlying structure of the expansions in different regimes.

%%%%%%%%%%%%%%%%%%%%%%%%%%%%%%%%%%%%%%%%%%%%%%%%%%%%
\subsection{General formalism}
\label{sec:general_formalism}
%%%%%%%%%%%%%%%%%%%%%%%%%%%%%%%%%%%%%%%%%%%%%%%%%%%%

Currently, we consider the emission of a gluon with energy $\omega$ from a parent parton with energy $E$ in the soft limit, i.e. $\omega \ll E$. The soft limit is used in this section because it is much more clear and readable. For a description beyond the strictly soft limit we refer to App.~\ref{app:Kernels_finitez}, which includes novel results.

Our starting point is the definition of the spectrum of medium-induced gluons~\cite{Baier:1996kr,Zakharov:1997uu,Wiedemann:2000za,Arnold:2002ja},
\begin{equation}
\label{eq:medspec-coord}
    \omega \frac{\rmd I}{\rmd \omega} = \frac{2\alpha_s C_R}{\omega^2}\rmR \int_0^\infty \rmd t_2 \int^{t_2}_0 \rmd t_1\, \partial_\x \cdot \partial_\y \big[{\cal K}(\x,t_2; \y,t_1) - {\cal K}_0(\x,t_2; \y,t_1) \big]_{\x=\y=0} \,,
\end{equation}
where $C_R$ is the Casimir color factor of the emitting particle ($C_R = C_F$ for a quark and $C_R = N_c$ for a gluon).\footnote{This expression can be derived directly from the fully $z$ dependent spectrum in Eq.~\eqref{eq:Spectrum_def_finitez}, see the discussion in App.~\ref{app:Kernels_finitez}.} 
%\kt{Let us first explain the various elements entering this equation.}
The three-point correlator ${\cal K}$ solves the Schr{\"o}dinger equation
\begin{equation}
    \label{eq:schrodinger}
    \left[i\partial_t + \frac{\partial_\x^2}{2\omega} + i v(\x,t) \right]{\cal K}(\x,t; \y,t_0)= i \delta(t - t_0)\delta(\x-\y) \,,
\end{equation}
where the potential $v(\x,t)$ describes scatterings in a thermal or quasi-particle like background,\footnote{Throughout, we adopt a shorthand notation, so that $\int_p  =  \int  \frac{\rmd^4 p}{(2\pi)^4}$, $\int_{\p}  =  \int  \frac{\rmd^2 \p}{(2\pi)^2}$, and $\int_\x  =  \int \rmd^2 \x$.}
\begin{equation}\label{eq:Def_v_potential}
    v(\x,t) = \int_\q \sigma(\q,t) \left(1 - \rme^{i \q \cdot \x} \right) \,.
\end{equation}
Here $\sigma(\q,t) = N_c n(t) \rmd^2 \sigma_{\rm el}/\rmd^2\q$ is proportional to the in-medium elastic scattering cross section, where $n(t)\sim T^3$ is the density of scattering centers.\footnote{We include the number density of the scattering centers $n(t)$ into $v(\x,t)$ and $\sigma(\q,t)$ similarly to the previous works in Refs.~\cite{Mehtar-Tani:2019tvy,Mehtar-Tani:2019ygg,Barata:2021wuf}.} The color factor $N_c$ appears because, in this limit, only the emitted gluon picks up transverse momentum in the medium.
The potential can be extracted from an effective theory that accounts for both large and small momentum exchanges with the medium~\cite{Moore:2021jwe}. In the main part of this paper, we will use the Gyulassy-Wang potential~\cite{Gyulassy:1993hr} that both contains a hard Coulomb-tail and implements screening in the infrared $\mu^2 \gg \q^2$,
\begin{equation}
\label{eq:gw-potential}
    \frac{\rmd^2 \sigma_{\rm el}}{\rmd^2 \q} = \frac{g^4}{(\q^2 + \mu^2)^2}\,,
\end{equation}
where $\mu$ is a screening mass of the order of the Debye mass of a thermal medium. We also provide the spectrum in App.~\ref{app:htl} with the LO hard thermal loop (HTL) potential~\cite{Aurenche:2002pd}. Also, in Eq.~\eqref{eq:schrodinger} we neglect quark and gluon thermal masses, which corresponds to taking the high-energy limit ($E \gg m_{q,g}^4/\hat q \sim \omegaBH$), see e.g. Refs.~\cite{Arnold:2008iy,Schlichting:2021idr} for further discussion. 

When the medium is not present, $v = 0$, one recovers the propagation of a single parton in vacuum, $\Kc(\x,t_2;\y,t_1)  \equiv  \Kc_0(\x - \y,t_2 - t_1)$ with $\Kc_0(\x,t) =  \frac{\omega}{2\pi i \, t} \exp[i\,\omega \x^2/(2t)]$. To only capture medium effects, the vacuum term is explicitly subtracted in Eq.~\eqref{eq:medspec-coord}.

The emission spectrum in Eq.~\eqref{eq:medspec-coord} is the result of a path integral formalism in which arbitrarily many soft and hard scatterings are included. It does, however, not account for longitudinal momentum ($\sim$ energy) exchange with the medium. Having written the expression as a spectrum we also implicitly assume that the creation of the initial parton is factorized from the induced process (for example it was created in a highly virtual vacuum process). Finally, the medium averages leading to the simple form of the three-point correlator, as in Eq.~\eqref{eq:schrodinger}, assumes independent scatterings on the medium. This parametrically holds if the size of the potential is much smaller than the mean free path, i.e. $\mu^{-1} \ll \lambda$, where the typical exchanged momentum is $|\delta \bm q|\sim \mu$~\cite{Baier:1996kr}.

We should also note that Eq.~\eqref{eq:medspec-coord} emerges as the result of a momentum integral of the differential spectrum $\rmd I/(\rmd\omega\,\rmd^2\bm k)$ in the soft limit \cite{Salgado:2003gb,Barata:2021wuf}, with no kinematical constraint on the transverse momentum ($\bm k$) integral similarly to Refs.~\cite{Arnold:2002ja,Arnold:2002zm}. A more careful treatment of the kinematics would be important especially if one is interested in emissions inside or out of a given cone \cite{Salgado:2003gb,Andres:2020vxs,Takacs:2021bpv}.

Let us now cast the equation for the spectrum in an equivalent form. On many occasions it is more practical to work in transverse momentum space,
\begin{equation}
    {\cal K}(\p,t_2;\p_0,t_1) = \int_{\x,\y} \, \rme^{-i \p \cdot \x + i \p_0 \cdot \y} {\cal K}(\x,t_2;\y,t_1) \,.
\end{equation}
The vacuum propagator $\cK_0$ then becomes a plane wave, i.e. $\Kc_0(\p,t) = \exp[ -i\, \p^2t/(2 \omega)]$.
In this representation, the solution to the Schr\"odinger equation~\eqref{eq:schrodinger} can be written as the recursive equation
\begin{align}
    \label{eq:oe-mom}
    {\cal K}(\p,t_2;\p_0,t_1) &= (2\pi)^2 \delta(\p-\p_0){\cal K}_0(\p; t_2-t_1) \nn
    &- \int_{t_1}^{t_2}\rmd s \int_\q\, {\cal K}_0(\p;t_2-s) v(\q,s) {\cal K}(\p-\q,s; \p_0,t_1) \,,
\end{align}
where now
\begin{equation}
    \label{eq:v-mom}
    v(\q,s) = (2\pi)^2 \delta(\q)\Sigma(s) - \sigma(\q,s)  \,,
\end{equation}
ensures probability conservation. Here, $\Sigma(s) \equiv \int_\q \, \sigma(\q,s)$ can be interpreted as the inverse of the (local) mean free path $\mfp$ along a trajectory of a propagating parton, or
\begin{equation}
    \label{eq:mfp}
    \mfp(s) = \frac{1}{\Sigma(s)} \,.
\end{equation}
In these expressions, we have assumed that the integral over the elastic scattering cross section exists. In many cases, e.g. for the HTL potential \cite{Aurenche:2002pd}, one needs to introduce an IR regulator. However, $v(\q,s)$ in Eq.~\eqref{eq:v-mom} is not sensitive to this IR regulation and therefore the expansion in Eq.~\eqref{eq:oe-mom} is well-defined. We have provided a further discussion of the HTL potential in App.~\ref{app:htl}.

The medium-induced spectrum now reads
\begin{equation}
\label{eq:medspec-mom}
    \omega \frac{\rmd I}{\rmd \omega} = \frac{2\alpha_s C_R}{\omega^2}\rmR \int_0^\infty \rmd t_2 \int^{t_2}_0 \rmd t_1 \int_{\p,\q}\, \p\cdot \q\,\big[{\cK}(\p,t_2; \q,t_1)- (2\pi)^2 \delta(\p-\q) {\cal K}_0(\p,t_2-t_1) \big] \,,
\end{equation}
The vacuum contribution can then be removed by inserting Eq.~\eqref{eq:oe-mom} into Eq.~\eqref{eq:medspec-mom}, yielding
\begin{equation}
    \omega \frac{\rmd I}{\rmd \omega} = \frac{4 \alpha_s C_R}{\omega} \rmR \,i \int_0^L \rmd t_2 \int_0^{t_2} \rmd t_1 \int_{\p,\p_0,\q} \, \frac{\p \cdot \p_0}{\p^2} v(\q,t_2) \Kc (\p-\q,t_2;\p_0,t_1) \,,
\end{equation}
where we regulated the integral over the latter time coordinate using an adiabatic turn-off at infinity (see also Ref.~\cite{Andres:2020kfg}). The other time integrals are limited by the extent of the medium $L$.
Noticing, that
\beq
    \int_{\p}\, \frac{p^i}{\p^2} v(\p-\k,s) = \frac{k^i}{\k^2} \Sigma(\k^2,s) \,,
\eeq
where $\Sigma(\k^2,s) = \int_{\q}\, \sigma(\q,s)\Theta(\q^2-\k^2)$,\footnote{Also $\Sigma(0,s) = \Sigma(s)$, consistent with the definition in Eq.~\eqref{eq:v-mom}.} we obtain
\begin{equation}
\label{eq:medspec-simplified}
    \omega \frac{\rmd I}{\rmd \omega} 
    = \frac{4 \alpha_s C_R}{\omega} \rmR \,i \int_0^L \rmd t_2 \int_0^{t_2} \rmd t_1 \int_{\p,\p_0} \, \Sigma(\p^2,t_2)\frac{\p\cdot\p_0}{\p^2} \Kc (\p,t_2;\p_0,t_1)\,.
\end{equation}
While the above results are valid for any medium potential, in this work we will focus on the GW scattering potential, defined in Eq.~\eqref{eq:gw-potential}. In this case, we find that
\begin{equation}
    \label{eq:Sigma-GW}
    \Sigma(\k^2,s) = \frac{\hat q_0(s)}{\k^2+\mu^2} \,,
\end{equation}
where $\qhat_0(s) =  4\pi \alpha_s^2 N_c n(s)$ is a measure of the scattering density. Currently, we consider a medium of constant density, $n(s)  =  n_0$.

The spectrum, given by Eq.~\eqref{eq:medspec-coord} or Eq.~\eqref{eq:medspec-simplified}, can be evaluated using numerical techniques \cite{Zakharov:2004vm,CaronHuot:2010bp,Feal:2018sml,Ke:2018jem,Andres:2020vxs,Schlichting:2020lef} or by employing analytical approximations. As we will see the approximate approaches rely on expanding the problem as a series, which will give the true answer at infinite order. The different series have different radii of convergence, and none of them will alone converge for all $L$ and $\omega$, meaning more than one have to be employed. In most cases, however, the first order expansion is sufficient to provide an accurate approximation of the all order result. In the following we discuss three well-defined approaches that together provide an accurate description of the true problem for all $L$ and $\omega$, called: the opacity expansion, the resummed opacity expansion, and the improved opacity expansion. We will derive these, discuss their limits and their regions of validity. 

We also point out that the medium parameters for the numerical evaluations in Figs.~\ref{fig:PhaseSpace_dIdw_OE}-\ref{fig:PhaseSpace_dIdw_IOE} (right) are chosen to maximally separate the relevant scales and to illustrate the main features of the spectrum. They are also similar to the ones used in phenomenological studies~\cite{Takacs:2021bpv,Barata:2021wuf,Caucal:2021cfb}. It is worth pointing out that, although this particular choice violates the assumption of non-overlapping scattering centers and should be treated with care, changing the values of the parameters would not alter the qualitative picture of separating different regimes in the $(\omega,t)$ plane.

%%%%%%%%%%%%%%%%%%%%%%%%%%%%%%%%%%%%%%%%%%%%%%%%%%%%
\subsection{Opacity expansion (OE)}
\label{sec:oe}
%%%%%%%%%%%%%%%%%%%%%%%%%%%%%%%%%%%%%%%%%%%%%%%%%%%%

The opacity expansion of the spectrum arises when inserting Eq.~\eqref{eq:oe-mom} directly into Eq.~\eqref{eq:medspec-simplified}, and was developed in Refs.~\cite{Gyulassy:2000er,Wiedemann:2000za}.\footnote{To be precise, expanding our formulas order by order in opacity reproduces the expansion defined in Ref.~\cite{Wiedemann:2000za}, which reproduces Ref.~\cite{Gyulassy:2000er} in the ``incoherent'' limit.} The truncation of this series at a given order $n$ in the medium scattering potential gives the $N = n$ term, which is by definition proportional to $(L/\lambda)^n$ (see Eq.~\eqref{eq:general-OE}). Physically this means, at $N=n$ one counts $n$ number of scatterings (both with and without momentum exchange) on the full elastic potential. The relevant energy scales that arise are $\bar\omega_c = \frac12\mu^2L$, and $\frac{L}{\lambda}\bar\omega_c = \frac12\hat q_0L^2$ as discussed in Sec.~\ref{sec:heuristic}. A general formula for the spectrum at any order is derived in App.~\ref{app:general-formula}, and with finite-$z$ corrections in App.~\ref{app:Kernels_finitez}. These results are used in the following calculations, and we will refer to the appendices for more details.

%%%%%%%%%%%%%%%%%%%%%%%%%%%%%%%%%%%%%%%%%%%%%%%%%%%%
\paragraph{First order ($N = 1$):}
%%%%%%%%%%%%%%%%%%%%%%%%%%%%%%%%%%%%%%%%%%%%%%%%%%%%
The spectrum at first order of opacity is well known~\cite{Gyulassy:1999zd,Wiedemann:2000za}. Since Eq.~\eqref{eq:medspec-simplified} already includes at least one scattering, we obtain the $N = 1$ term by replacing the full propagator $\Kc$ by the vacuum one. We then find,
\begin{align}
    \omega \frac{\rmd I^{N=1}}{\rmd \omega} = 8 \pi \abar \frac{L}{\lambda}\frac{\omegacbar}{\omega} \int_\p\, \tilde\Sigma(\p^2) \,\rmR \,i \int_0^1 \rmd t_1 \int_0^{t_1} \rmd t_0 \,\rme^{-i \p^2(t_1-t_0)}\,,
\end{align}
where we have switched to dimensionless integration variables by defining $\p^2  \to \p^2 L/(2 \omega)$ and $t  \to  t/L$, and where $\tilde\Sigma(\p^2) = (\p^2+\omegacbar/\omega)^{-1}$. This expression can also be obtained from the general $N = n$ result in  Eq.~\eqref{eq:general-OE}. After simplifications, the spectrum becomes
\begin{align}\label{eq:dIdomega-N=1}
    \omega \frac{\rmd I^{N=1}}{\rmd \omega} = 
    2 \abar\frac{L}{\lambda} \frac{\omegacbar}{\omega} \int_0^\infty \rmd p \, \frac{1}{p+\frac{\omegacbar}{\omega}}\frac{p-\sin{p}}{p^2} \,,
\end{align}
where $\abar= \alpha_s C_R/\pi$ and $\lambda = \mu^2/\qhat_0$. It also agrees with Eq.~(6.7) in Ref.~\cite{Wiedemann:2000za} (see also in Ref.~\cite{Salgado:2003gb}). We recognize the dependence on the medium opacity $L/\lambda$ and the ratio $\omegacbar/\omega$. The remaining integral can be done analytically, but the resulting expression is not very illuminating. However, the limiting behavior can readily be extracted,
\begin{equation}
    \label{eq:n1-limits}
    \omega \frac{\rmd I^{N=1}}{\rmd \omega}\simeq 
    \begin{cases}
        2\abar \frac{L}{\mfp} \left(\ln\frac{\omegacbar}{\omega}-1+\gamma_E\right), &\qquad \text{for } \omega\ll\omegacbar\,, \\[0.2pt]
        \frac{\pi}{2}\abar \frac{L}{\mfp} \frac{\omegacbar}{\omega}, &\qquad \text{for } \omega\gg\omegacbar\,.
    \end{cases}
\end{equation}
This agrees well with the heuristic discussion in Sec.~\ref{sec:heuristic}. In particular, we identify a logarithmic behavior $\sim  \ln \frac{\omegacbar}{\omega}$ in the infrared. Notice the different expansion structures in the soft $\sim \bar\alpha\frac{L}{\lambda}$ and in the hard $\sim \bar\alpha\frac{L}{\lambda}\frac{\bar\omega_c}{\omega}$ limits, which we will come back to.

%%%%%%%%%%%%%%%%%%%%%%%%%%%%%%%%%%%%%%%%%%%%%%%%%%%%
\paragraph{Second order ($N = 2$):}
%%%%%%%%%%%%%%%%%%%%%%%%%%%%%%%%%%%%%%%%%%%%%%%%%%%%
The calculation for $N = 2$ follows in a similar way, leading to
\begin{align}
    \omega \frac{\rmd I^{N=2}}{\rmd \omega}
    &= -8 \pi \abar \left(\frac{L}{\lambda}\right)^2 \frac{\omegacbar}{\omega}
    \int_{\p_2,\p_1} \, \tilde\Sigma(\p_2^2)\frac{\p_2\cdot\p_1}{\p_2^2} \tilde v(\p_2-\p_1)\nn
    &\times \rmR \,i\int_0^1 \rmd t_2 \int_0^{t_2} \rmd t_1 \int_0^{t_1} \rmd t_{0}\,
    \rme^{-i \p_2^2(t_2-t_1)}\rme^{-i \p_1^2(t_1-t_0)}\,,
\end{align}
with dimensionless integration variables, and where $\tilde v(\p) = (2\pi)^2\delta(\p) - \frac{\omegacbar}{\omega}\tilde\sigma(\p)$. In the GW model, $\tilde\sigma(\p)=4 \pi/(\p^2+\frac{\omegacbar}{\omega})^2$.
After inserting $\tilde v$, doing the time integrals and simplifying this can be written as
\begin{equation}
\label{eq:dIdomega-N=2}
    \omega \frac{\rmd I^{N=2}}{\rmd \omega}= -4\bar \alpha \left(\frac{L}{\lambda}\right)^2\frac{\omegacbar}{\omega}\left[\Ic_1\left(\frac{\omegacbar}{\omega}\right)-\frac{\omegacbar}{\omega}\Ic_2\left(\frac{\omegacbar}{\omega}\right)\right]\,,
\end{equation}
where we have defined the integrals
\begin{align}
   \Ic_1\left(\frac{\omegacbar}{\omega}\right)&= \int_0^\infty \rmd p \, \frac{1}{p+\frac{\omegacbar}{\omega}} \frac{1-\cos p -\frac{p}{2}\sin p}{p^3}\,,\\
    \Ic_2\left(\frac{\omegacbar}{\omega}\right)&= \int_0^\infty \rmd p_2 \int_0^\infty \rmd p_1 \, \frac{p_1}{p_2+\frac{\omegacbar}{\omega}} \frac{1}{\left[\left(p_2+p_1+\frac{\omegacbar}{\omega}\right)^2-4 p_2 p_1\right]^{3/2}}\nn
    &\times\frac{1}{p_2-p_1} \left[\frac{1}{p_1^2}\left(1-\cos{p_1}\right)-\frac{1}{p_2^2}\left(1-\cos{p_2}\right)\right]\,.
\end{align}
The $\Ic_1$ integral can be done analytically, but $\Ic_2$ is more complicated. It can be shown that it is much smaller than $\Ic_1$ in the soft limit. In the hard limit, $\Ic_1$ and $\Ic_2$ cancel at the order of $\mathcal O(\frac{\omegacbar}{\omega})$, leaving a positive contribution going as $\mathcal O(\frac{\omegacbar}{\omega})^2$. In summary,
\begin{align}
    \label{eq:n2-limits}
    \omega \frac{\rmd I^{N=2}}{\rmd \omega}\simeq 
    \begin{cases}
        -\abar\left(\frac{L}{\lambda}\right)^2 , & \text{for } \omega\ll\omegacbar\,, \\[0.5pt]
        \sim \abar \left(\frac{L}{\lambda}\right)^2\left(\frac{\omegacbar}{\omega}\right)^2, & \text{for } \omega\gg\omegacbar\,.
    \end{cases}
\end{align}
We notice that the $N = 2$ is proportional to $\bar\alpha(\frac{L}{\lambda})^2$ in the soft limit, and goes like $\bar\alpha(\frac{L}{\lambda}\frac{\bar\omega_c}{\omega})^2$ in the hard limit. This immediately implies that $N = 2$ is always subleading to $N = 1$ if the medium is dilute $L \ll \lambda$ or if the emission is hard $\omega \gg \frac{L}{\lambda}\bar\omega_c$. Given the structure of the expansion, we expect the previous statement to hold at arbitrary $N = n$ order. This is in agreement with the earlier, heuristic observation in Refs.~\cite{Gyulassy:2000fs,Arnold:2008iy}. 

\begin{figure}
    \centering
     \includegraphics[width=0.45\textwidth]{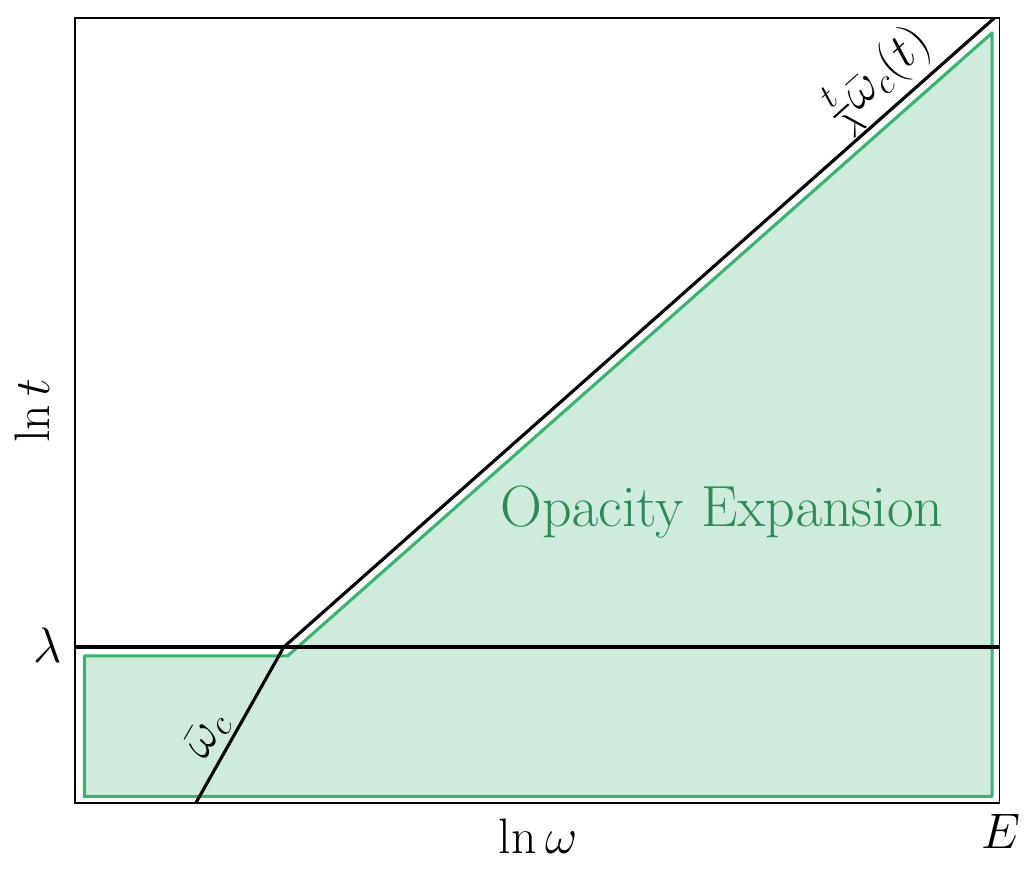} 
     \includegraphics[width=0.45\textwidth]{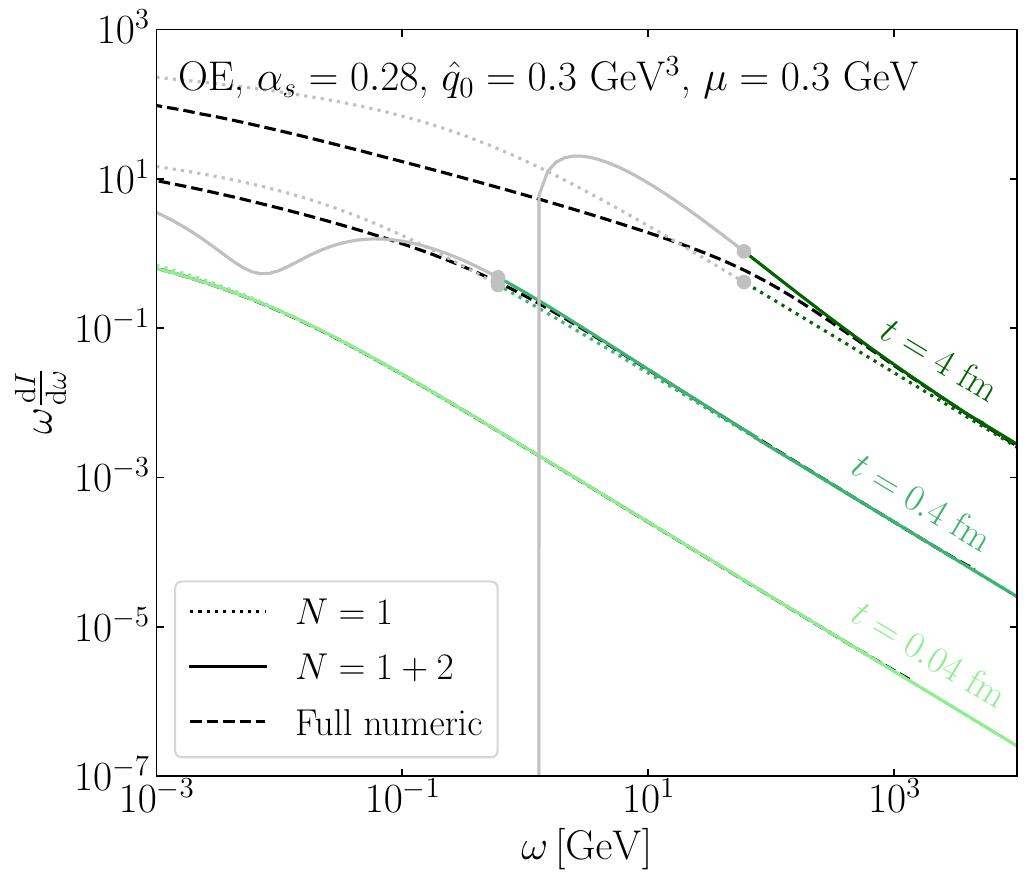} 
     \caption{{\it Left:} The sketch of the region of validity (and convergence) of the opacity expansion for different propagation length $t$ and emission energy $\omega$. {\it Right:} The induced emission spectrum for gluons in the opacity expansion. The gray part of the curves denotes regions, where the expansion is not valid. Using the parameters presented, $\lambda = 0.06$ fm. The full numeric solution is also presented with dashed lines.}
     \label{fig:PhaseSpace_dIdw_OE}
\end{figure}

Based on the limits for $N = 1$ and $N = 2$, given by Eqs.~\eqref{eq:n1-limits} and \eqref{eq:n2-limits}, in the regimes where the expansion holds the all order OE spectrum is expected to take the form
\begin{equation}
    \omega \frac{\rmd I}{\rmd \omega} =
    \begin{cases}
        \mathlarger{\bar\alpha\sum_{n=1}^\infty \left(\frac{L}{\lambda}\right)^n h_n\left(\frac{\omega}{\bar \omega_c} \right)}\,, & \omega \ll \omegacbar\,, \\
        \mathlarger{\bar\alpha\sum_{n=1}^\infty \left(\frac{L}{\lambda}\frac{\bar\omega_c}{\omega}\right)^n \tilde h_n\left(\frac{\bar\omega_c}{\omega} \right)}\,, & 
        \omega \gg \omegacbar\,,
    \end{cases} 
\end{equation}
where the OE coefficients $h_n,\tilde h_n$ are finite and can be calculated order by order. Note that we have not strictly proven this for all orders, although our $N = 1$ and $N = 2$ results strongly indicate this structure. In the soft limit $\omega \ll \omegacbar$, the OE expansion converges rapidly, defining the expected ``naive'' radius of convergence $L/\lambda < 1$. However, in the hard limit $\omega \gg \bar\omega_c$ there is convergence even if the medium is big, provided $\frac{L}{\lambda}\frac{\omegacbar}{\omega}<1$. The full region of convergence is shown in green in the left panel of Fig.~\ref{fig:PhaseSpace_dIdw_OE}. Outside of this region we expect higher orders to grow uncontrollably and hence the OE is not valid when truncated at any finite order.

The resulting spectrum from Eqs.~\eqref{eq:dIdomega-N=1} and \eqref{eq:dIdomega-N=2} is shown in the right panel of Fig.~\ref{fig:PhaseSpace_dIdw_OE} for different propagation lengths (labeled with $t$). For short lengths $t < \lambda$ the OE is valid for all $\omega$. For $t > \lambda$, the OE is only valid if $\frac{t}{\lambda}\frac{\omegacbar}{\omega}<1$ (see also the green region on the left panel). We note that the $N = 2$ correction becomes important at $t > \lambda$ and $\omega \approx \frac{t}{\lambda}\bar\omega_c$ (the latter constraint shown as bullets in the figure). For larger media, the grey bullets, representing the minimal energy for achieving convergence, moves to higher values, and the truncated OE series at $\omega$ smaller than this becomes ill-defined. This can be seen, for instance, in the upper line in Fig.~\ref{fig:PhaseSpace_dIdw_OE} (right) for a medium length of $t=4$ fm.
We have also compared to a full numerical evaluation of the spectrum from Ref.~\cite{Andres:2020vxs,Andres:2022}. The figure shows that this indeed is well approximated by the OE in its region of validity, as we have argued. Our curves are not expected to hold in the limit $\omega\ll1$ GeV, where several important effects were not taken into account such as thermal masses, realistic 2-2 elastic scatterings, and other non-perturbative effects. We still plot the curves down to very small $\omega$ to compare the different expansion schemes.

%%%%%%%%%%%%%%%%%%%%%%%%%%%%%%%%%%%%%%%%%%%%%%%%%%%%
\subsection{Resummed opacity expansion (ROE)}
\label{sec:roe}
%%%%%%%%%%%%%%%%%%%%%%%%%%%%%%%%%%%%%%%%%%%%%%%%%%%%

Next, we turn to dense media, $L \gg  \lambda$, where multiple scattering have to be accounted for. However, as depicted to the left in Fig.~\ref{fig:regimes-illustration} and discussed in Sec.~\ref{sec:heuristic}, for soft emissions with short formation times, a single scattering still gives the leading contribution to the spectrum. This defines the so-called Bethe--Heitler regime named after the QED analogue of this process. A qualitative argument of this regime was first given in Ref.~\cite{Wiedemann:2000za} and later developed in Ref.~\cite{Andres:2020kfg}, see also in Ref.~\cite{Schlichting:2021idr} which coined the name ``resummed opacity expansion'' (ROE).\footnote{In Ref.~\cite{Baier:1996kr}, authors derive the opacity expansion from the all-order formula Eq.~\eqref{eq:medspec-coord} in a similar way as we did by expanding $\cK$. Accidentally, in one of their intermediate steps in Sec. 4, they kept the virtual interactions resummed, which corresponds to our ROE. Back then, however, they did not realize the importance of that formula and they expanded it to reproduce the OE.}

Here, we provide for the first time a consistent framework for dealing with an expansion of the {\it real} scatterings with the medium, whereby real we mean interactions with a finite transverse momentum exchange. At the same time, an all-order resummation of the corresponding {\it virtual} interactions, with zero transverse momentum exchange, is performed.

Dividing Eq.~\eqref{eq:oe-mom} by the vacuum propagator leaves us with
\begin{align}
    \frac{{\cal K}(\p,t;\p_0,t_0)}{{\cal K}_0(\p;t-t_0)} &= (2\pi)^2 \delta(\p-\p_0) - \int_{t_0}^{t}\rmd s\int_\q \, v(\q,s) \frac{{\cal K}(\p-\q,s;\p_0,t_0)}{{\cal K}_0(\p;s-t_0)}\,.
\end{align}
Next, taking a derivative with respect to the latest time results in 
\begin{equation}
    \frac{\partial}{\partial t} \frac{{\cal K}(\p,t;\p_0,t_0)}{{\cal K}_0(\p;t-t_0)} + \Sigma(t) \frac{{\cal K}(\p,t;\p_0,t_0)}{{\cal K}_0(\p;t-t_0)} = \int_\q \,\sigma(\q) \frac{{\cal K}(\p-\q,t;\p_0,t_0)}{{\cal K}_0(\p;t-t_0)} \,,
\end{equation}
where $\Sigma(t)  =  \int_\q \, \sigma(\q,t)$, as before.
This motivates defining the elastic Sudakov factor,
\begin{equation}
    \label{eq:sudakov}
    \Delta(t,t_0) \equiv \rme^{-\int_{t_0}^{t} \rmd s\, \Sigma(s)}=\rme^{- (t-t_0)\Sigma} \,,
\end{equation}
where the last equality holds for media with constant density.\footnote{For medium potentials with unscreened soft divergences, such as the HTL potential, one has to modify this prescription to include an IR regulator. We refer to App.~\ref{app:htl} for a further discussion.} This represents the probability of no elastic scattering occurring between times $t_0$ and $t$.
Integrating out the time, we arrive at a slightly modified iterative equation
\begin{align}
    \label{eq:eoe-mom}
    {\cal K}(\p,t;\p_0,t_0) &= (2\pi)^2 \delta(\p-\p_0)\Delta(t,t_0){\cal K}_0(\p; t-t_0) \nn
    &+ \int_{t_0}^{t}\rmd s \frac{\Delta(t,t_0)}{\Delta(s,t_0)} \int_\q\, {\cal K}_0(\p;t-s) \sigma(\q,s) {\cal K}(\p-\q,s; \p_0,t_0) \,.
\end{align}
Compared to the standard opacity expansion in Eq.~\eqref{eq:oe-mom}, this looks very similar. However, the expansion is not in the potential $v(\q,s)$, which contains both a ``real'' and a ``virtual'' part, but in the potential scattering $\sigma(\q,s)$ which comes from the term that provides a finite momentum transfer in the process. The virtual contributions, where no net momentum was exchanged,  are accounted for to all orders in the Sudakov factor. This is why this expansion referred as a {\it resummed} opacity expansion. 

The relevant scale that appears at high opacity is $\omegaBH = \frac12\mu^2\lambda$, as we discussed in Sec.~\ref{sec:heuristic}. Interestingly, at low opacity $L \ll \lambda$, the scale changes to $\bar\omega_c = \frac12\mu^2L$, which we recognize from the opacity expansion. In this regime the ROE is actually equivalent to the OE if one gathers all the terms up to the same order in $(\frac{L}{\lambda})^n$. However, as we will see, the terms are reshuffled in the ROE compared to the OE. 

It is possible to reach a general formula for the resummed opacity expansion at arbitrary order. This was done in App.~\ref{app:general-rOE}, and we refer to that section for detailed calculations.

%%%%%%%%%%%%%%%%%%%%%%%%%%%%%%%%%%%%%%%%%%%%%%%%%%%%
\paragraph{First order ($N_r = 1$):}
%%%%%%%%%%%%%%%%%%%%%%%%%%%%%%%%%%%%%%%%%%%%%%%%%%%%
The first order can be obtained from Eq. \eqref{eq:general-rOE} with $n = 1$, and reads
\begin{align}
    \label{eq:nr1-1}
    \omega \frac{\rmd I^{N_r=1}}{\rmd \omega}
    = - 8 \pi \abar \frac{L}{\lambda} \frac{\omegacbar}{\omega}\int_{\p} \, \tilde\Sigma(\p^2)\, \rmI\,T\left(\p^2-i\frac{L}{\mfp} \right)\,,
\end{align}
in re-scaled, dimensionless variables. Equation~\eqref{eq:nr1-1} corresponds to the formula (Eq.~(4.6)) in Ref.~\cite{Andres:2020kfg}, but is here derived more rigorously. Here, we have defined the function
\begin{align}
    \label{eq:t-function}
    T(x)=\int_0^1 \rmd t_1 \int_0^{t_1} \rmd t_0 \, \rme^{-ix(t_1-t_0) } = \frac{1-ix-\rme^{-i x}}{x^2}\,.
\end{align}
The real and imaginary parts of $T(\p^2-i\chi)$ are given in Eq. \eqref{eq:T-real-imaginary}. After doing the angular integral this becomes
\begin{align}
    \label{eq:nr1-2}
    \omega \frac{\rmd I^{N_r=1}}{\rmd \omega}
    = -2 \abar \frac{L}{\lambda} \frac{\omegacbar}{\omega}\int_0^\infty \rmd p \, \frac{1}{p+\frac{\omegacbar}{\omega}}\, \rmI \,T\left( p - i \frac{L}{\mfp}\right) \,.
\end{align}
At low opacity $L \ll \mfp$, the function $T(p - i L/\mfp)$ becomes
\begin{equation}
    \label{eq:tfunc-small}
    \left. -\rmI\,T(p)\right|_{L \ll \mfp} = \frac{p-\sin(p)}{p^2} \,,
\end{equation}
making it equivalent to the OE result in Eq.~\eqref{eq:dIdomega-N=1}. The limiting behavior in the relevant limits of Eq.~\eqref{eq:nr1-2} can be extracted, leading to
\begin{align}
\label{eq:nr1-limit}
    \omega \frac{\rmd I^{N_r=1}}{\rmd \omega}\simeq 
    \begin{cases}
        2 \abar\frac{L}{\lambda} \left(\ln\left(\frac{\omegacbar}{\omega}\right)-1+\gamma_E \right)-\abar \left(\frac{L}{\lambda}\right)^2\left(1-\pi \frac{\omega}{\omegacbar}\right), & \text{for } \omega\ll\omegacbar\,, \\[0.5pt]
        \frac{\pi \abar}{2} \frac{L}{\lambda} \frac{\omegacbar}{\omega}-\frac{\pi}{6}\abar \left(\frac{L}{\lambda}\right)^2\frac{\omegacbar}{\omega} , & \text{for } \omega\gg\omegacbar\,.
    \end{cases}
\end{align}
At leading order in $\mathcal O(\frac{L}{\lambda})$ this is the same as the $N = 1$ opacity expansion, presented in Eq.~\eqref{eq:n1-limits}. However, in contrast to the OE, subleading ``$N=2$''-like terms $\sim (\frac{L}{\lambda})^2$ appear, which only will be relevant when compared to higher-order contributions at $N_r = 2$. 

In the high opacity limit $L \gg \lambda$, we have to extract the relevant limit of $T(p - i\chi)$ in a careful way, yielding
\begin{equation}
    \label{eq:tfunc-large}
    \left.-\rmI\, T(p- i\chi) \right|_{L\gg \mfp} \simeq \frac{p}{(L/\mfp)^2 + p^2} \,.
\end{equation}
Changing the integration variable to $q = p \mfp/L$, we observe that $\omegaBH$ replaces $\omegacbar$ as the relevant scale, and Eq.~\eqref{eq:nr1-2} becomes
\begin{align}
    \omega \frac{\rmd I^{N_r=1}}{\rmd \omega} &\simeq 
    2 \abar\frac{L}{\lambda} \frac{\omegaBH}{\omega} \int_0^\infty \rmd q \, \frac{1}{q+\frac{\omegaBH}{\omega}}\frac{q}{1+q^2}= 2 \abar\frac{L}{\lambda} \frac{\omegaBH}{\omega} \frac{\frac{\pi}{2}+\frac{\omegaBH}{\omega}\ln\left(\frac{\omegaBH}{\omega}\right)}{1+\left(\frac{\omegaBH}{\omega}\right)^2}\,.
\end{align}
Finally, one can extract the soft and hard limits of this expression, which are given by
\begin{align}\label{eq:nr1-limit-highop}
    \omega \frac{\rmd I^{N_r=1}}{\rmd \omega}\simeq 
    \begin{cases}
    2 \abar\frac{L}{\lambda} \left(\ln\left(\frac{\omegaBH}{\omega}\right)+\frac{\pi}{2}\frac{\omega}{\omegaBH}\right), & \text{for } \omega\ll\omegaBH\,,\\[0.5pt]
    \pi \abar \frac{L}{\lambda} \frac{\omegaBH}{\omega}, & \text{for } \omega\gg\omegaBH \,.
    \end{cases}
\end{align}
The soft limit agrees with the heuristic discussion in Sec.~\ref{sec:heuristic}. Strikingly, we see that the behavior in the soft and hard limit takes exactly the same form as for $N = 1$ except that $\omegacbar$ has been replaced by $\omegaBH$ (note that $\bar\omega_c(L = \lambda) = \omegaBH$). 

%%%%%%%%%%%%%%%%%%%%%%%%%%%%%%%%%%%%%%%%%%%%%%%%%%%%
\paragraph{Second order ($N_r = 2$):}
%%%%%%%%%%%%%%%%%%%%%%%%%%%%%%%%%%%%%%%%%%%%%%%%%%%%
The second order is found from Eq.~\eqref{eq:general-rOE} with $n = 2$, and reads
\begin{align}
    \omega \frac{\rmd I^{N_r=2}}{\rmd \omega}
    &= 8 \pi \abar \left(\frac{L}{\lambda}\right)^2 \left(\frac{\omegacbar}{\omega}\right)^2 \int_{\p_2,\p_1} \, \tilde\Sigma(\p_2^2)\frac{\p_2\cdot\p_1}{\p_2^2} \tilde \sigma(\p_2-\p_1)\nn
    &\times \frac{1}{\p_2^2-\p_1^2}
    \left(\rmR \,T(\p_1^2-i\chi)-\rmR \,T(\p_2^2-i\chi)\right)\,.
\end{align}
After going to polar coordinates and doing the angular integrals, this becomes
\begin{align}
    \omega \frac{\rmd I^{N_r=2}}{\rmd \omega} 
    &=4 \abar \left(\frac{L}{\lambda}\right)^2\left(\frac{\omegacbar}{\omega}\right)^2 
    \int_0^\infty \rmd p_2 \int_0^\infty \rmd p_1 \,
    \frac{1}{p_2+\frac{\omegacbar}{\omega}} \frac{p_1}{\left[\left(p_1+p_2+\frac{\omegacbar}{\omega}\right)^2-4 p_1 p_2\right]^{3/2}}\nn
    &\times \frac{1}{p_2-p_1}\left(\rmR \,T(p_1-i\chi)-\rmR \,T(p_2-i\chi)\right)\,.
\end{align}
We study this expression separately in the low- and high-opacity limits.

In the low opacity limit $L  \ll  \lambda$, the spectrum becomes
\begin{align}
    \omega \frac{\rmd I^{N_r=2}}{\rmd \omega} &\simeq 4 \abar \left(\frac{L}{\lambda}\right)^2\left(\frac{\omegacbar}{\omega}\right)^2 
    \int_0^\infty \rmd p_2 \int_0^\infty \rmd p_1 \,
    \frac{1}{p_2+\frac{\omegacbar}{\omega}} \frac{p_1}{\left[\left(p_1+p_2+\frac{\omegacbar}{\omega}\right)^2-4 p_1 p_2\right]^{3/2}}\nn
    &\times \frac{1}{p_2-p_1}\left(\frac{1-\cos{p_1}}{p_1^2}-\frac{1-\cos{p_2}}{p_2^2}\right)\,,
\end{align}
where again the only relevant energy scale is $\omegacbar$, as it is in the OE.
The double momentum integral can be recognized as $\mathcal{I}_2$ from $N = 2$ of the opacity expansion. The soft and hard limits are
\begin{align}
\label{eq:nr2-limit}
    \omega \frac{\rmd I^{N_r=2}}{\rmd \omega}\simeq 
    \begin{cases}
        \pi\abar\left(\frac{L}{\lambda}\right)^2\frac{\omega}{\omegacbar}, & \qquad\text{for } \omega\ll\omegacbar \,, \\[0.5pt]
        \frac{\pi}{6} \abar \left(\frac{L}{\lambda}\right)^2 \frac{\omegacbar}{\omega}, & \qquad\text{for } \omega\gg\omegacbar \,.
    \end{cases}
\end{align}
Summing up the two first orders of the ROE and OE we see that $N_r = 1 + 2$ agrees with $N = 1 + 2$, but only when keeping the subleading $\sim (L/\lambda)^2$ terms at order $N_r = 1$. As mentioned before, the opacity expansion is arranged so that the order $N = n$ only contains terms where the opacity scales as $\sim\chi^n$, where $\chi = L/\lambda$. The resummed opacity expansion also includes all of the same terms, but they are spread out over different orders of the expansion due to the resummation contained in the Sudakov factor. The orders $N_r < n$ do contain terms going as $\chi^n$. To get the right term at order $\chi^n$ in the ROE one therefore has to keep the subleading corrections going as $\chi^n$ at all previous orders of the expansion. For this reason the opacity expansion is more convenient to use in the low opacity limit, as it does not mix orders of opacity.

\begin{figure}
    \centering
     \includegraphics[width=0.45\textwidth]{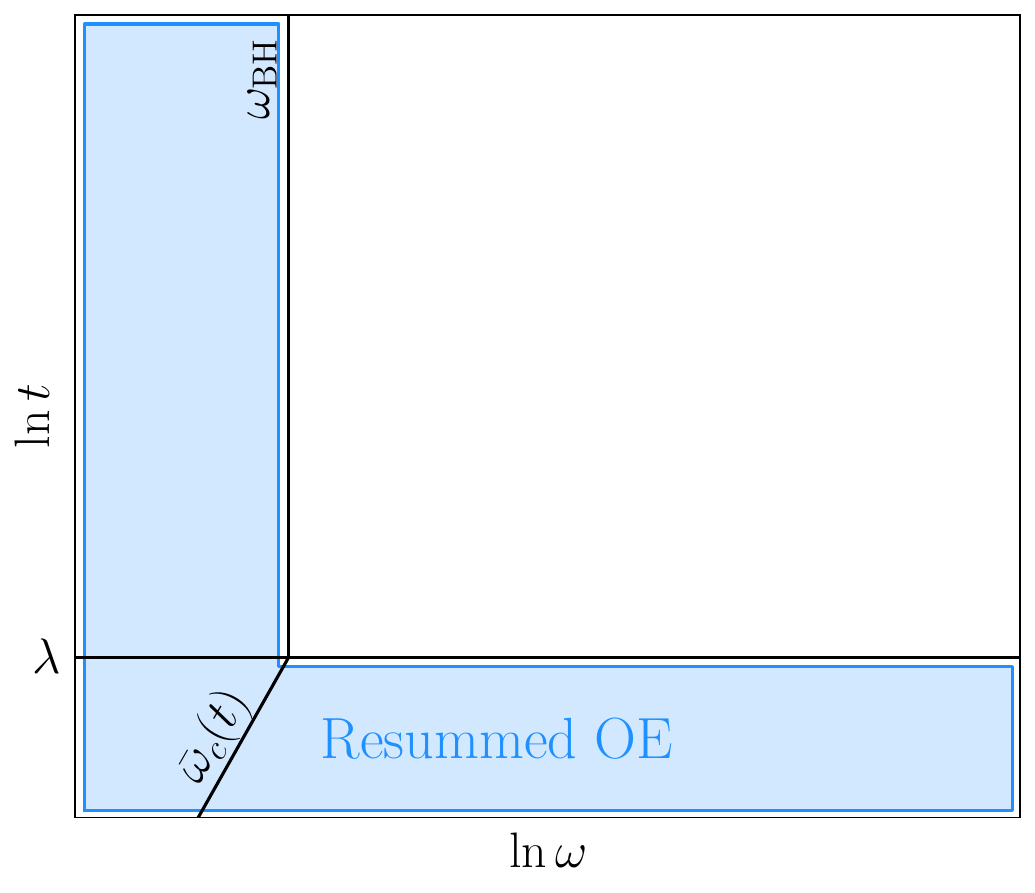} 
     \includegraphics[width=0.45\textwidth]{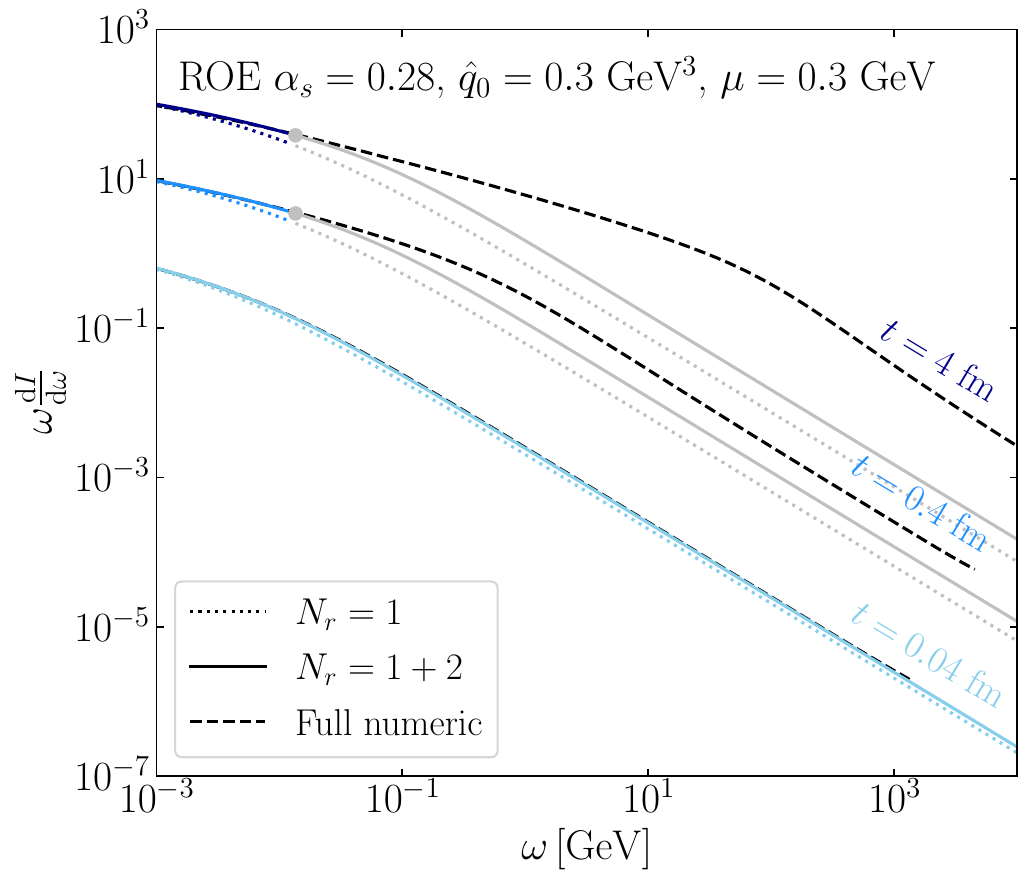} 
     \caption{{\it Left:} The sketch of region of validity (and convergence) of the resummed opacity expansion for different propagation length $t$ and emission energy $\omega$. {\it Right:} The induced emission spectrum for gluons in the resummed opacity expansion. The gray part of the curves denotes regions, where the expansion is not valid. With the parameters presented, $\lambda = 0.06$ fm. The full numeric solution is also presented with dashed lines.}
     \label{fig:PhaseSpace_dIdw_ROE}
\end{figure}

In the high opacity limit $L \gg  \mfp$, we get
\begin{align}
    \omega \frac{\rmd I^{N_r=2}}{\rmd \omega} &\simeq 4 \abar \frac{L}{\lambda}\left(\frac{\omegaBH}{\omega}\right)^2
    \int_0^\infty \rmd p_2 \int_0^\infty \rmd p_1 \,
    \frac{1}{p_2+\frac{\omegacbar}{\omega}} \frac{p_1}{\left[\left(p_1+p_2+\frac{\omegacbar}{\omega}\right)^2-4 p_1 p_2\right]^{3/2}}\nn
    &\times \frac{p_2+p_1}{(1+p_2^2)(1+p_1^2)}\,.
\end{align}
The soft and hard limits of this expression are given by
\begin{align}
    \omega \frac{\rmd I^{N_r=2}}{\rmd \omega}\simeq 
    \begin{cases}
        \pi\abar\frac{L}{\lambda}\frac{\omega}{\omegaBH}, & \qquad\text{for } \omega\ll\omegaBH\,,\\
        \pi \abar \frac{L}{\lambda} \frac{\omegaBH}{\omega}, & \qquad\text{for } \omega\gg\omegaBH\,,
    \end{cases}
\end{align}
where similarly to $N_r = 1$, the relevant scale is now $\omegaBH$. Both $N_r = 1$ and 2 goes as $\sim \bar\alpha\frac{L}{\lambda}$, however, in the soft limit $N_r = 1$ dominates, while in the hard limit $\rmd I^{N_r=2} \sim \rmd I^{N_r=1}$. This shows that ROE is quickly convergent if $\omega \ll \omegaBH$, while the expansion appears to break down for harder emissions. We expect this structure to appear to all orders in $N_r = n$. The resulting validity of the expansion is shown in the left of Fig.~\ref{fig:PhaseSpace_dIdw_ROE} in blue.
Based on our findings, the expansion scheme for the ROE at high opacity $L \gg \lambda$ is
\begin{equation}
\label{eq:ROE_Expansion}
    \omega\frac{\rmd I}{\rmd\omega}=\bar\alpha\frac{L}{\lambda}\sum_{n=0}^\infty f_n\left(\frac{\omega}{\omegaBH}\right)\,,
\end{equation}
where $f_n$ is a finite function that can be obtained order by order for $\omega \ll \omegaBH$.

The resulting spectrum is shown in the right of Fig.~\ref{fig:PhaseSpace_dIdw_ROE} for different propagation lengths. For short times ($t < \lambda$), the ROE is valid for all $\omega$ and it gives the same spectrum as the OE (compare to the right panel of Fig.~\ref{fig:PhaseSpace_dIdw_OE}). For longer propagation the ROE is only valid if $\omega < \omegaBH$, which is denoted with bullets in the figure. Outside of the valid region, the curves turn to gray (see also the left panel). Based on the figure, $N_r = 2$ has negligible contribution to the spectrum until $t \approx \lambda$ or $\omega \approx \omegaBH$. Again, the dashed line represents the full numerical evaluation of the spectrum from Ref.~\cite{Andres:2020vxs,Andres:2022} which is well approximated by the ROE in its region of validity.

%%%%%%%%%%%%%%%%%%%%%%%%%%%%%%%%%%%%%%%%%%%%%%%%%%%%
\subsection{Improved opacity expansion (IOE)}
\label{sec:ioe}
%%%%%%%%%%%%%%%%%%%%%%%%%%%%%%%%%%%%%%%%%%%%%%%%%%%%

The final expansion scheme we consider is the improved opacity expansion, introduced in Refs.~\cite{Mehtar-Tani:2019tvy,Mehtar-Tani:2019ygg,Barata:2020sav,Barata:2020rdn,Barata:2021wuf}. We saw that the ROE at high opacity does not work for energies much higher than $\omegaBH$. This makes sense since at higher energies the formation time $\tform$ of the emission becomes bigger than the mean free path, implying that the parton will scatter many times on the medium. The main motivation for the improved opacity expansion is to resum multiple soft scatterings, while account perturbatively for rare, hard scatterings. This is achieved by introducing a scale $Q^2$ that separates soft and hard scatterings in the scattering potential, 
\begin{equation}
\label{eq:v_IOE}
    v(\x,t) \approx v_\text{\tiny HO}(\x,t) + \delta v(\x,t)\,,
\end{equation}
where $ v_\text{\tiny HO}(\x,t) =  \frac{\qhat(t)}{4} \x^2$ and $\delta v(\x,t)  =  \frac{\qhat_0(t)}{4} \x^2 \ln{\frac{1}{Q^2\x^2}}$. Equation~\eqref{eq:v_IOE} is the $\mu|\bm x| \ll 1$ expansion of Eq.~\eqref{eq:Def_v_potential} with the GW potential. The first term is referred to as the harmonic oscillator approximation (HO), where the jet quenching parameter is
\begin{equation} 
    \label{eq:qhat}
    \qhat(t) = \qhat_0(t) \ln{\frac{Q^2}{\mu_*^2}} \,,
\end{equation}
where $\mu_\ast^2  =  \frac{\mu^2}{4}\rme^{-1+2\gamma_E}$ for the GW potential. The logarithm in $\hat q$ comes from the fact that the typical exchanged momentum $\langle \bm k^2\rangle = L\, \int_{\bm q}\bm q^2\sigma(\bm q)$ is divergent and thus it has to be regulated resulting in the leading logarithmic form in Eq.~\eqref{eq:qhat} (see in Ref.~\cite{Arnold:2009mr}). As long as $Q^2/\mu_\ast^2 \gg 1/(Q^2\bm x^2)$, the HO term dominates over $\delta v$, and the latter can be treated as a perturbation. This provides a big advantage, since the multiple scattering in the HO approximation can be resummed analytically.

The separation scale $Q^2$ has to be fixed in a meaningful way to not to interfere with the expansion~\cite{Barata:2020sav}. A natural choice that achieves this is evaluating $Q^2$ at the typical transverse momentum of the emission $\bm k^2 \sim \hat q \tform$, that yields
\begin{equation}
\label{eq:Running_Q2}
   Q_r^2(\omega) = \sqrt{\omega\hat q (\omega)} \,,
\end{equation}
which constitutes an implicit equation for $Q^2_r(\omega)$, and for $\hat q(\omega)\equiv \hat q( Q_r(\omega))$, see Eq.~\eqref{eq:qhat}.\footnote{Equation~\eqref{eq:Running_Q2} has a solution only if $\omega>2e\frac{\mu^4_\ast}{\hat q_0} \simeq 0.925\omegaBH$. When this is satisfied, and $L>\lambda$, then $Q_r^2>\mu^2_\ast$ by default. This is the necessary condition for the convergence of the IOE. The IOE therefore breaks down for $\omega\lesssim\omegaBH$.}

The IOE corresponds to expanding the full medium solution for $\Kc(\x;\y)$ around the harmonic oscillator, in contrast to the conventional opacity expansion where one expands around the vacuum solution. It can be cast in the iterative equation
\begin{align}
    \label{eq:ioe-coord}
    {\cal K}(\x,t_2;\y,t_1) &= {\cal K}_\text{\tiny HO}(\x,t_2;\y,t_1) \nn
    &- \int_{t_1}^{t_2}\rmd s \int_\z\, {\cal K}_\text{\tiny HO}(\x,t_2;\z,s) \delta v(\z,s) {\cal K}(\z,s; \y,t_1) \,.
\end{align}
Here, $\Kc_\text{\tiny HO}(\x;\y)$ is itself the solution to an iterative equation, namely
\begin{align}
    \Kc_\text{\tiny HO}(\x,t_2;\y,t_1) &= \Kc_0(\x-\y,t_2-t_1) \nn
    &- \int_{t_1}^{t_2} \rmd s\, \int_\z \, \Kc_0(\x-\z,t_2-s) v_\text {\tiny HO}(\z,s)  \Kc_\text{\tiny HO}(\z,s;\y,t_1) \,.
\end{align}
The formal solution for ${\cal K}_\text{\tiny HO}(\x,t_2; \y,t_1)$ can also be cast as a path integral, namely
\begin{equation}
    {\cal K}_\text{\tiny HO}(\x,t_2; \y,t_1) = \int_{\r(t_1)=\y}^{\r(t_2)=\x} {\cal D} \r\,  \rme^{i\int_{t_1}^{t_2}\rmd s\, \big[ \frac{\omega}{2} \dot \r^2 +i  v_\text{\tiny HO}(\r,s) \big] } \,,
\end{equation}
which has a well-known analytical solution in a static medium,
\begin{equation}\label{eq:ho-propagator}
    \mathcal{K}_\text{\tiny HO}\left(\boldsymbol{x}, t_{2} ; \boldsymbol{y}, t_{1}\right)=
    \frac{\omega \Omega}{2 \pi i \,\sin (\Omega \Delta t)} 
    \rme^{\frac{i \omega \Omega}{2 \sin(\Omega\Delta t) }\left[\cos(\Omega\Delta t) \, (\x^2+\y^2)-2 \boldsymbol{x} \cdot \boldsymbol{y}\right]}\,,
\end{equation}
where $\Delta t  \equiv  t_2 - t_1$ and $\Omega  = \frac{1-i}{2} \sqrt{\hat{q}(\omega)/\omega}$ is the characteristic oscillator frequency.

Inserting this expansion into the equation for the medium-induced spectrum Eq.~\eqref{eq:medspec-coord} separates it into two parts,
\begin{align}
    \label{eq:ioe-ho}
    \omega \frac{\rmd I^\text{\tiny HO}}{\rmd \omega} &= \frac{2\alpha_s C_R}{\omega^2}\rmR \int_0^\infty \rmd t_2 \int^{t_2}_0 \rmd t_1\, \partial_\x \cdot \partial_\y \big[{\cal K}_\text{\tiny HO}(\x,t_2; \y,t_1) - {\cal K}_0(\x,t_2; \y,t_1) \big]_{\x=\y=0} \,,\\
    \label{eq:ioe-nhos}
    \omega \frac{\rmd I^\text{\tiny IOE}}{\rmd \omega} &= -\frac{2\alpha_s C_R}{\omega^2}\rmR \int_0^\infty \rmd t_2 \int^{t_2}_0 \rmd s \int_0^{s}\rmd t_1 \, \int_\z\,  \nn
    &\times\partial_\x \cdot \partial_\y [{\cal K}_\text{\tiny HO}(\x,t_2;\z,s) \delta v(\z,s) {\cal K}(\z,s; \y,t_1)]_{\x=\y=0}\,.
\end{align}
The first term gives rise to the well-known HO spectrum, while the second constitutes an expansion in hard splittings around the harmonic oscillator. The IOE spectrum can be simplified further, giving
\begin{align}
    \label{eq:ioe-contrib}
    \omega \frac{\rmd I^\text{\tiny IOE}}{\rmd \omega} = 
    \frac{2\abar}{\omega}\rmR \,i \int_0^L \rmd t_2 \int^{t_2}_0 \rmd t_1 \int_\x \,\rme^{-i \frac{\omega \Omega}{2}\tan{(\Omega(L-t_2))\x^2}}\delta v(\x)\frac{\x}{\x^2}\cdot\partial_\y {\cal K}(\x,t_2; \y,t_1)\vert_{\y=0}\,,
\end{align}
where $\Kc(\x;\y)$ should be iterated using Eq.~\eqref{eq:ioe-coord} in order to generate higher orders of the expansion. 
A general formula for the improved opacity expansion at arbitrary order can also be derived, which was done in Sec.~\ref{app:general-IOE}, see also in Ref.~\cite{Barata:2020sav}.

%%%%%%%%%%%%%%%%%%%%%%%%%%%%%%%%%%%%%%%%%%%%%%%%%%%%
\paragraph{Harmonic oscillator (HO):}
%%%%%%%%%%%%%%%%%%%%%%%%%%%%%%%%%%%%%%%%%%%%%%%%%%%%
The harmonic oscillator approximation resums all coherent soft scatterings during the formation of the emission. The relevant scale is 
\begin{equation}
    \omega_c\equiv\frac{1}{2}\hat q(\omega_c) L^2\,,
\end{equation}
where the scale in the jet quenching parameter is set to $\omega_c$. This scale was already identified in Sec.~\ref{sec:heuristic}. The HO approximation is expected to be valid for $\omegaBH \ll \omega \ll \omega_c$.

We derive here the familiar harmonic oscillator spectrum in a new way. The vacuum contribution can easily be subtracted by inserting Eq.~\eqref{eq:ioe-coord} into Eq.~\eqref{eq:medspec-coord}, which gives
\begin{align}
    \omega \frac{\rmd I^\text{\tiny HO}}{\rmd \omega} = &-\frac{2\alpha_s C_R}{\omega^2}\rmR \int_0^\infty \rmd t_2 \int^{t_2}_0 \rmd s \int_0^{s} \rmd t_1 \int_\z \nn
    &\times\partial_\x \cdot \partial_\y \big[{\cal K}_0(\x,t_2; \z,s)v_\text{\tiny HO}(\z,s) {\cal K}_\text{\tiny HO}(\z,s; \y,t_1) \big]_{\x=\y=0}\,.
\end{align}
Using the fact that $\int_s^\infty \rmd t_2 \,\partial_\x{\cal K}_0(\x,t_2; \z,s)\vert_{\x=0}  =  -i\frac{\omega}{\pi}\frac{\z}{\z^2}$, the spectrum becomes
\begin{align}
    \omega \frac{\rmd I^\text{\tiny HO}}{\rmd \omega} = 
    &\frac{\abar \qhat(\omega)}{2\omega}\rmR \, i\int_0^L \rmd t_2 \int^{t_2}_0 \rmd t_1 \int_\z \,\z \cdot \partial_\y {\cal K}_\text{\tiny HO}(\z,t_2; \y,t_1)\vert_{\y=0}\,.
\end{align}
This can be further simplified by using that
\begin{align}
    \int_\z \, \z \cdot \bdel_\y \Kc_\text{\tiny HO}(\z,t_2;\y,t_1)|_{\y=0} &= -\frac{(\omega \Omega)^2}{2\pi  \sin^2 (\Omega (t_2-t_1))} \int_\z\, \z^2 \rme^{i \frac{\omega \Omega}{2} \z^2\cot( \Omega (t_2-t_1))} \,, \nn
    &=\frac{2}{\cos^2( \Omega(t_2-t_1))} \,.
\end{align}
The time integration can be now be dealt with straightforwardly, yielding
\beq
    \int_0^L \rmd t_2 \int_0^{t_2} \rmd t_1\, \frac{1}{\cos^2(\Omega(t_2-t_1))} = -\frac{ \ln \cos \Omega L}{\Omega^2} \,,
\eeq
and thus the spectrum becomes
\begin{align}
\label{eq:wdIdw_HO}
    \omega \frac{\rmd I^\text{\tiny HO}}{\rmd \omega} 
    &= 2\bar \alpha\, \ln \left| \cos \Omega L \right| \,,
\end{align}
which is the familiar BDMPS-Z spectrum~\cite{Baier:1996kr,Zakharov:1996fv}. The limits of this are
\begin{align}\label{eq:ho-limits}
    \omega \frac{\rmd I^\text{\tiny HO}}{\rmd \omega}\simeq 
    \begin{cases}
        \abar\sqrt{\frac{2\omega_c}{\omega}}, & \text{for } \omega\ll\omega_c\,, \\[0.7pt]
        \frac{\abar}{6}  \left(\frac{\omega_c}{\omega}\right)^2, & \text{for } \omega\gg\omega_c\,.
    \end{cases}
\end{align}
The soft limit agrees with the discussion in Sec.~\ref{sec:heuristic}, while the hard limit is subleading compared to the OE $N = 1$ in Eq.~\eqref{eq:n1-limits}. Defining $\hat q$ with a logarithm extends the region of validity, which was also found in Ref.~\cite{Arnold:2009mr}, leading to the curved $\omega_c(t)$ line in Fig.~\ref{fig:PhaseSpace_PhysicalProcesses}. 

%%%%%%%%%%%%%%%%%%%%%%%%%%%%%%%%%%%%%%%%%%%%%%%%%%%%
\paragraph{Next-to-harmonic oscillator (NHO):}
%%%%%%%%%%%%%%%%%%%%%%%%%%%%%%%%%%%%%%%%%%%%%%%%%%%%
Using the definition in Eq.~\eqref{eq:ioe-contrib} and the results of Sec.~\ref{app:general-IOE}, the first order of the improved opacity expansion can be written as
\begin{align}
    \omega \frac{\rmd I^\text{\tiny NHO}}{\rmd \omega}
    &= \frac{2 \abar}{\pi} \frac{L}{\lambda}\frac{\omegacbar}{\omega}
    \rmR \int_0^1 \rmd s \int_\u\, \frac12  \ln{\left(\frac{\omega}{\omegacbar}\frac{\mu^2}{2Q^2}\frac{1}{\u^2}\right)}
    \rme^{\frac{i}{2} f(s)\u^2}\,,
\end{align}
where we have defined the function $f(s) = \sigma\sqrt{\omega_c/\omega}[\cot(\sigma s \sqrt{\omega_c/\omega}) - \tan(\sigma (1-s) \sqrt{\omega_c/\omega} )]$, and $\sigma =  \frac{1-i}{\sqrt{2}}$. After doing the $\u$ integral this becomes
\begin{align}
    \omega \frac{\rmd I^\text{\tiny NHO}}{\rmd \omega}
    = 2 \abar \frac{L}{\lambda}\frac{\omegacbar}{\omega}
    \rmR \,i\int_0^1 \rmd s  \,
    \frac{1}{f(s)}
    \left[1-\gamma_E+\ln{\left(-i \frac{\omega}{\omegacbar}\frac{\mu_*^2}{Q^2}f(s)\right)}\right]\,.
\end{align}
The limits of this expression can readily be extracted. In the soft limit, $\omega  \ll  \omega_c$, we have $f(s)  \to  2 i \sigma\sqrt{\omega_c/\omega}$ while in the hard limit, $\omega  \gg  \omega_c$, it becomes $f(s)  \to  1/s$. These simplifications make it possible to do the last time integration. Hence, the extracted limiting behavior is,
\begin{align}
\label{eq:wdIdw_NHO}
    \omega \frac{\rmd I^\text{\tiny NHO}}{\rmd \omega}\simeq 
    \begin{cases}
        \abar \sqrt{\frac{2\omega_c}{\omega}}
        \, \frac{1}{2\ln Q^2/\mu_\ast^2}\left(\frac{\pi}{4} + \gamma_E + \ln{\left(\frac{\sqrt{\qhat \omega}}{\sqrt{2}Q^2}\right)}\right), & \qquad \text{for } \omega\ll\omega_c\,, \\
        \frac{\pi \abar}{2} \frac{L}{\lambda} \frac{\omegacbar}{\omega}, & \qquad \text{for } \omega\gg\omega_c\,.
    \end{cases}
\end{align}
In the soft limit $\omega \ll \omega_c$, N$^n$HO terms will take the form of the HO by using $Q_r$, and thus 
\begin{equation}
    \omega \frac{\rmd I}{\rmd \omega} = \abar \sqrt{\frac{\hat q(\omega) L^2}{\omega}} \left(1 + \frac{1}{2} \frac{a_0}{\ln Q_r^2/\mu^2_\ast} + \mathcal O\left(\frac{1}{\ln Q_r^2/\mu_\ast^2}\right)^2 \right)\,,
\end{equation}
where we added the HO term, and used Eq.~\eqref{eq:Running_Q2}. The choice of $Q = Q_r(\omega)$ is effective, when the medium is big enough $L\gg\lambda$. It is clear that the expansion parameter of the IOE is $\ln^{-1}(Q_r^2/\mu_\ast^2) \ll 1$ in the soft limit. Therefore, N$^n$HO terms can be absorbed into an effective jet transport parameter,
\begin{equation}
\label{eq:qhateff}
    \hat q_\text{eff}(Q^2) = \hat q_0 \ln\left(\frac{Q_r^2}{\mu^2_\ast}\right) \left[ 1 + \frac{a_0}{\ln Q_r^2/\mu_\ast^2} + \frac{a_1}{\ln^2 Q_r^2/\mu_\ast^2} + \dots \right]\,.
\end{equation}
The coefficients $a_0 = 1.016$ and $a_1 = 0.316$ of the expansion and higher-order terms up to N$^2$HO were found in Ref.~\cite{Barata:2020sav}.

In the hard limit of Eq.~\eqref{eq:wdIdw_NHO}, one can see that the IOE reproduces the hard limit of $N = 1$ in the OE from Eq.~\eqref{eq:n1-limits}. Furthermore, it is bigger than the HO contribution in Eq.~\eqref{eq:wdIdw_HO} and thus NHO dominates for $\omega\gg\omega_c$. 

\begin{figure}
    \centering
     \includegraphics[width=0.45\textwidth]{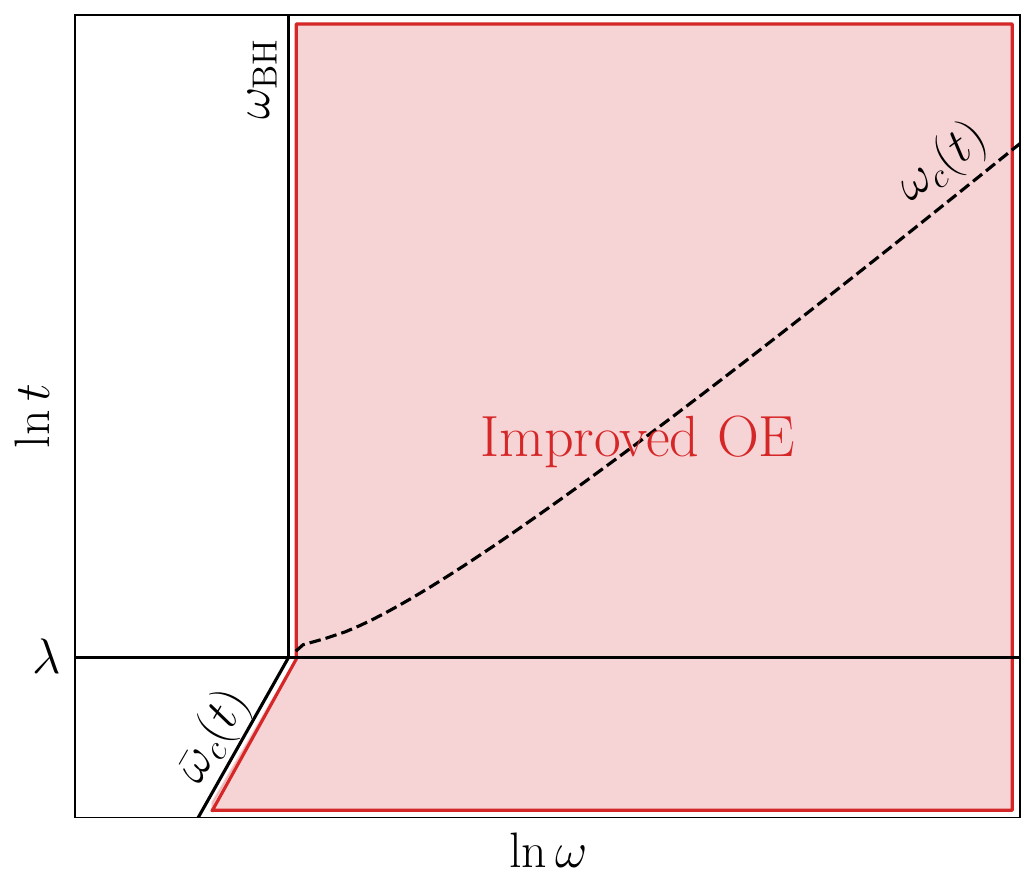} 
     \includegraphics[width=0.45\textwidth]{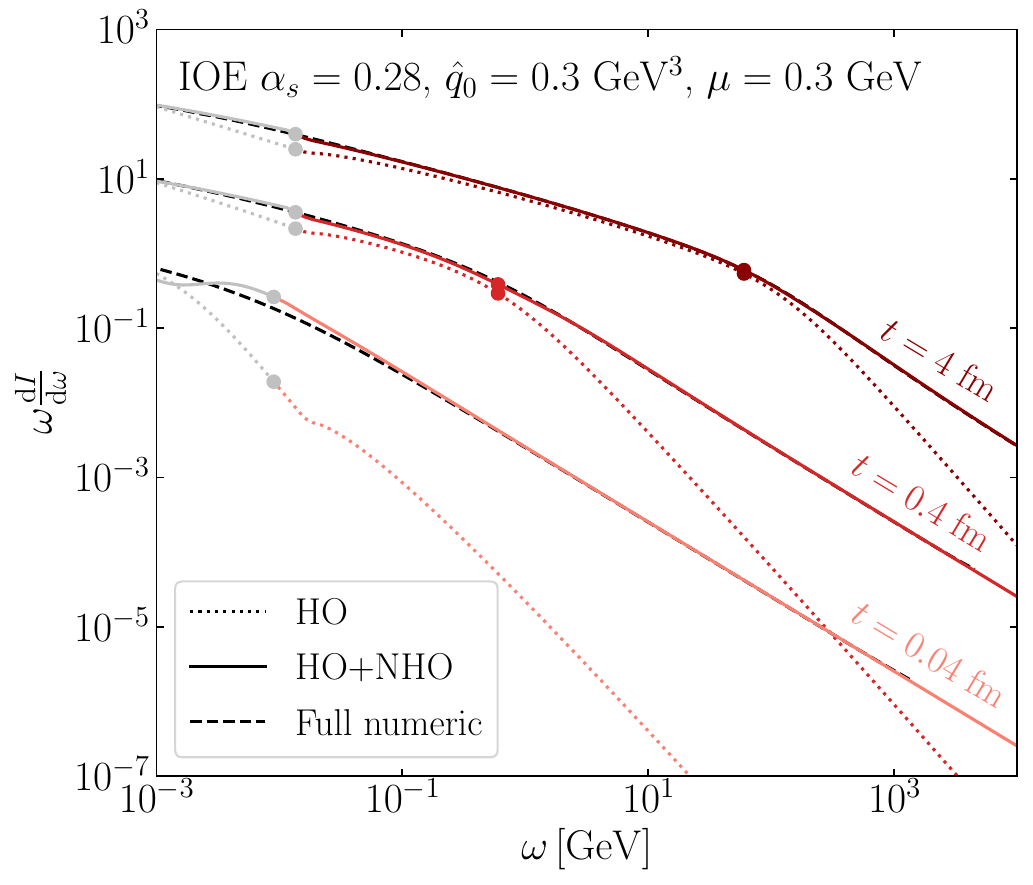} 
     \caption{{\it Left:} The sketch of region of validity (and convergence) of the improved opacity expansion for different propagation length $t$ and emission energy $\omega$. {\it Right:} The induced emission spectrum for gluons in the improved opacity expansion. The gray part of the curves denotes regions where the expansion is not valid. Using the parameters presented, $\lambda = 0.06$ fm. The full numeric solution is also presented with dashed lines.}
     \label{fig:PhaseSpace_dIdw_IOE}
\end{figure}

As mentioned above, the HO is meaningful if $L > \lambda$ and $\omega > \omegaBH$. Furthermore, the IOE is expected to converge if $v_\text{\tiny HO} > \delta v$ or equivalently $\omega > \omegaBH$. However, as the hard limits of the first order of the IOE and the OE are equal, and the OE is valid down to $\omegacbar$, it is reasonable to assume that also the IOE is valid down to $\omegacbar$. Therefore, the region of validity will be extended to all $L$ and $\omega > \min(\omegaBH,\omegacbar)$, as shown in the left of Fig.~\ref{fig:PhaseSpace_dIdw_IOE} in red. 

The spectra obtained with the IOE from Eq.~\eqref{eq:wdIdw_HO} and Eq.~\eqref{eq:wdIdw_NHO} are shown in the right panel of Fig.~\ref{fig:PhaseSpace_dIdw_IOE} for different propagation lengths. At early times $t < \lambda$, the HO approximation is highly suppressed, due to the absence of multiple scattering. However, the contribution from the NHO makes the total agree with the hard limit of $N = 1$ OE and $N_r = 1$ ROE (cf. Figs.~\ref{fig:PhaseSpace_dIdw_OE} and \ref{fig:PhaseSpace_dIdw_ROE}). 
The deviation close to $\omega\approx\bar\omega_c$ (gray bullet) arises since $Q_r$ was chosen to reproduce the HO spectrum which is strictly valid for $L\gg\lambda$. 
For later times $t > \lambda$, both the HO and NHO will give sizable contributions, where the HO dominates if $\omega < \omega_c$ (red bullets) and NHO dominates if $\omega > \omega_c$. The HO approximation breaks down if $\omega < \omegaBH$ (gray bullets). The dashed line is the full numerical solution from Refs.~\cite{Andres:2020vxs,Andres:2022} and the IOE well captures it in its region of validity.

%%%%%%%%%%%%%%%%%%%%%%%%%%%%%%%%%%%%%%%%%%%%%%%%%%%%
\subsection{Summary of the regimes and the induced emission spectrum}
\label{sec:summary_spectra}
%%%%%%%%%%%%%%%%%%%%%%%%%%%%%%%%%%%%%%%%%%%%%%%%%%%%

In this section, we have presented three distinct perturbative expansions (OE, ROE, and IOE) that provide different ways of calculating the induced emission spectrum $\omega \frac{\rmd I}{\rmd \omega}$ in their respective regimes of convergence. 
The OE and IOE are expansion schemes that were first developed in previous works, while the ROE is rigorously derived in this section for the first time. Their regions of validity are sketched previously in Figs. \ref{fig:PhaseSpace_dIdw_OE}--\ref{fig:PhaseSpace_dIdw_IOE}, and at least one of the expansions is valid at every point in the phase space $(\omega,t)$. Here $\omega$ is the emitted energy and $t$ is the propagated length ($L$ is the maximal length of the medium and $E$ is the energy of the emitting particle). As a consequence, our description of the spectrum is complete in the full phase space, as one can always use one of the expansions to reach an approximation of the true spectrum, and one can reach better accuracy by including higher orders. Note that the expansions are overlapping: for $L < \lambda$ both OE and ROE are valid, and for $\omega > \omega_c$ both IOE the OE can be used.

\begin{figure}
    \centering
    \includegraphics[width=\textwidth]{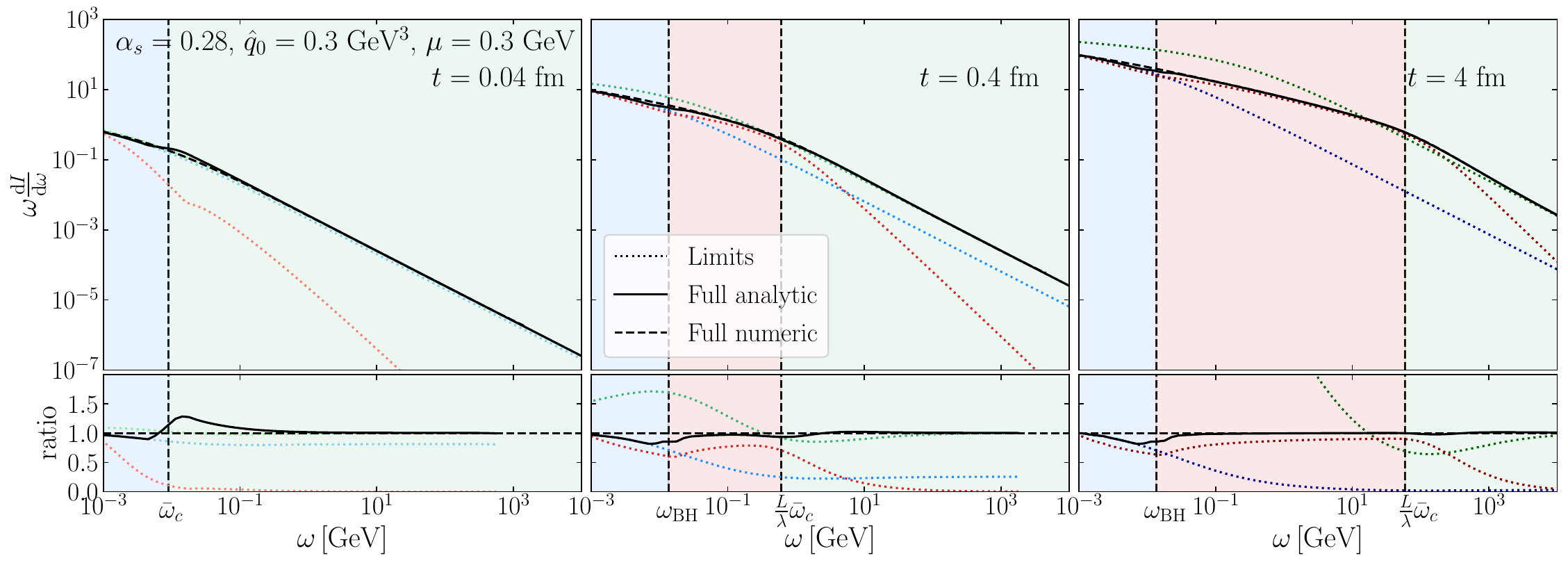}
    \caption{Summary of the induced emission spectrum for gluons, combined from the three expansion schemes at different propagation length (different panels). The black lines are our final forms from Eq.~\eqref{eq:spectrum_Full} that uses $N_r=1$ and HO+NHO. The shaded areas denote the leading scattering process and the corresponding dotted lines are the limiting $N = 1$, HO and $N_r = 1$ contributions. The dashed lines are the numerical solution of Eq.~\eqref{eq:medspec-coord} from Ref.~\cite{Andres:2020vxs}.}
    \label{fig:spectra-numerics}
\end{figure}

The results presented obtained so far within the unified resummation framework are valid in both dilute and dense regimes and can be systematically improved to arbitrary high order in the expansions. For practical purposes, however, a handy and efficient interpolation formula such suffice to capture the relevant features to high precision. This would be very useful for other applications, such as resumming multiple emissions in Secs.~\ref{sec:multiple} and \ref{sec:numerics}. To describe the spectrum in the whole phase space, we use (to first order)
\begin{equation}
\label{eq:spectrum_Full}
    \frac{\rmd I^{\rm Full}}{\rmd\omega}=
    \begin{cases}
        \frac{\rmd I^{\rm ROE}}{\rmd\omega}\,, & \omega<\min(\omegaBH,\bar\omega_c(t))\,,\\
        \frac{\rmd I^{\rm IOE}}{\rmd\omega}\,, & {\rm otherwise}\,.
    \end{cases}
\end{equation}
Based on Figs.~\ref{fig:PhaseSpace_dIdw_OE}--\ref{fig:PhaseSpace_dIdw_IOE} (and the all order expansion formulas), the first-order terms already capture the most important effects. We stress that this is arguable the most straightforward interpolation scheme. However, it turns out that it gives a good description in almost the whole phase space, deviating maximally 30\% from the exact numerical results around the Bethe-Heitler energy in the dilute regime, see Fig.~\ref{fig:spectra-numerics} (left, lower panel).\footnote{The interpolation can be further inspected in Fig.~11, where we plot the spectrum on a semilog scale. The exact details of matching the ROE regime with the IOE involves a smooth interpolation function, that avoids blowing up the logarithmic dependence of $\hat q$. This is described in detail in App.~\ref{app:Numeric_impl}.}

To summarize, the limiting behaviour of the spectrum in different regions of the phase space is
\begin{equation}\label{eq:spectrum_SoftLimit_lowop}
    \left.\omega\frac{\rmd I}{\rmd\omega} \right|_{L\ll \mfp} = \begin{cases}
        2\abar \frac{L}{\mfp} \left(\ln\frac{\omegacbar}{\omega}-1+\gamma_E\right)\,, & \quad\text{for } \omega \ll \omegacbar \,,\\
        \frac{\pi}{2}\abar \frac{L}{\mfp} \frac{\omegacbar}{\omega}\,, &\quad \text{for }  \omegacbar \ll \omega \,,
    \end{cases}
\end{equation}
for $L \ll \mfp$, and 
\begin{equation}\label{eq:spectrum_SoftLimit_highop}
    \left. \omega \frac{\rmd I}{\rmd\omega} \right|_{L\gg \mfp} = \begin{cases}
       2 \abar\frac{L}{\lambda} \ln\left(\frac{\omegaBH}{\omega}\right)\,, &\quad  \text{for } \omega \ll \omegaBH \,,\\
        \abar\sqrt{\frac{2\omega_c}{\omega}}\,, &\quad  \text{for } \omegaBH \ll \omega \ll \omega_c \,,\\
       \frac{\pi}{2} \abar\frac{L}{\lambda} \frac{\omegacbar}{\omega}\,, &\quad  \text{for }  \omega_c \ll \omega \,,
    \end{cases}
\end{equation}
for $L \gg \mfp$.
This agrees with the formulas from the heuristic discussion in Sec.~\ref{sec:heuristic}. 
In Fig.~\ref{fig:spectra-numerics}, we evaluated Eq.~\eqref{eq:spectrum_Full} (black curve) up to $N_r = 1$ and HO+NHO for different times. The dotted curves are the limits of $N_r = 1$, HO and $N = 1$, shown in blue, red and green respectively. The regions are shaded with the same colors as in Fig.~\ref{fig:PhaseSpace_PhysicalProcesses}, visualizing the regions of the distinct scattering processes. 
At the transition point $\min(\omegaBH,\bar\omega_c(t))$, the spectrum is not completely smooth, and the difference is expected to vanish as one goes to higher orders in the perturbative expansion. We defined a switching function that makes the transition smoother, which is described in App.~\ref{app:Numeric_impl}.

In Fig.~\ref{fig:spectra-numerics}, we have also plotted the full numerical spectra from Ref.~\cite{Andres:2020vxs,Andres:2022} with dashed lines. The excellent agreement with our curves corroborates the validity of our formula in Eq.~\eqref{eq:spectrum_Full}, where we used $N_r=1$ and HO+NHO in the plot. In the ratio panel, one can see, how the different regions are adding up to give an overall very accurate description of the full spectrum. Moreover, as we saw in Figs.~\ref{fig:PhaseSpace_dIdw_OE}--\ref{fig:PhaseSpace_dIdw_IOE}, by including higher orders e.g. $N_r=2$, the description becomes more accurate, smoothing the transition around $\omegaBH$. We leave the study of these higher-order corrections and the corresponding uncertainties for future studies.

%%%%%%%%%%%%%%%%%%%%%%%%%%%%%%%%%%%%%%%%%%%%%%%%%%%%
%%%%%%%%%%%%%%%%%%%%%%%%%%%%%%%%%%%%%%%%%%%%%%%%%%%%
\section{Resumming multiple emissions in the medium}
\label{sec:multiple}
%%%%%%%%%%%%%%%%%%%%%%%%%%%%%%%%%%%%%%%%%%%%%%%%%%%%
%%%%%%%%%%%%%%%%%%%%%%%%%%%%%%%%%%%%%%%%%%%%%%%%%%%%

Section~\ref{sec:spectrum} presents a theoretical framework consisting of different perturbative expansions (namely the opacity expansion (OE), the resummed OE (ROE), and the improved OE (IOE)) to describe the full phase space ($\omega,t$) of the medium-induced gluon emission spectrum. The emitted energy $\omega$ is limited by the energy of the emitter $\omega \ll E$, and the propagation time (length) $t$ is in turn limited by the medium length $t < L$. 
Our effective formalism accounts for arbitrarily many scatterings that can be arbitrarily hard or soft. Finally, Eq.~\eqref{eq:spectrum_Full} describes the emission spectrum up to arbitrary precision and recovers the full solution of Eq.~\eqref{eq:medspec-coord} that has only been achieved numerically before~\cite{CaronHuot:2010bp,Feal:2018sml,Andres:2020vxs}. 

It is now time to explore what consequences the full induced spectrum instills on a parton propagating through the medium. In this section, we will present analytical solutions of the evolution equation of the medium-induced cascade. We present the inclusive gluon energy distribution for different medium lengths, and we will focus on gluons for transparency. The numerical solutions are presented in Sec.~\ref{sec:numerics}. Vacuum emissions belong outside of the scope of the current work.

%%%%%%%%%%%%%%%%%%%%%%%%%%%%%%%%%%%%%%%%%%%%%%%%%%%%
\subsection{The necessity of multiple emissions}
%%%%%%%%%%%%%%%%%%%%%%%%%%%%%%%%%%%%%%%%%%%%%%%%%%%%

Multiple emissions have to be taken into account whenever the multiplicity of gluons is large. We define the multiplicity of gluons above the energy $\omega$ in terms of the spectrum $\rmd I/\rmd \omega$, as 
\begin{equation}
    N(\omega) = \int_\omega^\infty \rmd \omega' \frac{\rmd I}{\rmd \omega'} \,.
\end{equation}
The upper limit of the integral is taken to infinity because we will currently assume that the energy of the emitter $E$ is much larger than the largest available medium energy scale. Also, for our current purposes, it suffices to consider the leading behavior of the spectrum in the various scattering regimes presented in Fig.~\ref{fig:PhaseSpace_PhysicalProcesses}. 

At low opacity $L\ll \lambda$ and starting from $\omega < \omegaBH$, we find
\begin{equation}
    N(\omega) \simeq \bar\alpha\frac{L}{\lambda}\ln^2\frac{\bar\omega_c}{\omega}+\frac{\pi\bar\alpha}{2}\frac{L}{\lambda} \,.
\end{equation}
where $\omegacbar= \frac12\mu^2 L$ and we have only kept the leading terms.
The maximal multiplicity in the hard regime, see the second term, is always small for perturbative splittings with $\bar\alpha\ll 1$. Furthermore, the multiplicity in the soft (Bethe--Heitler) regime, given by the first term, becomes large only at very small energies, i.e. $\omega < \rme^{\sqrt{1/(\bar \alpha L/\lambda)}} \omegacbar$. We can therefore safely neglect multiple emissions at low opacities.

For dense media, $L\gg \mfp$, we discuss the three pertinent cases. As usual, we will denote $\omega_c =\frac12\hat q L^2$. Then, for  $\omega \gg \omega_c$ (rare hard scattering regime), we find
\begin{equation}
    \label{eq:mult-dense-1}
    N(\omega) \simeq \bar \alpha \frac{L}{\mfp} 
    \frac{\omegacbar}{\omega} \,,
\end{equation}
where we again neglected subleading terms. Hence hard emissions, described by the OE expansion, can safely be considered to be rare.

Next, for $\omegaBH \ll \omega \ll \omega_c$ we get
\begin{equation}
    \label{eq:mult-dense-2}
    N(\omega) \simeq 2^{\frac32} \bar \alpha \sqrt{\frac{ \omega_c}{\omega}} \,,
\end{equation}
where we introduced the leading behavior of the spectrum and neglected the multiplicity from hard emissions, following the discussion above. The multiplicity becomes large $N(\omega)\gg 1$ at energies $\omega \ll \bar \alpha^2 \omega_c$ and thus multiple emission becomes dominant.
For large enough medium length $L\gg \mfp$, there is a significant phase space allowing for multiple emission ($\bar \alpha_s^2 \omega_c(t) > \omegaBH$), resulting in the power enhancement. 

Finally, for soft gluon energies $\omega < \omegaBH \ll \omega_c$, the multiplicity is 
\begin{equation}
    \label{eq:mult-dense-4}
    N(\omega) \simeq \bar \alpha \frac{L}{\mfp} \ln^2 \frac{\omegaBH}{\omega} + 2^{\frac32}\bar \alpha \sqrt{\frac{ \omega_c}{\omegaBH}} \,,
\end{equation}
where, again, only the leading terms from each regime were kept. 
The second term in Eq.~\eqref{eq:mult-dense-4} scales as $\sim(L/\mfp)\sqrt{\hat q/\hat q_0}$. Based on the discussion above, this term is already large and the multiplicity continues to grow only logarithmically for small $\omega$, and therefore multiple emissions are going to happen.

%%%%%%%%%%%%%%%%%%%%%%%%%%%%%%%%%%%%%%%%%%%%%%%%%%%%
\subsection{Resummation of multiple emissions}
%%%%%%%%%%%%%%%%%%%%%%%%%%%%%%%%%%%%%%%%%%%%%%%%%%%%

Considering multiple emissions in a medium poses a tremendous theoretical challenge. The situation is quite analogous to the description of multiple gluon emissions in vacuum. Similar to QCD jets, the main challenge when considering medium effects lies in dealing with intricate interference effects between subsequent emissions, see e.g. Refs.~\cite{Arnold:2015qya,Arnold:2020uzm,Arnold:2021pin}. 
However, when considering multiple soft emissions, that occur quasi-instantaneously, these effects can safely be neglected \cite{Arnold:2002zm,Blaizot:2013vha}.\footnote{We can also extend this logic for the semi-hard emissions which are included in our formalism, since they are rare occurrences and therefore the resummation has no effect, see Sec.~\ref{sec:analytic_solutions_eveq}.}

In order to clarify the framework that we work in, let us briefly recall the main arguments for neglecting interference effects for a set of multiple induced emissions. For the time being, we stick to emissions in the HO region which dominate the multiplicity. The typical time it takes an emission to form, often referred to as a \textit{formation time} (or in some works branching time), of a soft gluon is $\tform \sim \sqrt{\omega/\hat q}$. This time is much smaller than the extent of the medium $\tform \ll L$ as long as $\omega \ll \omega_c$. 

Another relevant quantity is the time between two subsequent emissions. This is related to the no emission probability (or Sudakov factor). 
For a leading particle with energy $E$, the first emission is produced at time $\tform_1$, with energy $\omega_1$. A second, strongly ordered emission ($\omega_2\ll\omega_1$) of the original parton forms much quicker $\tform_2\ll\tform_1$. The time between the two emissions $\tsplit$ can be estimated with $ \int_{\tform_1}^{\tsplit} \rmd t \int_{\omega_2}^{\omega_1} \rmd \omega \,\frac{\rmd I}{\rmd \omega \rmd t} \sim 1$, that is basically the probability of not having emissions between $\omega_2<\omega<\omega_1$, resulting in
\begin{equation}
    \tsplit \sim \tform_1+\frac{1}{\bar \alpha} \tform_2 \,.
\end{equation}
Hence, our rough estimate implies that typically $\tsplit \gg \tform_2$ (for $\bar \alpha \ll 1$). Therefore, the formation of emissions is short compared to the time that separates emissions and thus emissions form independently. This motivates the resummation of multiple independent emissions in terms of a rate equation. 

Similar analysis can be done for hard emissions $\omega\gg\omega_c$, for which $t_{\rm split}$ is very long due to the unlikeliness of hard scatterings, and therefore the emissions are formed independently. 

For soft emissions $\omega\ll\omegaBH$, $t_{\rm split}\sim \tform_1+ \frac{1}{\bar\alpha}\ln^{-2}\frac{\lambda}{\tform_{2}}$ and thus emissions form independently. Close to the boundary in cases where $t_{\rm split}\approx \tform_1+\tform_2$, a more complicated structure appears in terms of resummation, as emissions might overlap. In this case, interference effects between the two emissions have to be included. A similar thing happens in vacuum for wide angle soft emissions, which result in angular ordering and in non-global effects for which the resummation has been understood just recently~\cite{Banfi:2021xzn}. We will use the rate equation to account for emissions with any $\omega$. However, it will not necessarily account correctly for interference among them and further study is needed in the future.

One question still remains open, namely the choice of the time scale used in the rate equation. In case of two emissions, the second emission experiences a shorter medium, of the scale $\sim L-t_{\rm f1}-t_{\rm split}$. We know, however, that in the soft limit $\omega\ll\omega_c$, the formation time is $t_{\rm f}\ll L$, and therefore, the length degradation should not matter for a large medium~\cite{Arnold:2002zm,Blaizot:2013vha}. For small media, or for emissions with comparable formation times, these corrections can become significant. It is an unresolved question how to incorporate these corrections into a rate equation see e.g. Ref.~\cite{Arnold:2020uzm}.\footnote{See also Sec. 4 in Ref.~\cite{Barata:2021byj}, where modifications of the rate due to finite formation time were studied.} However, as argued above, the corrections to the rate coming from finite-size effects can be treated in a perturbative fashion. While these issues merit further studies, perhaps within a Monte Carlo approach, we consider them to go beyond our present scope and we assume that all emissions experience the same length $L$. This matches the approximation in most of the current energy-loss models.

In this section, we will focus on the single-inclusive energy distribution of partons carrying energy $x E$ after traveling length $t$ in the medium, where $E$ is the initial energy. It is defined as 
\begin{equation}
    D(x,t) \equiv x \frac{\rmd N}{\rmd x} \,.
\end{equation}
The formalism can easily be extended to account for parton flavors, see e.g. \cite{Mehtar-Tani:2018zba}, but for now, we restrict our attention to a pure gluon cascade.

In Sec.~\ref{sec:spectrum} we focused on emissions of soft gluons with energies $\omega \ll E$. Now we will consider generic splitting processes where a parton with flavor index $a = q,g$ and initial energy $E$ shares its energy with two daughter partons, with energies $zE$ and $(1-z)E$ and flavor indices $b$ and $c$, respectively, for $0 < z <1$. The spectrum of such splittings $\rmd I_{ba}/\rmd z$ is given by Eq.~\eqref{eq:Spectrum_def_finitez}. The general features of Fig.~\ref{fig:PhaseSpace_PhysicalProcesses} remain the same with the substitution $\omega \to z(1-z)E$.\footnote{Hence the upper limit of $\omega$ in Fig.~\ref{fig:PhaseSpace_PhysicalProcesses} should now be $E/4$.} For further details, see the discussion in App.~\ref{app:Kernels_finitez}.

As was shown in Ref.~\cite{Blaizot:2012fh,Blaizot:2013vha}, for sufficiently soft emissions, with formation times much smaller than the medium length, interference effects are suppressed and one can consider multiple emissions as occurring independently. The evolution equation for the energy distribution, that accounts for an arbitrary number of induced emissions, is given by
\begin{align}
    \label{eq:Evolution_eq}
    \frac{\partial}{\partial t}D(x,t)&= 
    \int_{x}^1 \rmd z \, \cK\left(z,\frac{x}{z}E,t\right)D\left(\frac{x}{z},t\right)-\int_0^1 \rmd z  \,z \,\cK\left(z,x E,t\right)D(x,t) \,.
\end{align}
The initial condition is a single gluon carrying energy $E$, hence $D(x,0) = \delta(1-x)$.
The splitting kernel $\cK(z,E,t)$ is the rate of emissions off a particle with energy $E$,
\begin{equation}\label{eq:Emission_Rate}
    \cK(z,E,t)= 2\left.\frac{\rmd I_{gg}}{\rmd z\rmd t}\right|_{E}\,,
\end{equation}
and the rate with full $z$-dependence for the $g\to gg$ splitting can be found in Eq.~\eqref{eq:Spectrum_def_finitez}.\footnote{In the soft limit, the kernel is closely related to the spectrum discussed in the previous section or, more precisely, the rate $\rmd I/(\rmd \omega \rmd t)$, calculated in App.~\ref{app:rate}. Importantly, for the gluon splitting kernel, the divergences in $z \to 0$ and $z\to 1$ are folded together in the limit $\omega \to 0$, hence the additional symmetry factor in Eq.~\eqref{eq:Emission_Rate}.}
The first term in Eq.~\eqref{eq:Evolution_eq} is a real emission describing an emitted gluon with energy fraction $x$ (gain term), while the second is a virtual emission that does not change the energy of the emitter (loss term). Both terms contribute to cancelling out the apparent divergence at $z\to 1$.
The evolution equation conserves the total energy contained in the spectrum,
\begin{equation}
    \label{eq:consv_energy}
    \int_0^1 \rmd x \, D(x,t) = 1 \,,
\end{equation}
which can be confirmed directly from \eqref{eq:Evolution_eq}.

The leading parametric behavior of the splitting kernels can be derived by taking appropriate limits and is presented in Sec.~\ref{sec:spectrum}, cf. Eqs.~\eqref{eq:n1-limits}, \eqref{eq:nr1-limit-highop}, and \eqref{eq:ho-limits}. This results in,
\begin{equation}\label{eq:Rate_SoftLimit_lowop}
    \left.\cK(z,E,t) \right|_{t\ll \mfp} = \begin{cases}
        \frac{2\abar}{z(1-z)} \frac{1}{\mfp} \ln \left(\frac{\omegacbar(t)}{z(1-z)E} \right) & \quad\text{for } z(1-z)E \ll \omegacbar(t) \,,\\
        \frac{\abar \pi}{2} \frac{\qhat_0 t}{[z(1-z)]^2E} &\quad \text{for }  \omegacbar(t) \ll z(1-z)E \,,
    \end{cases}
\end{equation}
for $t \ll \mfp$, and 
\begin{equation}\label{eq:Rate_SoftLimit_highop}
    \left. \cK(z,E,t) \right|_{t\gg \mfp} = \begin{cases}
        \frac{2\abar}{z(1-z)} \frac{1}{\mfp} \ln \left(\frac{\omegaBH}{z(1-z)E} \right) &\quad  \text{for } z(1-z)E \ll \omegaBH \,,\\
        \abar \sqrt{\frac{\hat q}{[z(1-z)]^3 E}} &\quad  \text{for } \omegaBH \ll z(1-z)E \ll \omega_c(t) \,,\\
        \frac{\abar \pi}{2}\frac{\qhat_0 t}{[z(1-z)]^2 E} &\quad  \text{for }  \omega_c(t) \ll z(1-z)E \,,
    \end{cases}
\end{equation}
for $t \gg \mfp$, where $\omegacbar(t) = \frac{1}{2}\mu^2t$ and $\omega_c = \frac12 \hat q t^2$. For the analytical estimates in this section we neglect the running of $\hat q$, but this will be included in the numerics presented in Sec.~\ref{sec:numerics}.

\begin{figure}
    \centering
    \includegraphics[width=\textwidth]{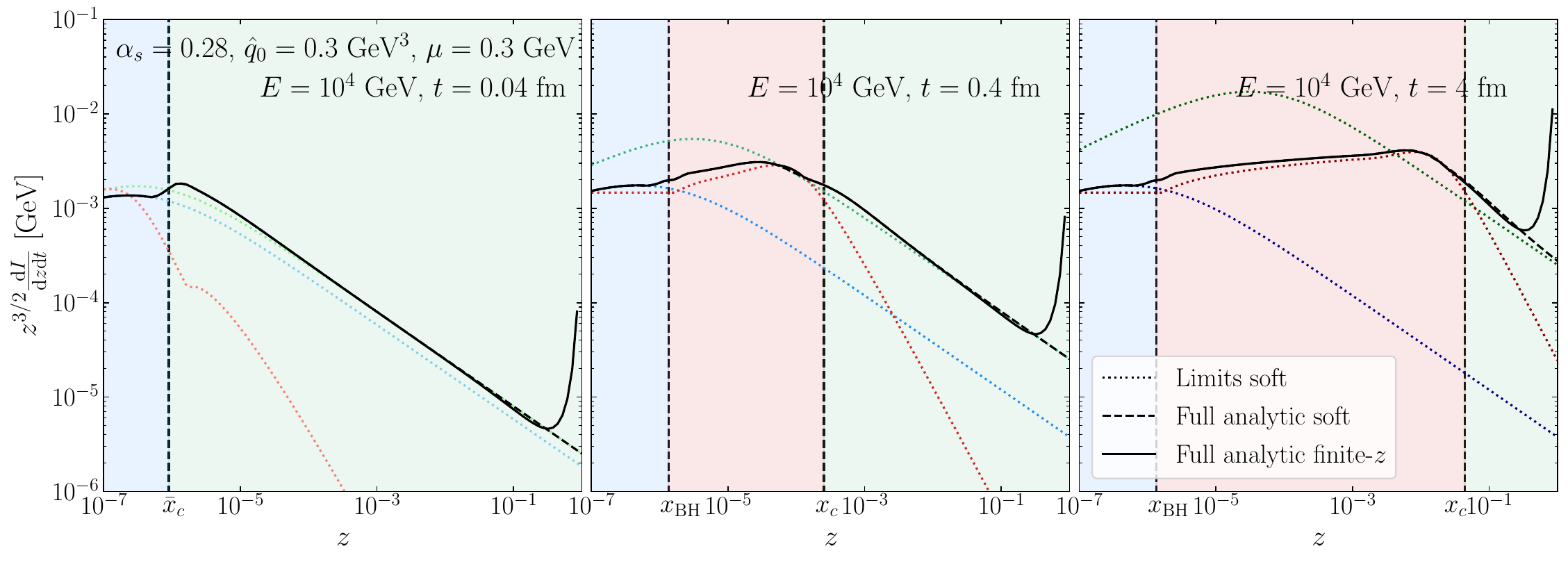}
    \includegraphics[width=\textwidth]{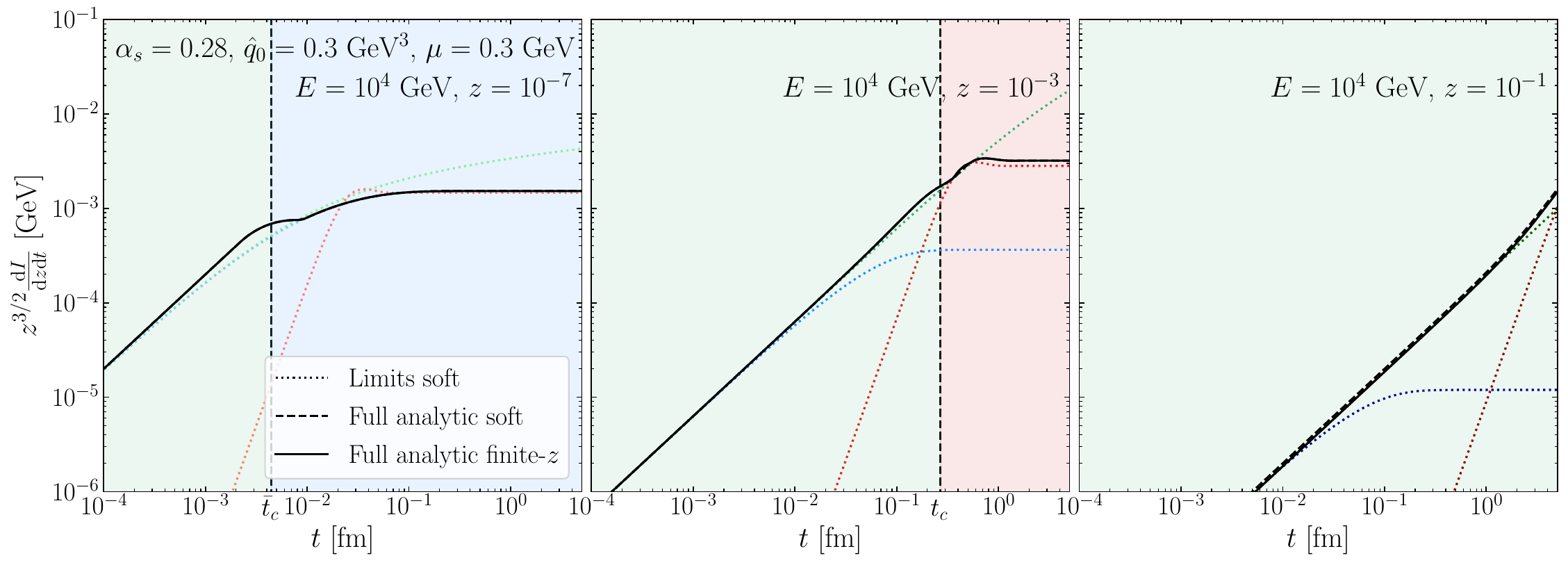}
    \caption{The medium-induced rate from Eqs.~\eqref{eq:Emission_Rate}, and \eqref{eq:spectrum_Full} for gluons for different emitted energy and propagation time (black dashed lines). The black solid lines include finite-$z$ corrections. The dotted lines are the $N_r = 1$, HO and $N = 1$ lines according to Fig.~\ref{fig:spectra-numerics}. The color shading corresponds to Fig.~\ref{fig:PhaseSpace_PhysicalProcesses} i.e. denoting the dominant scattering processes. The parameters correspond to $\lambda = 0.06$ fm.}
    \label{fig:wtdIdwdt_t}
\end{figure}

The rate $\rmd I/(\rmd z \rmd t)$ is plotted in Fig.~\ref{fig:wtdIdwdt_t}, in a similar manner to the spectrum in Fig.~\ref{fig:spectra-numerics}, and rescaled by a factor $z^{3/2}$ to highlight the behavior at small $z$. The panels in the upper part show the $z$-dependence for three different $t$ (early, mid-, and late times), and has a very similar structure to that of the spectrum. The color shading corresponds to Fig.~\ref{fig:PhaseSpace_PhysicalProcesses}, i.e. the dominant scattering processes. The dotted lines are the $N_r = 1$, HO and $N = 1$ lines from Sec.~\ref{sec:spectrum} as in Fig.~\ref{fig:spectra-numerics}, where $\omega \to zE$ was used. Similarly, the dashed line is the full solution in the soft limit with $\omega\to zE$. The solid black lines are from  Eq.~\eqref{eq:Full_emission_kernel} and they contain finite-$z$ corrections from App.~\ref{app:Kernels_finitez}. The finite-$z$ corrections change the rate for hard emissions around $z\sim1$. The panels in the lower part show the time dependence of the rate for a fixed emitted energy.

%%%%%%%%%%%%%%%%%%%%%%%%%%%%%%%%%%%%%%%%%%%%%%%%%%%%%
\subsection{Analytic solutions of the evolution equation}
\label{sec:analytic_solutions_eveq}
%%%%%%%%%%%%%%%%%%%%%%%%%%%%%%%%%%%%%%%%%%%%%%%%%%%%%

The evolution equation \eqref{eq:Evolution_eq} is readily solved by numerical evaluation, which will be discussed in Sec.~\ref{sec:numerics}. Here we will discuss limiting cases where analytical solutions can be found. We can find such solutions at early times (considering only one emission) and at late times (considering many soft emissions). These are by now well-known limiting cases. Finally, we also consider the novel case of the evolution equation at intermediate times, where both rare, hard emissions and multiple, soft emissions can occur in sequence according to their respective allowed phase space of emissions, given in Fig.~\ref{fig:PhaseSpace_PhysicalProcesses}.

To simplify our discussion, in this section we will neglect the Bethe--Heitler regime. We will nevertheless include it in the full numerical solutions presented in Sec.~\ref{sec:numerics}.

\begin{figure}
    \centering
    \includegraphics[width=\textwidth]{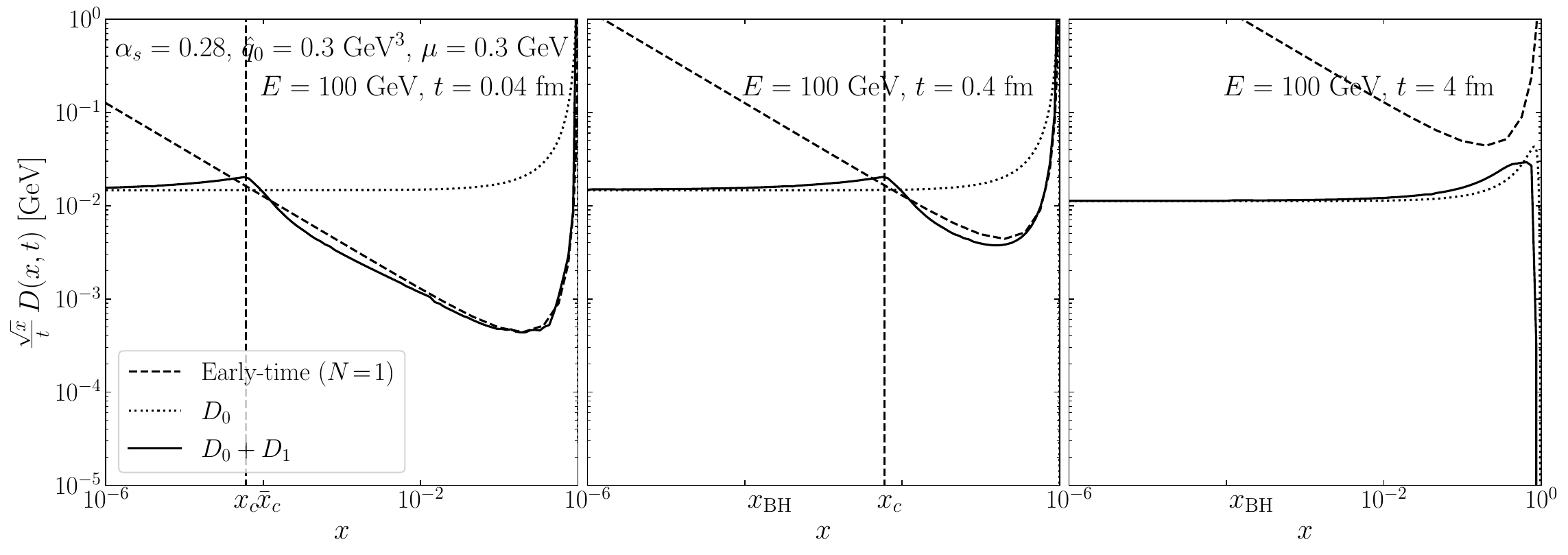}
    \caption{Analytic solutions of the energy distribution in different approximations (Eqs.~\eqref{eq:D_early0}, \eqref{eq:distr-coh} and \eqref{eq:evolution-eq-D1}). With these parameters $\lambda=0.06$ fm, and thus the different panels are $t<\lambda$, $\lambda <t < L_c$, and $t\lesssim L_c$.}
    \label{fig:analytic-D}
\end{figure}

\paragraph{Early time evolution:}
At an early stage of medium propagation, the leading parton has little time to interact with the medium which also translates into a small probability of splitting. Given our previous discussion, the natural medium scale to compare with is the mean free path $\mfp$. Hence, at $t \lesssim \mfp$ we consider a single splitting, leading to 
\begin{align}
    \label{eq:D_early0}
    D(x,t)&\simeq \delta(1-x)\left[1-\int_0^ t\rmd s \int_0^1 \rmd z \,  z \cK(z,xE, s) \right] + \int_0^ t\rmd  s  \int_x^1 \rmd z\,\cK\left(z,\frac{x}{z}E, s\right)\delta\left(1-\frac{x}{z}\right)\nn
    &= \int_0^t \rmd s \, x\cK(x,E,s) \,,
\end{align}
where we dropped the term proportional to $\delta(1-x)$, which is only important for energy conservation, cf. Eq.~\eqref{eq:consv_energy}. Using the results in Eq.~\eqref{eq:Rate_SoftLimit_lowop}, we find that
\begin{equation}
    \label{eq:D_early}
    D(x,t) \simeq
    \begin{cases}
        2 \abar \frac{t}{\lambda}\frac{1}{1-x}\ln\left(\frac{\omegacbar(t)}{x(1-x)E}\right) & \quad \text{for } x\ll\bar x_c\,,\\
        \frac{\pi \abar}{4} \frac{\qhat_0}{E}\frac{t^2}{x(1-x)^2} & \quad \text{for } \bar x_c \ll x \ll 1-\bar x_c\,,\\
    \end{cases}
\end{equation}
where we have defined
$\bar x_c  =  \omegacbar(t)/E$. For $x\ll\bar x_c$, corresponding to the single soft or Bethe--Heitler scattering regime, a characteristic $D \sim \ln 1/x$ structure appears that is similar to the DGLAP energy distribution in vacuum. On the contrary, for $x\gg\bar x_c$, corresponding to the single hard scattering regime, $D \sim 1/x$, and the two regimes are separated by $\bar x_c$. 
The limiting case of Eq.~\eqref{eq:D_early} is shown in Fig.~\ref{fig:analytic-D} with dashed lines for different times (different panels). Since we, as we move onward, will largely neglect the description of the infared regime, we only used the single hard scattering (green) contribution that is valid $x>\bar x_c$.

Formally, the early-time expansion breaks down when $t > \mfp$ which is also the characteristic time when multiple interactions with the medium become important. Finally, we do not expect the early time solution to hold for for $x<\abar^2\omega_c/E$ where multiple emissions play an important role.

%%%%%%%%%%%%%%%%%%%%%%%%%%%%%%%%%%%%%%%%%%%%%%%%%%%%%
\paragraph{Late time evolution:}
The evolution equation can be solved exactly if one assumes coherent scatterings dominate for all momentum fractions $x$ and all times $t$. This approximation is most sound when $t \gg  L_c$, as seen in Fig.~\ref{fig:PhaseSpace_PhysicalProcesses}. The analytical solution neglects the Bethe--Heitler, which will make the solution less reliable for very small $x<\omegaBH/E$. We will call this solution $D_0$ as it serves as a baseline for subsequent calculations. It is the solution to the evolution equation
\begin{align}
    \label{eq:ev_equation-D0}
    \frac{\partial}{\partial t} D_0(x,t)&= 
    \int_{x}^1 \rmd z \, \cK_{\rm coh}\left(z,\frac{x}{z}E,t\right)D_0\left(\frac{x}{z},t\right)-\int_0^1 \rmd z  \,z \cK_{\rm coh}\left(z,x E,t\right)D_0(x,t) 
\end{align}
where $\cK_{\rm coh}$ is the soft limit of the harmonic oscillator regime, see the middle term in Eq.~\eqref{eq:Rate_SoftLimit_highop}.
The solution from Ref.~\cite{Blaizot:2013hx} is
\begin{equation}
    \label{eq:distr-coh}
    D_0(x,\tau) = \frac{\tau}{\sqrt{x}(1-x)^{3/2}} \rme^{-\pi \frac{\tau^2}{1-x}} \,,
\end{equation}
where we dropped the $\delta(x)$ zero mode term, which is only important for energy conservation. We have also defined the re-scaled evolution variable which absorbs the energy scale,
\begin{equation}
    \label{eq:tau}
    \tau = \bar \alpha \sqrt{\frac{\hat q_0}{E}}t \,.
\end{equation}
The solution has the $D_0\sim \tau/\sqrt x$ shape characteristic of the turbulent cascade. Equation~\eqref{eq:distr-coh} is shown in Fig.~\ref{fig:analytic-D} with dotted lines for different times. 

%%%%%%%%%%%%%%%%%%%%%%%%%%%%%%%%%%%%%%%%%%%%%%%%%%%%%
\paragraph{Intermediate time evolution:}
To reach an approximate solution for intermediate times it is useful to cast the evolution in different variables.
Following Ref.~\cite{Blaizot:2015jea}, and defining $\xi = x/z$ in the gain term and $\xi =  xz$ in the loss term the evolution can be rewritten as
\begin{equation}
    \label{eq:ev_equation2}
    \frac{\partial}{\partial \tau} D(x,\tau)=\int_x^1\rmd\xi \,P(x,\xi,\tau)D(\xi,\tau)-D(x,\tau)\int_0^x\rmd\xi \,P(\xi,x,\tau)\,,
\end{equation}
where $P(x,\xi,\tau) = \frac{1}{\abar}\sqrt{\frac{E}{\qhat_0}} \frac{x}{\xi^2}\cK(\frac{x}{\xi},\xi E,t)$ and $\tau$ is defined in \eqref{eq:tau}. This form of the evolution equation clearly shows the cancellation of divergences between the gain and loss terms at $\xi  \to  x$.

At early times, $t < \mfp$, the emissions are governed by single interactions with the medium. In the regime $\mfp  <  t  <  L_c$ you also have to take into account that coherent emissions play a role in the soft regime $\omega < \omega_c(t)$ (neglecting Bethe--Heitler emissions). However, the region above $\omega_c(t)$ will contain emissions from single hard scatterings, see Eq.~\eqref{eq:Rate_SoftLimit_highop} (third line).
In that case the kernel will be modified to include both a soft and a hard component, divided by the $\omega_c(t)$ separation line, as follows
\begin{equation}
    P(x,\xi,\tau) = \theta_c(x,\xi,\tau) P_{\rm coh}(x,\xi,\tau) + \theta_h(x,\xi,\tau) P_{\rm hard}(x,\xi,\tau)\,,
\end{equation}
where the coherent and hard emission kernels were defined in Eq.~\eqref{eq:Rate_SoftLimit_highop}. After changing variables they become
\begin{align}\label{eq:rates_P}
    P_{\rm coh}(x,\xi,\tau)&=\sqrt{\frac{\xi}{x}} \frac{1}{(\xi-x)^{3/2}}  \qquad \text{for } x \ll x_c \text{ or } \xi-x_c \ll x \,,\nn
    P_{\rm hard}(x,\xi,\tau)&=\frac{\pi}{2\abar}  \, \frac{\xi}{x} \frac{1}{(\xi -x)^2} \tau 
    \qquad \text{for } x_c \ll x \ll \xi- x_c \,,
\end{align}
where $x_c \equiv \omega_c(t)/E= \tau^2/(2\abar^2)$ and we have assumed that $x_c \ll 1$ in the limits.
These conditions can be encoded in a set of Heaviside theta-functions to make sure that each kernel is used solely in its regime of validity, namely
\footnote{In the case where $x_c$ is not small the conditions are slightly more complicated, and the hard regime is encoded in $\theta_h(x,\xi,\tau) = \Theta\left(x-\frac{\xi}{2}\left(1-\sqrt{1-\frac{4 x_c}{\xi}}\right)\right)\Theta\left(\frac{\xi}{2}\left(1+\sqrt{1-\frac{4 x_c}{\xi}}\right)-x\right)$, and $
\theta_c = 1- \theta_h$.}
\begin{align}
    \theta_c(x,\xi,\tau) &= \Theta(x_c-x) + \Theta(x_c+x-\xi) - \Theta(x_c-x)\Theta(x_c+x-\xi) \,, \nn
    \theta_h(x,\xi,\tau) &= \Theta(x-x_c)\Theta(\xi-x_c-x) \,.
\end{align}
The now $\tau$-dependent separation line $x_c(\tau)$ distinguishes the different regimes. 

The full solution can be written as
\begin{equation}
    D(x,\tau) = D_0(x,\tau)+\delta D(x,\tau)\,,
\end{equation}
where $D_0(x,\tau)$ is a solution to the coherent, soft kernel defined in Eq.~\eqref{eq:distr-coh}, and $\delta D(x,\tau)$ is a correcting factor.

Inserting this into the evolution equation~\eqref{eq:ev_equation2}, we get
\begin{align}\label{eq:evolution-eq-D0-deltaD}
    \frac{\partial}{\partial \tau} D(x,\tau)&=\int_x^1\rmd\xi \,P(x,\xi,\tau)D_0(\xi,\tau)-D_0(x,\tau)\int_0^x\rmd\xi \,P(\xi,x,\tau)\nn
    &+\int_x^1\rmd\xi \,P(x,\xi,\tau)\delta D(\xi,\tau)-\delta D(x,\tau)\int_0^x\rmd\xi \,P(\xi,x,\tau)\,,
\end{align}
Taking into account that $\theta_c + \theta_h = 1$, one can rewrite the kernel as
\begin{align}
    P(x,\xi,\tau) &= P_{\rm coh}(x,\xi,\tau) + \delta P(x,\xi,\tau)\,,
\end{align}
where $\delta P \equiv(P_{\rm hard} - P_{\rm coh}) \theta_h $. Inserting the new kernel into Eq.~\eqref{eq:evolution-eq-D0-deltaD}, the term $\partial D_0/\partial \tau$ cancels, and we are left with an iterative formula for $\delta D$,
\begin{align}\label{eq:evolution-eq-deltaD}
    \delta D(x,\tau)&=  D_1(x,\tau)\nn
    &+\int^\tau_0 \rmd \sigma\int_x^1\rmd\xi \,P(x,\xi,\sigma)\delta D(\xi,\sigma)-\int^\tau_0 \rmd \sigma\delta D(x,\sigma)\int_0^x\rmd\xi \,P(\xi,x,\sigma)\,.
\end{align}
Here we have defined the leading term in the correction as
\begin{align}
    \label{eq:evolution-eq-D1}
    D_1(x,\tau)&=\int^\tau_0 \rmd \sigma \int_x^1\rmd\xi \,\delta P(x,\xi,\sigma)D_0(\xi,\sigma)-\int^\tau_0 \rmd \sigma D_0(x,\sigma)\int_0^x\rmd\xi \, \delta P(\xi,x,\sigma)\,,
\end{align}
which is given entirely in terms of known functions.
To capture the main modifications with respect to the purely coherent solution $D_0(x,\tau)$, it is sufficient to keep only $D_1(x,\tau)$. This is sound because $D_0(x,t=0)=\delta(1-x)$, implying that $\delta D(x,t=0)=0$. Therefore, all terms going as $\sim\delta D$ start out small. At later times, single hard emissions are rare, and thus $\delta D$ becomes less and less important.

Based on this discussion it is reasonable to assume that one can approximate the intermediate time solution by the sum of the two leading terms $D_0+D_1$. For this to be true it must be checked that it reproduces the correct behavior at early and late times. 
The early times expansion is
\begin{align}
    \lim_{\tau \to 0}(D_0(x,\tau)+D_1(x,\tau))
    &\simeq \int_0^\tau\rmd \sigma \int_x^1\rmd\xi \,P_{\rm coh}(x,\xi,\sigma) \delta(1-\xi)\nn
    &+ \int_0^\tau\rmd \sigma \int_x^1\rmd\xi \,\left[P_{\rm hard}(x,\xi,\tau)-P_{\rm coh}(x,\xi,\tau)\right] \delta(1-\xi)\nn
    &= \frac{\pi}{4 \abar} \frac{\tau^2}{x(1-x)^2}\,,
\end{align}
where we kept the leading term and ignored virtual terms containing $\delta(1-x)$. Therefore, the sum $D_0 + D_1$ reproduces the hard part of the early time expansion given in Eq.~\eqref{eq:D_early}. Moreover, at late times ($t > L_c$), the phase space for hard ($\omega > \omega_c$) emissions vanishes, and thus $D_1 \to 0$. Hence, at late times the intermediate time solution simply goes to the late time solution $D_0$. Consequently, we expect that the sum of the two first terms $D_0+D_1$ to provide a decent approximation of the true solution at all times. One can systematically calculate corrections to this solution by iterating Eq. \eqref{eq:evolution-eq-deltaD}. 

The early time solution Eq.~\eqref{eq:D_early0} (dashed), the soft limit of the HO approximation~\eqref{eq:distr-coh} (dotted), and the first correction $D_0+D_1$ are shown in Fig.~\ref{fig:analytic-D} with full lines for different times. For short lengths (left panel), $D_0+D_1$ closely resembles the early time solution, as expected. For late times $D_0+D_1$ reduces to $D_0$, as there is not any phase space for hard emissions left. The $D_0$ presents the small $x$ tail $D_0\sim\sqrt{x}$ characteristic for turbulence~\cite{Blaizot:2013hx}. At intermediate times we see that $D_0+D_1$ goes to $D_0$ at low $x$, while at high $x$ there is a suppression due to the lack of coherent scatterings at early times. Qualitatively, Fig.~\ref{fig:analytic-D} resembles the full numerical solution shown in Fig.~\ref{fig:D_t}. In order to compare the two figures, we have marked the value of $x_\text{\tiny BH} \equiv \omegaBH/E$ in Fig.~\ref{fig:analytic-D}. We have also not included the color coding in this figure since it does not include the physics from all the relevant regimes represented in Fig.~\ref{fig:PhaseSpace_PhysicalProcesses}.

%%%%%%%%%%%%%%%%%%%%%%%%%%%%%%%%%%%%%%%%%%%%%%%%%%%%
\section{Numerical evaluation of the medium cascade}
\label{sec:numerics}
%%%%%%%%%%%%%%%%%%%%%%%%%%%%%%%%%%%%%%%%%%%%%%%%%%%%

In Sec.~\ref{sec:spectrum} we presented an effective framework that describes medium-induced emissions up to arbitrary precision. By using this framework we showed in Sec.~\ref{sec:multiple} how different scattering processes contribute to multiple induced emissions. Based on the properties of the medium (e.g. length, mean free path), not all induced emissions are necessary to resum (or to consider many of them). For example, induced emissions from hard scatterings are not as important to resum as emissions from multiple soft scatterings. We developed a simple analytic model to include a single hard emission correction to the resummation of the multiple soft ones. In Sec.~\ref{sec:multiple}, we made several simplifications e.g. we simplified the emission kernel $\cK$, and neglected Bethe--Heitler emissions. It is important to understand how the energy distribution behaves also without using these simplifications and to understand the error introduced by employing them. That is the subject of this section.

\begin{figure}
    \centering
    \includegraphics[width=\textwidth]{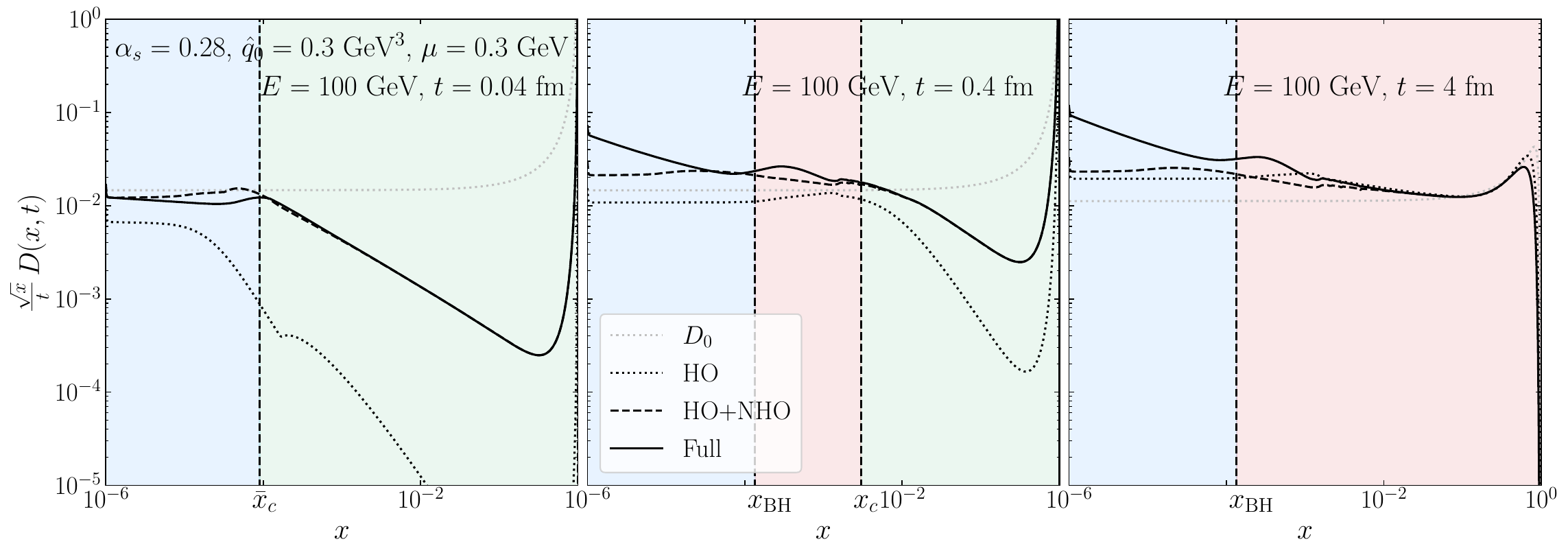}
    \caption{The energy distribution solution of the medium-induced emission evolution Eq.~\eqref{eq:Evolution_eq} at different times (different panels) and using different scattering mechanisms to calculate the emission rate in Eq.~\eqref{eq:Emission_Rate} (different line styles). The vertical dashed lines and the shaded areas separate different scattering regions, where on the first panel $t<\lambda$ and on the last panel $t>L_c$.}
    \label{fig:D_t}
\end{figure}

In this section, we evaluate Eq.~\eqref{eq:Evolution_eq} numerically using the kernel in our new framework. For more details about the numerical implementation, see App.~\ref{app:Numeric_impl} and the complementary code \cite{MedKernels:2022}. 
%(\url{https://github.com/adam-takacs/kernels.git}). 
The resulting energy distribution is presented in Fig.~\ref{fig:D_t}, where the different lines stem from using emission kernels $\cK(z,E,t)$ at varying levels of approximation. The three panels correspond to three different time stages in the evolution: $t<\lambda$, $\lambda <t < L_c$, and $t> L_c$. As a reminder, there are no rare hard splittings for $t>L_c$, which is evident on the rightmost panel of Fig.~\ref{fig:D_t}. 

The evolution starts at $t=0$, with a single gluon of energy distribution $D(x,0)=\delta(1-x)$ with $E=100$ GeV energy. We solely use gluons during the evolution for simplicity, however our formalism is valid for other flavors too. The kernels we use include finite-$z$ corrections, see App.~\ref{app:finitez-ioe} for more details. The vertical dashed lines in Fig.~\ref{fig:D_t} separate the regions where different scattering processes dominate, which are the same regions as in Fig.~\ref{fig:PhaseSpace_PhysicalProcesses}. The phase space for emissions is determined by comparing $z(1-z)E$ with $\omegacbar(t),\omegaBH$, and $\omega_c(t)$ in the relevant regions, as it is described in App.~\ref{app:Kernels_finitez}. The turbulent cascade solution $D_0(x,t)$ from Eq.~\eqref{eq:distr-coh} is also shown as a baseline, and was discussed in more detail in Sec.~\ref{sec:analytic_solutions_eveq}. 

It is important to note that vacuum emissions are not included in the current study, and thus many important effects (e.g. vacuum fragmentation, medium resolution, color coherence), which are essential if one wishes to compare to measurements, will not be discussed in this work. We refer the interested readers to Refs.~\cite{Mehtar-Tani:2012mfa,Kurkela:2014tla,Caucal:2018dla,Dominguez:2019ges,Barata:2021byj} for further details.

%%%%%%%%%%%%%%%%%%%%%%%%%%%%%%%%%%%%%%%%%%%%%%%%%%%%
\paragraph{Harmonic Oscillator:}
%%%%%%%%%%%%%%%%%%%%%%%%%%%%%%%%%%%%%%%%%%%%%%%%%%%%
A simple and much studied method of solving the evolution equation Eq.~\eqref{eq:Evolution_eq} is by using the harmonic oscillator (HO) approximation of the emission rate (see for example in Refs.~\cite{Schenke:2009gb,Blaizot:2013hx,Blaizot:2015jea,Caucal:2018dla,Caucal:2019uvr,Caucal:2020xad,Kutak:2018dim,Blanco:2020uzy}). The kernel is then simply given by
\begin{equation}
    \cK(z,E,t)\approx2\left.\frac{\rmd^2 I^{\text{\tiny HO}}}{\rmd z\rmd t}\right|_{E}\,,
\end{equation}
where the HO approximation was discussed in Sec.~\ref{sec:ioe} and in the numerics we used Eq.~\eqref{eq:HO_rate_finitez}. In this approximation, the induced emissions originate from multiple soft elastic scatterings with the medium, and we expect this process to dominate for $t\gg\lambda$ and $\xBH\ll x\ll x_c(t)$ (red region in Fig.~\ref{fig:PhaseSpace_PhysicalProcesses}). 

The energy distribution obtained by using the HO approximation is shown in Fig.~\ref{fig:D_t} with black dotted lines. The HO spectrum exhibits the well know turbulent cascade behavior~\cite{Blaizot:2013hx,Blaizot:2015jea}, resulting in a characteristic tail $D(x\ll1,t)\sim t/\sqrt x$, which can be seen on the plot as horizontal lines at small $x$. This is also a feature of the simplified analytic solution $D_0$. The turbulent cascade involves a constant flux of energy propagating in time to $x\to0$. The turbulence appears below $x<x_c(t)$, and the energy increases with time due to the time dependent emission phase space (see also $\omega_c(t)$ in Fig.~\ref{fig:PhaseSpace_PhysicalProcesses}). 

Above $x > x_c(t)$ the HO kernel switches from $\cK\sim1/z^{3/2}$ to $1/z^{3}$ and thus the energy distribution starts going as $D(x,t)\sim1/x^{2}$. This is especially visible in the middle panel of Fig.~\ref{fig:D_t}.

It is interesting to note that due to the running of $\hat q$ (see Eq.~\eqref{eq:qhat}), deviations from the pure $D\sim1/\sqrt x$ are expected in the HO solution. A consequence of this can be seen as a small kink on the dotted curves (most visible in the left panel at $x \approx 10^{-4}$). Below the kink, $\hat q(\omega)\to\hat q_0$ is used.

%%%%%%%%%%%%%%%%%%%%%%%%%%%%%%%%%%%%%%%%%%%%%%%%%%%%
\paragraph{Improved Opacity Expansion:}
%%%%%%%%%%%%%%%%%%%%%%%%%%%%%%%%%%%%%%%%%%%%%%%%%%%%
The IOE from Sec.~\ref{sec:ioe} makes it possible to extend the HO description to include rare hard scatterings, covering both red and green regions in Fig.~\ref{fig:PhaseSpace_PhysicalProcesses}. Here, the evolution equation is solved using the IOE kernel, which is
\begin{equation}
    \cK(z,E,t)\approx2\left.\frac{\rmd^2 I^{\rm IOE}}{\rmd z\rmd t}\right|_{E}\,.
\end{equation}
This is valid from early to late times, for energies above the Bethe--Heitler regime $x\gg\xBH$. 

The energy distribution obtained using the IOE (HO+NHO is used in the numerical implementation from Eqs.~\eqref{eq:HO_rate_finitez}~-~\eqref{eq:NHO_rate_finitez}) is shown in Fig.~\ref{fig:D_t} with black dashed lines. In the $x \ll x_c(t)$ region the HO dominates and $D(x,t)$ qualitatively does not change by including hard emissions. The turbulent tail $D\sim1/\sqrt{x}$ is still present. The offset between the HO and HO+NHO results originates from the effective contribution of the NHO term to $\hat q\approx\hat q_0\ln\frac{Q_r^2}{\mu_\ast^2}(1+1.016\ln^{-1}\frac{Q_r^2}{\mu_\ast^2})$. 

In the region $x> x_c(t)$, the distribution function changes from the purely HO result, due to the inclusion of the NHO corrections. The splitting function here behaves as $\cK\sim 1/z^2$, resulting in the distribution going as $D(x)\sim 1/x$, which is visible in the green region. Our analytic result from Sec.~\ref{sec:analytic_solutions_eveq}, given by $D_0+D_1$, includes a single hard emission in addition to the HO cascade. Comparing the analytic results in Fig.~\ref{fig:analytic-D} with the numerical ones in Fig.~\ref{fig:D_t} it is evident that the analytical result succeeds in capturing qualitatively the behavior induced by using the IOE kernel.

Several models are based mainly on the $N=1$ term, see Refs.~\cite{Idilbi:2008vm,Ovanesyan:2011xy,Majumder:2013re}, and also Refs.~\cite{Guo:2000nz,Wang:2013cia}. However, these models miss the soft scatterings that are present in a large medium with $L>\mfp$. Soft scatterings are crucial to include to achieve an accurate description of the energy distribution for $x < x_c(t)$. We would like to also emphasize that by numerically solving Eq.~\eqref{eq:Evolution_eq}, we also resum hard emissions and thus we consider the possibility of emitting arbitrarily many of them.

%%%%%%%%%%%%%%%%%%%%%%%%%%%%%%%%%%%%%%%%%%%%%%%%%%%%
\paragraph{Full solution:}
%%%%%%%%%%%%%%%%%%%%%%%%%%%%%%%%%%%%%%%%%%%%%%%%%%%%
Finally, we present the energy distribution using our new picture. It is given by solving the evolution equation \eqref{eq:Evolution_eq} using the full kernel
\begin{equation}\label{eq:Full_emission_kernel}
    \cK(z,E,t)=2\frac{\rmd^2 I^{\rm Full}}{\rmd z\rmd t}=2
    \begin{cases}
        \frac{\rmd^2 I^{\rm ROE}}{\rmd z\rmd t}\,, & {\rm if}\quad z(1-z)E<\min(\bar\omega_c(t),\omegaBH)\,,\\
        \frac{\rmd^2 I^{\rm IOE}}{\rmd z\rmd t}\,, & {\rm otherwise}\,.
    \end{cases}
\end{equation}
This covers the full emission phase space in ($z,t,E$) and was derived in detail in Sec.~\ref{sec:spectrum}. In practice we used $N_r = 1$ and HO+NHO terms, derived from Eqs.~\eqref{eq:nr1-limit}, \eqref{eq:nr1-limit-highop}, \eqref{eq:ho-limits} and \eqref{eq:wdIdw_NHO}. This provides an excellent approximation to the true result, as can be seen when comparing to the numerical evaluation of the kernel in Fig.~\ref{fig:spectra-numerics}. 

The resulting energy distribution is shown as full black lines in Fig.~\ref{fig:D_t}. One can see that the results obtained by solely using the IOE (dashed lines) agrees very well with the full kernel for $x\gg \xBH$, but they start to differ when $x \lesssim \xBH$. It is clear that all the difference between these two curves comes from Bethe--Heitler emissions. An increasing tail appears at low $x$, which becomes more and more important for later times, due to the logarithmic soft limit of the ROE kernel. In this region the ROE kernel is $\cK\sim\ln(z)/z$, resulting in the energy distribution going as $D\sim\ln\frac{1}{x}$, which is similar to the DGLAP evolution in vacuum.

A new, interesting bump also appears in the full solution close to $\xBH$, which warrants an explanation. First, the energy flux that brings quanta from $x=1$ to $x=0$ is not the same on the two sides of $\xBH$. If the energy transport is more efficient from the right (red region), that will result in a slowing of the flux when going from higher to lower $x$. The bump can then be understood as a sediment of energy building up around $\xBH$. Secondly, the emission kernel defined in Eq.~\eqref{eq:Full_emission_kernel} is not smooth around the transition point $\min(\bar\omega_c(t),\omegaBH)$. This discontinuity introduces additional uncertainties in the true behavior of the energy distribution around $\xBH$. This transition would smoothen by including more orders of the expansion, which should be studied in the future. This uncertainty could also be connected with the observed bump. In our numerical implementation we introduced a smoothing function to minimize this uncertainty, but it is still present. In addition, other effects like $2\to2$ scattering process and thermal masses are also important here~\cite{Schlichting:2020lef}, which we do not presently discuss. A thorough study of this region is needed where all these effects are included, which motivates future work. 

Bethe--Heitler emissions have a soft divergence and thus an IR regulator $\omega_\min$ has to be introduced. A similar regulator was introduced in the numerical differential equation solver. For more details, see App.~\ref{app:Numeric_impl}.

%%%%%%%%%%%%%%%%%%%%%%%%%%%%%%%%%%%%%%%%%%%%%%%%%%%%
%%%%%%%%%%%%%%%%%%%%%%%%%%%%%%%%%%%%%%%%%%%%%%%%%%%%
\section{Conclusions and outlook}
\label{sec:conclusions}
%%%%%%%%%%%%%%%%%%%%%%%%%%%%%%%%%%%%%%%%%%%%%%%%%%%%
%%%%%%%%%%%%%%%%%%%%%%%%%%%%%%%%%%%%%%%%%%%%%%%%%%%%

In this paper, we studied emissions induced by elastic scatterings on the quark gluon plasma. We considered the interplay of the relevant length scales of the problem, which are the propagation length, the mean free path between scatterings, and the formation time. From these considerations, we derived the emergent energy scales that separate the induced emission spectrum into regimes governed by different scattering processes. The emerging hierarchy of scales and related processes is illustrated in Fig.~\ref{fig:PhaseSpace_PhysicalProcesses}. 

We presented a new theoretical framework consisting of different perturbative expansions, namely the opacity expansion (OE), the improved OE (IOE), and the resummed OE (ROE), which we derived rigorously. Together these are suitable to describe the induced emission spectrum in different regimes. We showed that at least one of the expansions is valid in every phase space ($\omega,t$) point, where $\omega$ is the energy of the emitted gluon and $t$ is the propagation time (length) in the medium.\footnote{For the general case valid beyond the strictly soft limit $\omega$ refers to the reduced energy of the three-body evolution $\omega = z(1-z)E$, where $E$ is the initial energy of the emitter and $z$ is the momentum sharing fraction.} While these expansions are formally simply reorganizations of the multiple scattering series, it is also important to note that none of the expansions \textit{when truncated at any fixed order} is valid everywhere. Relying on the multiple approaches to obtain the spectrum, our composite framework can account for an arbitrary number of medium interactions that can be soft or hard. Most importantly, it is systematically improvable and can, in principle, describe the emission spectrum (and rate) up to arbitrary precision, which has only been achieved numerically before~\cite{CaronHuot:2010bp,Feal:2018sml,Andres:2020vxs}. 

In the current work, we have elucidated the convergence properties up to second order in the studied resummations. Identifying the expansion structure in the different regimes opens for the possibility of studying the accuracy of resummations in the medium. Finally, our new description provides a quick and efficient way to evaluate the induced emission spectrum (and rate), which is an essential ingredient of medium-induced cascades and jet quenching study. Our implementation is available in an online repository \cite{MedKernels:2022}.

In order to tackle multiple emissions, we also studied the in-medium energy distribution $D(x,t)$ within our new formalism. This is an essential ingredient of jet quenching phenomenology as it describes how the energy of a leading particle gets distributed within a cascade. 
Having first identified the conditions to generate multiple emissions in the medium, we demonstrated the separation between early, rare emissions -- generated mainly by a single momentum exchange with the medium -- and a following cascade of soft splittings. 
We also showed that emissions are formed independently up to suppressed terms, justifying a posteriori the formulation of the cascade via a rate equation. 
We developed new analytic tools to combine the resummation of multiple soft with rare hard emissions, and thus we showed how different scattering processes appear on the level of the energy distribution. Finally, using numerical evaluation based on the previously derived full splitting kernels, we showed more rigorously the transition effects between different scattering regions. We identified the importance of the time dependent phase space separation, multiple Bethe--Heitler emissions, and the running of $\hat q$, to mention some.

Even though this work can immediately be applied to jet quenching phenomenology one should also tackle other challenges, such as vacuum emissions and coherence effects, which we have neglected here and left for future studies. We also acknowledge the importance of soft emissions for which one should study thermal masses, energy changing $2\to2$ scatterings, and thermalization~\cite{Ghiglieri:2015ala,Schlichting:2020lef} to correctly describe the very infrared regime close to the thermal scale and below. Moreover, other non-perturbative effects such as expanding, inhomogeneous medium are also important, in which direction our framework is extendable~\cite{Casalderrey-Solana:2012evi,Adhya:2019qse,Caucal:2020xad,Adhya:2021kws}. 

%%%%%%%%%%%%%%%%%%%%%%%%%%%%%%%%%%%%%%%%%%%%%%%%%%%%
%%%%%%%%  ACKNOWLEDGMENTS 
%%%%%%%%%%%%%%%%%%%%%%%%%%%%%%%%%%%%%%%%%%%%%%%%%%%%
\acknowledgments
We thank Paul Caucal, Alexandre Falc\~ao, Aleksas Mazeliauskas, and S\"oren Schlichting for helpful discussions. We also thank Carlota Andres, Liliana Apolin\'ario, Fabio Dominguez, and Marcos Gonzalez Martinez for providing us with the numerical results for the spectrum computed in Ref.~\cite{Andres:2020vxs}. The work is supported by a Starting Grant from Trond Mohn Foundation (BFS2018REK01) and the University of Bergen.

%%%%%%%%%%%%%%%%%%%%%%%%%%%%%%%%%%%%%%%%%%%%%%%%%%%%
%%%%%%%%  APPENDIX 
%%%%%%%%%%%%%%%%%%%%%%%%%%%%%%%%%%%%%%%%%%%%%%%%%%%%
\appendix

%%%%%%%%%%%%%%%%%%%%%%%%%%%%%%%%%%%%%%%%%%%%%%%%%%%%
%%%%%%%%%%%%%%%%%%%%%%%%%%%%%%%%%%%%%%%%%%%%%%%%%%%%
\section{All order formulas for medium-induced spectrum in the soft limit}
\label{app:general-formula}
%%%%%%%%%%%%%%%%%%%%%%%%%%%%%%%%%%%%%%%%%%%%%%%%%%%%
%%%%%%%%%%%%%%%%%%%%%%%%%%%%%%%%%%%%%%%%%%%%%%%%%%%%

Here we present some general formulas for the different expansions, that aided us in calculating the spectrum in Sec.~\ref{sec:spectrum}. In all cases the full spectrum is given as the sum of all the terms $\omega \frac{\rmd I}{\rmd \omega} = \sum_n \omega \frac{\rmd I^{N=n}}{\rmd \omega}$. The expansion therefore only converges if the terms decrease sufficiently fast order by order $\omega \frac{\rmd I^{N=n+1}}{\rmd \omega}< \omega \frac{\rmd I^{N=n}}{\rmd \omega}$.

%%%%%%%%%%%%%%%%%%%%%%%%%%%%%%%%%%%%%%%%%%%%%%%%%%%%
\subsection{Opacity expansion}
\label{app:general-OE}
%%%%%%%%%%%%%%%%%%%%%%%%%%%%%%%%%%%%%%%%%%%%%%%%%%%%
Combining the formulas for the opacity expansion Eq.~\eqref{eq:oe-mom} and the spectrum Eq.~\eqref{eq:medspec-simplified} one can write the formula for the $n^{\rm th}$ term of the opacity expansion as\footnote{The gluon color factors appearing in $v$ are not trivial at $n^{\rm th}$ order and thus we refer the reader to App.~\ref{app:Kernels_finitez} for further discussion.}
\begin{align}
\label{eq:Spectrum_oe_general}
    &\omega \frac{\rmd I^{N=n}}{\rmd \omega}
    = (-1)^{n-1}\frac{4 \alpha_s C_R}{\omega}  \int_{\p_n,\dots,\p_1} \, \Sigma(\p_n^2)\frac{\p_n\cdot\p_1}{\p_n^2} v(\p_n-\p_{n-1})\dots v(\p_2-\p_1)\\
    &\times\rmR \,i\int_0^L \rmd t_n \int_0^{t_n} \rmd t_0 \int_{t_0}^{t_n} \rmd t_{n-1} \int_{t_0}^{t_{n-1}} \rmd t_{n-2}\dots \int_{t_0}^{t_2} \rmd t_1 \,
    \rme^{-i \frac{\p_n^2}{2\omega}(t_n-t_{n-1})}\dots \rme^{-i \frac{\p_1^2}{2\omega}(t_1-t_0)}\,,\nonumber
\end{align}
where $v(\p,s)$ is given in Eq.~\eqref{eq:v-mom} and $\Sigma(\p^2)$ in given in Eq.~\eqref{eq:Sigma-GW} in the GW model. It is useful to go to unitless integration variables, by defining $\sqrt{L/2 \omega}\p_k \to  \p_k$ and $\frac{t_k}{L} \to  t_k$,
\begin{align}\label{eq:general-OE}
    &\omega \frac{\rmd I^{N=n}}{\rmd \omega}
    = (-1)^{n-1}8 \pi \abar \left(\frac{L}{\lambda}\right)^n \frac{\omegacbar}{\omega}  \int_{\p_n,\dots,\p_1} \, \tilde\Sigma(\p_n^2)\frac{\p_n\cdot\p_1}{\p_n^2} \tilde v(\p_n-\p_{n-1})\dots \tilde v(\p_2-\p_1)\nn
    &\times \rmR \,i\int_0^1 \rmd t_n \int_0^{t_n} \rmd t_0 \int_{t_0}^{t_n} \rmd t_{n-1} \int_{t_0}^{t_{n-1}} \rmd t_{n-2}\dots \int_{t_0}^{t_2} \rmd t_1 \,
    \rme^{-i \p_n^2(t_n-t_{n-1})}\dots \rme^{-i \p_1^2(t_1-t_0)}\,.
\end{align}
Here we have used the unitless function
$\tilde v(\p) = (2\pi)^2\delta(\p) - \frac{\omegacbar}{\omega}\tilde\sigma(\p)$, where in the GW model 
\begin{align}\label{eq:unitless-GW-sigma-app}
    \tilde\sigma(\p)&=\frac{4 \pi}{\left(\p^2+\frac{\omegacbar}{\omega}\right)^2}\,,\nn
    \tilde\Sigma(\p^2)&=\frac{1}{\p^2+\frac{\omegacbar}{\omega}}\,.
\end{align}
The momentum and time integrals only depend on the unitless combination $\omegacbar/\omega$. This means that the spectrum will take the form $\omega \frac{\rmd I^{N=n}}{\rmd \omega} =  \abar \left(\frac{L}{\lambda}\right)^n h_n(\frac{\omega}{\omegacbar})$, for some function $h_n$. This would naively imply convergence for $L/\lambda<1$, but the exact for of the function $h_n$ must also be taken into account. In Sec.~\ref{sec:oe} we showed for $N = 1,2$ at low energy $\omega \ll \omegacbar$ that $h_n(\frac{\omega}{\omegacbar})$ is a finite function, meaning the series converges when $L/\lambda<1$. However, at high energy $\omega \gg \omegacbar$ the function takes the form $h_n(\frac{\omegacbar}{\omega}) \sim (\frac{\omegacbar}{\omega})^n\tilde h_n(\frac{\omegacbar}{\omega})$, where $\tilde h_n$ is finite, implying convergence when $\frac{L}{\lambda}\frac{\omegacbar}{\omega} < 1$. Proving this is true to all orders is deferred to future work.

One can also derive a corresponding formula for the rate $\omega \frac{\rmd I^{N=n}}{\rmd \omega \rmd L}$, by taking the length derivative of Eq.~\eqref{eq:general-OE}
\begin{align}\label{eq:general-OE-rate}
    &\omega \frac{\rmd I^{N=n}}{\rmd \omega \rmd L}
    = (-1)^{n-1}8 \pi \abar \frac{1}{L}\left(\frac{L}{\lambda}\right)^n \frac{\omegacbar}{\omega}  \int_{\p_n,\dots,\p_1} \, \tilde\Sigma(\p_n^2)\frac{\p_n\cdot\p_1}{\p_n^2} \tilde v(\p_n-\p_{n-1})\dots \tilde v(\p_2-\p_1)\nn
    &\times \rmR \,i \int_0^{1} \rmd t_0 \int_{t_0}^{1} \rmd t_{n-1} \int_{t_0}^{t_{n-1}} \rmd t_{n-2}\dots \int_{t_0}^{t_2} \rmd t_1 \,
    \rme^{-i \p_n^2(1-t_{n-1})}\rme^{-i \p_{n-1}^2(t_{n-1}-t_{n-2})}\dots \rme^{-i \p_1^2(t_1-t_0)}\,.
\end{align}

%%%%%%%%%%%%%%%%%%%%%%%%%%%%%%%%%%%%%%%%%%%%%%%%%%%%
\subsection{Resummed opacity expansion}
\label{app:general-rOE}
%%%%%%%%%%%%%%%%%%%%%%%%%%%%%%%%%%%%%%%%%%%%%%%%%%%%
One can also derive all order formulas for the resummed opacity expansion, by starting with Eq.~\eqref{eq:eoe-mom} and following the same procedure,
\begin{align}
\label{eq:Spectrum_roe_general}
    &\omega \frac{\rmd I^{N_r=n}}{\rmd \omega}
    = \frac{4 \alpha_s C_R}{\omega} \int_{\p_n,\dots,\p_1} \, \Sigma(\p_n^2)\frac{\p_n\cdot\p_1}{\p_n^2} \sigma(\p_n-\p_{n-1})\dots \sigma(\p_2-\p_1)\\
    &\times\rmR \,i\int_0^L \rmd t_n \int_0^{t_n} \rmd t_0 \int_{t_0}^{t_n} \rmd t_{n-1} \int_{t_0}^{t_{n-1}} \rmd t_{n-2}\dots \int_{t_0}^{t_2} \rmd t_1 \,
    \Delta(t_n,t_0) \rme^{-i \frac{\p_n^2}{2\omega}(t_n-t_{n-1})}\dots \rme^{-i \frac{\p_1^2}{2\omega}(t_1-t_0)}\,.\nonumber
\end{align}
The ROE and OE expansions are equivalent at infinite order at low opacity $L/\lambda \ll 1$, but at finite order the terms are mixed up. The terms containing a delta function in the potential $v$ in the OE are included in the Sudakov factor $\Delta$ in the ROE. One can see this by taking $\Delta \to 1$ and $v \to -\sigma$ in Eq.~\eqref{eq:Spectrum_roe_general} and Eq.~\eqref{eq:Spectrum_oe_general} respectively, in which case the expansions become exactly the same. This is however not a good approximation at any order.

After changing to unitless variables we arrive at
\begin{align}
\label{eq:Spectrum_ROE_general}
    &\omega \frac{\rmd I^{N_r=n}}{\rmd \omega}
    = 8 \pi \abar \left(\frac{L}{\lambda}\right)^n \left(\frac{\omegacbar}{\omega}\right)^n \int_{\p_n,\dots,\p_1} \, \tilde\Sigma(\p_n^2)\frac{\p_n\cdot\p_1}{\p_n^2} \tilde \sigma(\p_n-\p_{n-1})\dots \tilde \sigma(\p_2-\p_1)\\
    &\times \rmR \,i\int_0^1 \rmd t_n \int_0^{t_n} \rmd t_0 \int_{t_0}^{t_n} \rmd t_{n-1} \int_{t_0}^{t_{n-1}} \rmd t_{n-2}\dots \int_{t_0}^{t_2} \rmd t_1 \,
    \rme^{-\frac{L}{\lambda}(t_n-t_0)}\rme^{-i \p_n^2(t_n-t_{n-1})}\dots \rme^{-i \p_1^2(t_1-t_0)}\,,\nonumber
\end{align}
where we have used that in the static medium $\Delta(t_n,t_0) = \exp(-\frac{t_n-t_0}{\lambda})$. Again we use the unitless functions $\tilde\Sigma(\p^2)$ and $\tilde\sigma(\p)$ defined in Eq.~\eqref{eq:unitless-GW-sigma-app} for the GW model. In this case the time integrals can be done analytically
\begin{align}
    &i\int_0^1 \rmd t_n \int_0^{t_n} \rmd t_0 \int_{t_0}^{t_n} \rmd t_{n-1} \int_{t_0}^{t_{n-1}} \rmd t_{n-2}\dots \int_{t_0}^{t_2} \rmd t_1 \,
    \rme^{-\frac{L}{\lambda}(t_n-t_0)}\rme^{-i \p_n^2(t_n-t_{n-1})}\dots \rme^{-i \p_1^2(t_1-t_0)} \nn
    &=i^n\int_0^1 \rmd t_n \int_0^{t_n} \rmd t_0 \sum_{k=1}^{n}
    \frac{\rme^{-i \left(\p_k^2-i\frac{L}{\lambda}\right)(t_n-t_0)}}{\prod_{l \neq k}^n (\p_k^2-\p_l^2)}=\left(\cos{\frac{n \pi}{2}}+i \sin{\frac{n \pi}{2}}\right)\sum_{k=1}^{n}
    \frac{T\left(\p_k^2-i\frac{L}{\lambda}\right)}{\prod_{l \neq k}^n (\p_k^2-\p_l^2)}\,,
\end{align}
where we have defined the function
\begin{align}
    T(x)&=\int_0^1 \rmd t_n \int_0^{t_n} \rmd t_0 \,\rme^{-ix(t_n-t_0)}=\frac{1-ix-\rme^{-x}}{x^2}\,.
\end{align}
The reason this simplification was not possible in the OE is that the formula for the time integrals is only valid if $\p_k  \neq  \p_l$, while the OE has terms containing $\delta(\p_k-\p_{k-1})$. However, in the ROE there are no such deltas, so the formula is valid.
In the end the spectrum is 
\begin{align}\label{eq:general-rOE}
    \omega \frac{\rmd I^{N_r=n}}{\rmd \omega}
    &= 8 \pi \abar \left(\frac{L}{\lambda}\right)^n \left(\frac{\omegacbar}{\omega}\right)^n \int_{\p_n,\dots,\p_1} \, \tilde\Sigma(\p_n^2)\frac{\p_n\cdot\p_1}{\p_n^2} \tilde \sigma(\p_n-\p_{n-1})\dots \tilde \sigma(\p_2-\p_1)\nn
    &\times \sum_{k=1}^{n}
    \frac{\cos{\frac{n \pi}{2}}\rmR \,T(\p_k^2-i\chi)-\sin{\frac{n \pi}{2}}\rmI \,T(\p_k^2-i\chi)}{\prod_{l \neq k}^n (\p_k^2-\p_l^2)}\,.
\end{align}
The remaining momentum integrals will in the end be some function of $\omegacbar/\omega$ and opacity $\chi = L/\lambda$. As a reference, the real and imaginary parts of $T(\p^2-i\chi)$ are
\begin{align}\label{eq:T-real-imaginary}
    \rmR \,T(\p^2-i\chi)&=\frac{\chi(\p^4+\chi^2)+\p^4-\chi^2-\rme^{-\chi} \left((\p^4-\chi^2)\cos{\p^2}+2 \chi \p^2 \sin{\p^2}\right)}{(\p^4+\chi^2)^2}\nn
    -\rmI \,T(\p^2-i\chi)&=\frac{\p^2(\p^4+\chi^2)-2\chi \p^2 -\rme^{-\chi}\left((\p^4-\chi^2)\sin{\p^2}-2 \chi \p^2 \cos{\p^2}\right)}{(\p^4+\chi^2)^2}\,.
\end{align}
It is also possible to look at the limits of this function when the opacity $\chi$ is low or high. In the low opacity case $\chi \ll 1$ we have
\begin{align}\label{eq:general-rOE-low-op}
    \omega \frac{\rmd I^{N_r=n}}{\rmd \omega}
    &\simeq 8 \pi \abar \left(\frac{L}{\lambda}\right)^n \left(\frac{\omegacbar}{\omega}\right)^n \int_{\p_n,\dots,\p_1} \, \tilde\Sigma(\p_n^2)\frac{\p_n\cdot\p_1}{\p_n^2} \tilde \sigma(\p_n-\p_{n-1})\dots \tilde \sigma(\p_2-\p_1)\nn
    &\times\sum_{k=1}^{n}
    \frac{\cos{\frac{n \pi}{2}}(1-\cos{\p_k^2})+\sin{\frac{n \pi}{2}}(\p_k^2-\sin{\p_k^2})}{\p_k^4\prod_{l \neq k}^n (\p_k^2-\p_l^2)}\,.
\end{align}
In this case the momentum integrals will give a function of $\omegacbar/\omega$. As mentioned in Sec.~\ref{sec:roe} is it preferable to use the OE \eqref{eq:general-OE} at low opacity, as the order of opacity is mixed up in the ROE. 

In the high opacity case $\chi \gg 1$ you get
\begin{align}\label{eq:general-rOE-high-op}
    \omega \frac{\rmd I^{N_r=n}}{\rmd \omega}
    &\simeq 8 \pi \abar \frac{L}{\lambda} \left(\frac{\omegaBH}{\omega}\right)^n \int_{\p_n,\dots,\p_1} \, \tilde\Sigma(\p_n^2,\omegaBH)\frac{\p_n\cdot\p_1}{\p_n^2} \tilde \sigma(\p_n-\p_{n-1},\omegaBH)\dots \tilde \sigma(\p_2-\p_1,\omegaBH)\nn
    &\times\sum_{k=1}^{n}
    \frac{\cos{\frac{n \pi}{2}}+\sin{\frac{n \pi}{2}}\p_k^2}{(1+\p_k^4)\prod_{l \neq k}^n (\p_k^2-\p_l^2)}\,.
\end{align}
Here we see the emergence of the scale $\omegaBH=\frac12\mu^2\lambda$, which takes the place of $\omegacbar$ in the functions $\tilde \Sigma$ and $\tilde \sigma$, as emphasized in the above formula. The momentum integrals then only become some function of $\omegaBH/\omega$. Notice that this scales as $\sim L/\lambda$ at every order, meaning it can converge also when $L/\lambda>1$. 

In Sec.~\ref{sec:ioe} we calculated the high opacity limit to orders $N=1,2$. There it was clear that for low $\omega\ll \omegaBH$ the $N=2$ limit is subleading compared to the $N=1$ limit, while at high $\omega\gg \omegaBH$ the $N=2$ and $N=1$ limits are of the same order. This seems to imply that the expansion converges at low $\omega$, while it breaks down at high $\omega$. Again, proving this for all orders is deferred to future work.

Again the calculation for the rate $\omega \frac{\rmd I^{N_r=n}}{\rmd \omega\rmd L}$ is similar
\begin{align}\label{eq:general-rOE-rate}
    &\omega \frac{\rmd I^{N_r=n}}{\rmd \omega \rmd L}
    = 8 \pi \abar \frac{1}{L}\left(\frac{L}{\lambda}\right)^n \left(\frac{\omegacbar}{\omega}\right)^n \int_{\p_n,\dots,\p_1} \, \tilde\Sigma(\p_n^2)\frac{\p_n\cdot\p_1}{\p_n^2} \tilde \sigma(\p_n-\p_{n-1})\dots \tilde \sigma(\p_2-\p_1)\nn
    &\times \sum_{k=1}^{n}
    \frac{\cos{\frac{n \pi}{2}}\left[\chi-\rme^{-\chi}\left(\chi \cos{\p_k^2}-\p_k^2\sin{\p_k^2}\right)\right]+\sin{\frac{n \pi}{2}}\left[\p_k^2-\rme^{-\chi}\left(\p_k^2 \cos{\p_k^2}+\chi\sin{\p_k^2}\right)\right]}{(\p_k^4+\chi^2)\prod_{l \neq k}^n (\p_k^2-\p_l^2)}\,.
\end{align}

%%%%%%%%%%%%%%%%%%%%%%%%%%%%%%%%%%%%%%%%%%%%%%%%%%%%
\subsection{Improved opacity expansion}
\label{app:general-IOE}
%%%%%%%%%%%%%%%%%%%%%%%%%%%%%%%%%%%%%%%%%%%%%%%%%%%%
The spectrum for the improved opacity expansion can be written as the iterative equation
\begin{align}
    \omega \frac{\rmd I^{\rm IOE}}{\rmd \omega} &= -\frac{2\alpha_s C_R}{\omega^2}\rmR \int_0^\infty \rmd t_2 \int^{t_2}_0 \rmd t_1\int_{t_1}^{t_2}\rmd s \, \int \rmd^2 \z \nn
    &\times\partial_\x \cdot \partial_\y [{\cal K}_\text{\tiny HO}(\x,t_2;\z,s) \delta v(\z,s) {\cal K}(\z,s; \y,t_1)]_{\x=\y=0}\,,
\end{align}
where the zeroth order solution is the HO from Eq.~\eqref{eq:wdIdw_HO} has to be added. In the following we will make use of results from Ref.~\cite{Mehtar-Tani:2019tvy,Mehtar-Tani:2019ygg}. The harmonic oscillator propagator is
\begin{equation}
    \mathcal{K}_{\mathrm{HO}}\left(\boldsymbol{x}, t_{2} ; \boldsymbol{y}, t_{1}\right)=\frac{\omega}{2 \pi i S\left(t_{2}, t_{1}\right)} \exp \left(\frac{i \omega}{2 S\left(t_{2}, t_{1}\right)}\left[C\left(t_{1}, t_{2}\right) \boldsymbol{x}^{2}+C\left(t_{2}, t_{1}\right) \boldsymbol{y}^{2}-2 \boldsymbol{x} \cdot \boldsymbol{y}\right]\right)\,.
\end{equation}
The functions $S$ and $C$ are given implicitly by
\begin{equation}
\begin{aligned}
&{\left[\frac{\mathrm{d}^{2}}{\mathrm{~d}^{2} t}+\Omega^{2}(t)\right] S\left(t, t_{0}\right)=0, \quad S\left(t_{0}, t_{0}\right)=0, \quad \partial_{t} S\left(t, t_{0}\right)_{t=t_{0}}=1} \\
&{\left[\frac{\mathrm{d}^{2}}{\mathrm{~d}^{2} t}+\Omega^{2}(t)\right] C\left(t, t_{0}\right)=0, \quad C\left(t_{0}, t_{0}\right)=1, \quad \partial_{t} C\left(t, t_{0}\right)_{t=t_{0}}=0}\,,
\end{aligned}
\end{equation}
where the frequency $\Omega(t)$ is given by
\begin{equation}
\Omega(t)=\frac{1-i}{2} \sqrt{\frac{\hat{q}(t)}{\omega}}
\end{equation}
To continue it is useful to apply the formulas
\begin{align}
    \int_s^\infty \rmd t_2 \partial_\x {\cal K}_\text{\tiny HO}(\x,t_2;\z,s)\vert_{\x=0} &= \frac{-i \omega}{\pi} \frac{\z}{\z^2} \rme^{i \frac{\omega}{2} \Omega(s)^2 \frac{S(s,L)}{C(s,L)}\z^2}\nn
    \int_0^s \rmd t_1 \partial_\y {\cal K}_\text{\tiny HO}(\z,s;\y,t_1)\vert_{\y=0} &= \frac{-i \omega}{\pi} \frac{\z}{\z^2} \rme^{-i \frac{\omega}{2} \frac{C(0,s)}{S(0,s)}\z^2}\,.
\end{align}
Then the spectrum takes the following form
\begin{align}
\label{eq:IOE_subtracted}
    \omega \frac{\rmd I^{\rm IOE}}{\rmd \omega} = 
    \frac{2\abar}{\omega}\rmR \,i \int_0^L \rmd t_2 \int^{t_2}_0 \rmd t_1 \int \rmd^2 \x
    \,\rme^{-i \frac{\omega \Omega}{2}\tan{(\Omega(L-t_2))\x^2}}\delta v(\x)\frac{\x}{\x^2}\cdot\partial_\y {\cal K}(\x,t_2; \y,t_1)\vert_{\y=0}\,,
\end{align}
where we have used that for the brick medium $\qhat(t) = \Theta(L - t) \qhat$, the functions $S$ and $C$ are simply
\begin{equation}
    S\left(t_{2}, t_{1}\right)=\frac{1}{\Omega} \sin \Omega\left(t_{2}-t_{1}\right), \quad \text { and } \quad C\left(t_{2}, t_{1}\right)=\cos \Omega\left(t_{2}-t_{1}\right)\,.
\end{equation}
One can derive a formula for the IOE at arbitrary order, which was also explored in Ref.~\cite{Barata:2020sav}.
\begin{align}
    \omega \frac{\rmd I^{N_I=n}}{\rmd \omega}
    &= (-1)^{n-1}\frac{2 \abar}{\pi} \int\rmd^2\x_n \dots \rmd^2 \x_1 \, \frac{\x_n\cdot \x_1}{\x_n^2\x_1^2}\delta v(\x_n)\dots \delta v(\x_1)\nn
    &\times\rmR \,\int_0^L \rmd t_n \int_{0}^{t_n} \rmd t_{n-1} \dots \int_{0}^{t_2} \rmd t_1 \,
    \rme^{i\frac{\omega \Omega}{2}\left[\cot{(\Omega t_1)}\x_1^2-\tan{(\Omega(L-t_n))\x_n^2}\right]}\nn
    &\times\cK_\text{\tiny HO}(\x_n,t_n;\x_{n-1},t_{n-1})\dots \cK_\text{\tiny HO}(\x_2,t_2;\x_1,t_1)\,.
\end{align}
After changing to unitless variables by defining $\u_k = \sqrt{\mu^2\omega/(2\omegacbar)}\x_k$ and $s_k=\frac{t_k}{L}$,
\begin{align}\label{eq:IOE-spectrum-unitless}
    \omega \frac{\rmd I^{N_I=n}}{\rmd \omega}
    &= (-1)^{n-1}\frac{2 \abar}{\pi} \left(\frac{L}{\lambda}\right)^n\left(\frac{\omegacbar}{\omega}\right)^{n}
    \int\rmd^2\u_n \dots \rmd^2 \u_1 \, \frac{\u_n\cdot \u_1}{\u_n^2\u_1^2}\delta \tilde v(\u_n)\dots \delta \tilde v(\u_1)\nn
    &\times\rmR \,\int_0^1 \rmd s_n \int_{0}^{s_n} \rmd s_{n-1} \dots \int_{0}^{s_2} \rmd s_1 \,
    \rme^{i\frac{\sigma}{2}\sqrt{\frac{\omega_c}{\omega}}\left[\cot{(\sigma \sqrt{\frac{\omega_c}{\omega}} s_1)}\u_1^2-\tan{(\sigma \sqrt{\frac{\omega_c}{\omega}}(1-s_n))\u_n^2}\right]}\nn
    &\times\tilde\cK_\text{\tiny HO}(\u_n,s_n;\u_{n-1},s_{n-1})\dots \tilde\cK_\text{\tiny HO}(\u_2,s_2;\u_1,s_1)\,.
\end{align}
Here we have defined $\sigma=\frac{1-i}{\sqrt{2}}$ and the unitless functions
\begin{align}
\label{eq:K_HO_unitless}
    \delta \tilde v(\u)&= \frac12 \u^2 \ln{\left(\frac{\omega}{\omegacbar}\frac{\mu^2}{2 Q^2}\frac{1}{\u^2}\right)}\,,\nn
    \tilde\cK_\text{\tiny HO}(\u_2,s_2;\u_1,s_1)&=\frac{\sigma \sqrt{\frac{\omega_c}{\omega}}}{2\pi i \sin{\left(\sigma \sqrt{\frac{\omega_c}{\omega}} (s_2-s_1)\right)}}\nn
    &\times\rme^{\frac{i\sigma\sqrt{\frac{\omega_c}{\omega}}}{2\sin{\left(\sigma\sqrt{\frac{\omega_c}{\omega}}(s_2-s_1)\right)}}\left[\cos{\left(\sigma\sqrt{\frac{\omega_c}{\omega}}(s_2-s_1)\right)(\u_2^2+\u_1^2)-2\u_2\cdot \u_1}\right]}\,.
\end{align}
As the integrals only depend on $\sqrt{\frac{\omega_c}{\omega}}$ and $\frac{\omega}{\omegacbar}\frac{\mu^2}{2 Q^2}$ the IOE spectrum can be written as $\omega \frac{\rmd I^{N_I=n}}{\rmd \omega}  =  \frac{2 \abar}{\pi} \left(\frac{L}{\lambda}\right)^n\left(\frac{\omegacbar}{\omega}\right)^{n}f_n\left(\sqrt{\frac{\omega_c}{\omega}},\frac{\omega}{\omegacbar}\frac{\mu^2}{2 Q^2}\right)$ where the function $f_n$ is given by the integrals. 
The soft limit $\omega \ll \omega_c$ of the IOE expansion was discussed in detail in \cite{Barata:2020sav}, and also in Sec.~\ref{sec:ioe}.

In the hard limit $\omega \gg \omega_c$, the spectrum becomes
\begin{align}
    \omega \frac{\rmd I^{N_I=n}}{\rmd \omega}
    &\simeq (-1)^{n-1} \frac{2 \abar}{\pi} \left(\frac{L}{\lambda}\right)^n\left(\frac{\omegacbar}{\omega}\right)^{n}
    \int\rmd^2\u_n \dots \rmd^2 \u_1 \, \frac{\u_n\cdot \u_1}{\u_n^2\u_1^2}\delta \tilde v(\u_n)\dots \delta \tilde v(\u_1)\nn
    &\times \rmR \,\int_0^1 \rmd s_n \int_{0}^{s_n} \rmd s_{n-1} \dots \int_{0}^{s_2} \rmd s_1 \,
    \rme^{i\frac{\u_1^2}{2 s_1}}\nn
    &\times\tilde\cK_0(\u_n,s_n;\u_{n-1},s_{n-1})\dots \tilde\cK_0(\u_2,s_2;\u_1,s_1),
\end{align}
where the BDMPS propagator has gone to the vacuum propagator
\begin{equation}
    \tilde\cK_0(\u_2,s_2;\u_1,s_1)=\frac{1}{2\pi i (s_2-s_1)}
    \rme^{i\frac{i(\u_2-\u_1)^2}{2(s_2-s_1)}}\,.
\end{equation}
Notice that there is no remaining dependence on the BDMPS scale $\omega_c$. In Sec.~\ref{sec:ioe} we calculated the hard limit to first order, and the resulting expression \eqref{eq:wdIdw_NHO} is the same as the OE limit \eqref{eq:n1-limits} for high $\omega$. Whether this correspondence is also true at higher orders is an interesting question that will be explored in future work.

%%%%%%%%%%%%%%%%%%%%%%%%%%%%%%%%%%%%%%%%%%%%%%%%%%%%
%%%%%%%%%%%%%%%%%%%%%%%%%%%%%%%%%%%%%%%%%%%%%%%%%%%%
\section{General formulas for the emission rate in the soft limit}\label{app:rate}
%%%%%%%%%%%%%%%%%%%%%%%%%%%%%%%%%%%%%%%%%%%%%%%%%%%%

In this appendix, we gather the formulas relevant for computing the emission rate in the soft limit. The rate can be written as 
\begin{equation}
    \label{eq:rate-mom}
    \omega \frac{\rmd I}{\rmd \omega \rmd t} = \frac{4 \abar \pi}{\omega} \rmR \,i \int_0^t \rmd t_1 \int_{\p,\p_0} \, \Sigma(\p^2,t) \frac{\p \cdot \p_0}{\p^2} \Kc(\p,t; \p_0,t_1) \,,
\end{equation}
in momentum-space representation of the three-point function, and
\begin{equation}
    \omega\frac{\rmd I}{\rmd \omega \rmd t} = \frac{2 \abar}{\omega} \rmR \, i \int_0^t \rmd t_1 \int_\z \, v(\z,t) \frac{\z}{\z^2}\cdot \bdel_\y\left. \Kc(\z,t; \y,t_1)\right|_{\y=0} \,,
\end{equation}
in coordinate-space representation. Equation~\eqref{eq:rate-mom} can be employed directly to derive expressions for the rate in the OE and ROE, by simply inserting the expansions \eqref{eq:oe-mom} and \eqref{eq:eoe-mom}. We will not attempt at deriving higher-order corrections to these rates here, since they can be also be found for a medium with constant density by taking the appropriate derivative with respect to length on the expression for the spectrum. 

For the IOE, the harmonic oscillator spectrum is directly calculable, following the decomposition in Eqs.~\eqref{eq:ioe-ho} and \eqref{eq:ioe-nhos}. For the IOE rates, we find
\begin{align}
    \omega \frac{\rmd I^\ho}{\rmd \omega \rmd t} &= \frac{\abar}{2\omega} \rmR \,i \int_0^t \rmd t_1\int_\z \, \hat q(t) \z\cdot \bdel_\y \left. \Kc_\ho(\z,t;\y,t_1) \right|_{\y=0} \,,\\
    \omega\frac{\rmd I^\ioe}{\rmd \omega \rmd t} &= \frac{2 \abar}{\omega} \rmR \, i \int_0^t \rmd t_1 \int_\z \, \delta v(\z,t) \frac{\z}{\z^2}\cdot \bdel_\y\left. \Kc(\z,t; \y,t_1)\right|_{\y=0} \,,
\end{align}
where the three-point correlator $\Kc(\z;\y)$ is found from iterating \eqref{eq:ioe-coord}. As a cross-check, for a medium with constant density we obtain
\begin{equation}
    \omega \frac{\rmd I^\ho}{\rmd\omega \rmd t} = \abar x \, \rmR \, (i-1) \tan \left[ \frac{1-i}{2}xt \right] = \abar x \frac{\sinh \left(x t \right) - \sin \left(x t \right)}{\cosh \left(x t \right) + \cos \left(x t \right)} \,,
\end{equation}
where $x\equiv \sqrt{\hat q/ \omega}$, for the harmonic oscillator term. 

%%%%%%%%%%%%%%%%%%%%%%%%%%%%%%%%%%%%%%%%%%%%%%%%%%%%
%%%%%%%%%%%%%%%%%%%%%%%%%%%%%%%%%%%%%%%%%%%%%%%%%%%%
\section{Medium-induced spectrum and rate with finite-$z$ corrections}
\label{app:Kernels_finitez}
%%%%%%%%%%%%%%%%%%%%%%%%%%%%%%%%%%%%%%%%%%%%%%%%%%%%
%%%%%%%%%%%%%%%%%%%%%%%%%%%%%%%%%%%%%%%%%%%%%%%%%%%%
The process we study is a parton of energy $E$ splitting into two partons with energy $zE$ and $(1-z) E$. In the main text, we refer to the emitted energy $z E$ as $\omega$. However, we stress that this definition is only true in the soft limit. In the more general case we refer to $\omega$ as the reduced energy of the three-body evolution, that is $\omega=z(1-z)E$. This quantity is only equal to the emitted energy when $z\to0$, in which case it reduces to $\omega \simeq z E$. This appendix accounts for how our framework generalizes when considering all the finite-$z$ contributions, which means we strictly use the full definition $\omega=z(1-z)E$.

Reference~\cite{Barata:2021wuf} already has an implementation of the OE ($N = 1$) and IOE (HO+NHO) medium-induced emission spectrum in the strictly soft limit ($\omega  \ll  E$) for a homogeneous brick. We improve on this by keeping finite-$z$ terms, including the rates, and by including the OE and ROE expansions. The resulting code is available online \cite{MedKernels:2022}.

The starting equation for keeping finite-$z$ corrections can be found in Ref.~\cite{Mehtar-Tani:2019ygg},
\begin{align}\label{eq:Spectrum_def_finitez}
    \frac{\rmd I_{ba}^{\rm med}}{\rmd z}=\frac{\alpha_s}{\momega^2}P_{ba}(z){\rm Re}
    \int_0^\infty\rmd t_2\int^{t_2}_0\rmd t_1\,\bdel_{\bm x}\cdot\bdel_{\bm y}\left[\cK_{ba}(\bm x,t_2;\bm y,t_1)-\cK_0(\bm x,t_2;\bm y,t_1)\right]_{\bm x=\bm y=0}\,,
\end{align}
where the parent parton $a$ carrying energy $E$ splits into partons $b$ and $c$, carrying energy $zE$ and $(1-z)E$, respectively. It is the finite-$z$ analog of Eq.~\eqref{eq:medspec-coord}. The Altarelli--Parisi splitting functions are
\begin{align}
    P_{gq}(z)&=C_F\frac{1+(1-z)^2}{z}\,,\quad\quad P_{qq}(z)=P_{gq}(1-z)\,,\quad\quad \nn
    P_{gg}(z)&=C_A\frac{[1+z(1-z)]^2}{z(1-z)}\,,\quad\quad P_{qg}(z)=N_fT_F\left[z^2+(1-z)^2\right]\,,
\end{align}
which are valid for $0 < z < 1$. In the soft limit $z \ll 1$, for quarks the splitting function reduces to $P_q(z) \approx \frac{2C_F}{z}$, while for gluons (where $1-z\ll1$ also has to be included) $P_g(z) \approx \frac{C_A}{z(1-z)} \approx \frac{2C_A}{z}$, where in the last step the $1 - z$ contribution has been folded to $z$ with the additional factor of 2.
The three-point correlator $\cK(\x;\y)$ satisfies the following Schr\"odinger-like equation
\begin{align}\label{eq:Schrodinger_eq}
    \left[i\frac{\partial}{\partial t} + \frac{\bdel^2_{\bm x}}{2\momega}+iv_{ba}(\bm x,t)\right]\cK_{ba}(\bm x,t;\bm y,t_0)=i\delta(t-t_0)\delta(\bm x-\bm y)\,,
\end{align}
where the potential $v_{ba}$ describes the splitting induced by partons scattering with the medium,
\begin{align}\label{eq:Def_vba_potential}
    v_{ba}(\bm x,t)=
    \frac{C_{cba}}{2N_c}v(\bm x,t)+\frac{C_{acb}}{2N_c}v(z\bm x,t)+\frac{C_{bac}}{2N_c}v((1-z)\bm x,t)\,,    
\end{align}
where $C_{ijk} \equiv  C_i + C_j - C_k$ and $C_i$ is the Casimir operator squared for particle $i$ and $v(\x,t)$ is defined in Eq.~\eqref{eq:Def_v_potential}. 
In the soft limit, $v_g(\bm x,t) \approx v_q(\bm x,t) \approx \frac{C_{b,c}}{N_c}v(\bm x,t)$, where the soft emission is always a gluon ($C_b$ or $C_c = N_c$). Surprisingly, this shows that the potential is sensitive to the emitted gluon's and not the emitter's color in the soft limit because $v$ is proportional with $N_c$ by definition. This was observed previously in the opacity expansion and was explained heuristically in~\cite{Wiedemann:2000za}. For quarks in the $z\to1$ limit, the potential goes to $v_q(\bm x,t)\approx\frac{C_F}{N_c}v(\bm x,t)$, where the color factor compensates the $N_c$ in $v$.

In momentum space, we find
\begin{align}
    v_{ba}(\q,t) &= \frac{C_{cba}}{2N_c} v(\bm q,t) + \frac{C_{acb}}{2N_c}\frac{1}{z^2} v\left(\frac{\bm q}{z},t\right) + \frac{C_{bac}}{2N_c}\frac{1}{(1-z)^2} v\left(\frac{\bm q}{1-z},t\right) \,,
\end{align}
where $v(\bm q,t) = (2\pi)^2\delta(\bm q)\Sigma(t) - \sigma(\bm q,t)$, and $\Sigma(t)  =  \Sigma(0,t)  =  \int_\q \sigma(\q,t)$. In the soft limit, $v_g(\bm q,t) \approx v_q(\bm q,t) \approx v(\bm q,t)$, which only becomes apparent once the explicit form of $\sigma(\bm q)$ is used in $v(\frac{\bm q}{z})$.
It is at this point worth extending the definition of the interaction potential to also include an argument defining the screening mass, i.e.
\begin{equation}
    \sigma(\q,t,\mu) \equiv \frac{4\pi\hat q_0(t)}{(\q^2 + \mu^2)^2} \,,
\end{equation}
for the GW model (see Eq.~\eqref{eq:gw-potential}), and similarly for the inverse mean free path $\Sigma(\p^2,t)  \to  \Sigma(\p^2,t,\mu)$, where $\Sigma(\p^2,t,\mu)  =  \int_\q \sigma(\q,t,\mu) \Theta(\q^2 - \p^2)$. Then, $\frac{1}{z^2}\sigma(\frac{\p}{z},t,\mu)  =  z^2 \sigma(\p,t,z \mu)$.

Further simplifications can be made following the discussion in Sec.~\ref{sec:general_formalism}. For the spectrum in momentum space representation, generalizing Eq.~\eqref{eq:medspec-simplified} to finite-$z$, we arrive at
\begin{equation}
    \label{eq:finitez-momspec-1}
    \frac{\rmd I_{ba}}{\rmd z} = \frac{2\alpha_s}{\momega} P_{ba}(z)\, \rmR \,i \int_0^L \rmd t_2 \int_0^{t_2} \rmd t_1 \int_{\p,\p_0} \, \Sigma_{ba}(\p^2,t_2)  \frac{\p\cdot \p_0}{\p^2} \Kc_{ba}(\p,t_2;\p_0,t_1) \,,
\end{equation}
where 
\begin{equation}
    \label{eq:finitez-sigmaba}
    \Sigma_{ba}(\q^2,t) = \frac{C_{cba}}{2N_c} \Sigma(\q^2,t,\mu) + \frac{C_{acb}}{2N_c} z^2 \Sigma(\q^2,t,z\mu) + \frac{C_{bac}}{2N_c} (1-z)^2 \Sigma(\q^2,t,(1-z)\mu) \,,
\end{equation}
and the three-point function in momentum-space representation is found through the implicit equation
\begin{align}
    \Kc_{ba}(\p,t;\p_0,t_0) &= (2\pi)^2 \delta(\p-\p_0) \Kc_0(\p;t-t_0) \nn
    & - \int_{t_0}^t \rmd s \int_\q \, \Kc_0(\p;t-s) v_{ba}(\q,s) \Kc_{ba}(\p-\q,s;\p_0,t_0) \,.
\end{align}
Then, analogously to the derivations in App.~\ref{app:rate}, the rate at finite-$z$ reads
\begin{equation}
    \label{eq:rate-mom-z}
    \frac{\rmd I_{ba}}{\rmd z \rmd t} = \frac{2 \alpha_s}{\momega} P_{ba}(z) \rmR \,i \int_0^t \rmd t_1 \int_{\p,\p_0} \, \Sigma_{ba}(\p^2,t) \frac{\p \cdot \p_0}{\p^2} \Kc_{ba}(\p,t; \p_0,t_1) \,.
\end{equation}
Similar manipulations in coordinate-space representation will be done directly in the IOE section below.

To simplify the expressions below we also introduce the shorthand that accounts for the recurring combinations of color and $z$ factors, see e.g. in Eq.~\eqref{eq:finitez-sigmaba}. Hence, we have
\begin{equation}
    \label{eq:permutation-operator}
    \sum_{p=1}^3 \Cc_p z_p^2 f(z_p^2 x) = \frac{C_{cba}}{2N_c} f(x) + \frac{C_{acb}}{2N_c} z^2 f(z^2 x) + \frac{C_{bac}}{2N_c} (1-z)^2 f((1-z)^2 x) \,,
\end{equation}
that runs over the three cyclic permutations of $\{ a, b,c\}$. Here $\Cc_p=\left[\frac{C_{cba}}{2N_c},\frac{C_{acb}}{2N_c},\frac{C_{bac}}{2N_c}\right]$ and $z_p=[1,z,(1-z)]$, with $p$ running from $1$ to $3$. In the soft limit this expression simply becomes $\sum_p\Cc_p z_p^2 f(z_p^2 x) \to f(x)$.

%%%%%%%%%%%%%%%%%%%%%%%%%%%%%%%%%%%%%%%%%%%%%%%%%%%%%%%%%%%%%
\subsection{Opacity Expansion}
%%%%%%%%%%%%%%%%%%%%%%%%%%%%%%%%%%%%%%%%%%%%%%%%%%%%%%%%%%%%%

From Eq.~\eqref{eq:finitez-momspec-1}, and following Sec.~\ref{sec:oe}, we find the $N = 1$ contribution of the OE at finite-$z$ to be,
\begin{equation}
    \frac{\rmd I^{N=1}_{ba}}{\rmd z}
    =\frac{2\alpha_s}{\pi} \frac{P_{ba}(z)}{z(1-z)} \frac{L}{\lambda} \frac{\mu^2 L}{2 E} \, \sum_p \Cc_p z_p^2 \,\Ic_{N=1}(z_p^2 y) \,,
\end{equation}
where $\lambda = \mu^2/\hat q_0$ is the mean free path, $y = \frac{\bar\omega_c}{\momega}  =  \frac{\mu^2 L}{2z(1-z)E}$, and the relevant integral is\footnote{The integrals are available with trigonometric integral functions
\begin{align}
    \int_0^\infty\frac{\rmd u}{u^2}\frac{u-\sin u}{u+y}&=\frac{1}{y^2}\left[y(\gamma_E-1+\ln y)+\pi\sin^2\frac{y}{2}-{\rm Ci}(y)\sin y+{\rm Si}(y)\cos y\right]\,,\nonumber \\
    \int_0^\infty\frac{\rmd u}{u}\frac{1-\cos u}{u+y}&=\frac{1}{2y}\left[2(\gamma_E+\ln(y))-2\cos(y){\rm Ci}(y)+\sin(y)(\pi-2{\rm Si}(y))\right]\,,\nonumber
\end{align}
where Euler Gamma $\gamma_E$, ${\rm Ci}(z) = -\int_z^\infty\rmd t/t\cos t$, and ${\rm Si}(z) = \int_0^t\rmd t/t\sin t$.}
\begin{equation}
    \Ic_{N=1}(y) = \int_0^\infty\frac{\rmd u}{u^2}\frac{u-\sin u}{u+y} = \begin{cases}
        \frac{\pi}{4}\,, & \text{for } y \ll 1 \,,\\
        \frac{1}{y}\left(\gamma_E + \ln y \right)\,, & \text{for } y \gg 1 \,.
    \end{cases}
\end{equation}
We have also extracted the asymptotic behaviors for future convenience. The soft limit of a hard emission ($\omegacbar \ll zE \ll E$) reduces to $\frac{\rmd I}{\rmd z} = \bar\alpha\frac{\pi}{2}\frac{L}{\lambda}\frac{\bar\omega_c}{z^2E}$, which is in agreement with Eq.~\eqref{eq:n1-limits}, with $\omega \to zE$. One could also use $P_g(z) \approx \frac{C_A}{z(1-z)}$ and keep $z(1 - z)$, and then an extra 1/2 factor will appear to not double count both contributions $z,1 - z \ll 1$. The rate can be found directly from Eq.~\eqref{eq:rate-mom-z} and is given by
\begin{equation}
    \label{eq:N1_rate_finitez}
    \frac{\rmd I^{N=1}_{ba}}{\rmd z\rmd t} = \frac{2\alpha_s}{\pi}\frac{P_{ba}(z)}{z(1-z)} \frac{1}{\lambda}\frac{\mu^2 t}{2 E} \, \sum_p \Cc_p z_p^2 \,\tilde\Ic_{N=1}(z_p^2 y) \,,
\end{equation}
where 
\begin{equation}
    \tilde\Ic_{N=1}(y) = \int_0^\infty\frac{\rmd u}{u}\frac{1-\cos u}{u+y} = \begin{cases}
        \frac{\pi}{2}\,, & \text{for } y \ll 1 \,,\\
        \frac{1}{y} \left(1+ \gamma_E + \ln y \right)\,, & \text{for } y \gg 1 \,.
    \end{cases} 
\end{equation}

%%%%%%%%%%%%%%%%%%%%%%%%%%%%%%%%%%%%%%%%%%%%%%%%%%%%%%%%%%%%%
\subsection{Resummed Opacity Expansion}
%%%%%%%%%%%%%%%%%%%%%%%%%%%%%%%%%%%%%%%%%%%%%%%%%%%%%%%%%%%%%

The finite-$z$ potential in Eq.~\eqref{eq:Def_vba_potential} introduces an additional complication for the ROE. In Sec.~\ref{sec:roe}, we separated and resummed the zero momentum exchange mode, while expanding in real scatterings. In Eq.~\eqref{eq:Def_vba_potential}, however, this separation is more complicated because additional zero modes appear in the real terms in the $z,1 - z \to 0$ limits. Therefore, we make sure to explicitly subtract the zero mode terms in real scatterings
\begin{equation}
    v_{ba}(\bm p,t)=(2\pi)^2\sum_p\Cc_p\left[(1-f(z_p))\delta(\bm p)\Sigma(t)-\int_{\bm q}(\delta(\bm p-z_p\bm q)-f(z_p)\delta(\bm p))\sigma(\bm q)\right]\,.
\end{equation}
We introduced $f(z)$ arbitrary function, that goes to $1$ in the soft limit $z\to0$ (and $1 - z \to 0$). In this paper, we make the choice $f(z) = 1 - z^2$.
The resummed opacity expansion involves the Sudakov factor of the no elastic scattering probability,
\begin{equation}
    \Delta_{ba}(t_2,t_1) = \exp \left[- \int_{t_1}^{t_2} \rmd s\, \hat\Sigma(s,z)\right] \,,
\end{equation}
here $\hat\Sigma(s,z) = [\frac{C_{cba}}{2N_c} + z^2\frac{C_{acb}}{2N_c} + (1 - z)^2\frac{C_{bac}}{2N_c}]\Sigma(s)$. In the soft limit $\hat\Sigma \to \Sigma$, and thus the Sudakov goes to Eq.~\eqref{eq:sudakov}.
The expansion reads
\begin{align}
    \Kc_{ba}(\p,t;\p_0,t_0) &= (2\pi)^2 \delta(\p-\p_0) \Delta_{ba}(t_2,t_1) \Kc_0(\p;t_2-t_1)\nn&- \int_{t_1}^{t_2} \rmd s \int_q \, \Delta_{ba}(t_2,s) \Kc_0(\p;t_2-s) \hat\sigma_{ba}(\q,s) \Kc_{ba}(\p-\q,s;\p_0,t_0) \,,
\end{align}
where $\hat\sigma_{ba}(\bm q,s) = \sum_p\Cc_p\frac{1}{z_p^2}\sigma(\frac{\bm q}{z_p},t) - \sum_p\Cc_pf(z_p)\Sigma(s)$.

The first order ($N_r = 1$) can be read directly off from Eq.~\eqref{eq:finitez-momspec-1}, and reads
\begin{align}
    \frac{\rmd I_{ba}^{N_r=1}}{\rmd z} &= \frac{2\alpha_s}{\momega} P_{ba}(z)\, \rmR \,i \int_0^L \rmd t_2 \int_0^{t_2} \rmd t_1 \int_{\p} \, \Sigma_{ba}(\p^2,t_2) \Kc_0(\p;t_2-t_1) \Delta_{ba}(t_2,t_1) \,,\nn
    &= \frac{2\alpha_s}{\pi} \frac{P_{ba}(z)}{z(1-z)} \frac{L}{\mfp} \frac{\mu^2 L}{2 E} 
    \, \sum_p \Cc_p z_p^2 \Ic_{N_r=1}(z_p^2 y) \,,
\end{align}
where $y  =  \frac{\mu^2 L}{2\momega}$ and, defining $\hat \chi  \equiv  \hat \Sigma L$,
\begin{equation}
    \Ic_{N_r=1}(y) = - \int_0^\infty \frac{\rmd u}{u+y}\, \rmI T(u - i \hat \chi) \,,
\end{equation}
and $T(u)  =  (1 - i u  -  \rme^{-i u})/u^2$. The imaginary part can also be written explicitly, as 
\begin{equation}
    -\rmI\,T(u-i \hat \chi) = \frac{u[u^2+\hat \chi(\hat \chi -2)]+\left[2u\hat \chi \cos u-(u^2-\hat \chi^2)\sin u \right]\rme^{-\hat \chi}}{[u^2+ \hat \chi^2]^2} \,.
\end{equation}
In the big medium limit ($\hat\chi \gg 1$), our formula reproduces Eq.~\eqref{eq:nr1-limit-highop}, $\frac{\rmd I}{\rmd z} = \frac{\bar\alpha}{z}\frac{L}{\lambda} \ln\frac{\mu^2\lambda}{2zE}$.

The rate follows directly from Eq.~\eqref{eq:rate-mom-z}, and reads
\begin{equation}
\label{eq:Nr1_rate_finitez}
    \frac{\rmd I_{ba}^{N_r=1}}{\rmd z \rmd t} = \frac{2\alpha_s}{\pi} \frac{P_{ba}(z)}{z(1-z)} \frac{1}{\mfp} \frac{\mu^2 t}{2 E} 
    \, \sum_p \Cc_p z_p^2 \tilde\Ic_{N_r=1}(z_p^2 y) \,,
\end{equation}
with the relevant integral being,\footnote{
The integral is analytical using ${\rm Ei}(x) = -\int_{-x}^\infty\rmd t \rme^{-t}/t$,
\begin{align}
    \int_0^\infty\rmd u\frac{u(\rme^{\chi}-\cos u)-\chi\sin u}{(u+y)(u^2+\chi^2)}&=\frac{\rme^{-\chi}}{2(\chi^2+y^2)}\bigg[
    \pi x\rme^\chi-\pi\chi(\cos y-\sin y)-2y(\cos(y){\rm Ci}(y)+\sin(y){\rm Si}(y))\nn
    &\left.+2\chi(\cos (y){\rm Si}(y)-\sin (y){\rm Ci}(y))+2y\rme^\chi\left(\ln\frac{y}{\chi}+{\rm Ei}(-x)\right)\right] \,. \nonumber
\end{align}
}
\begin{equation}
    \tilde\Ic_{N_r=1}(y) = - \int_0^\infty \frac{\rmd u}{u+y}\, \rmI \tilde T(u - i \hat \chi) \,,
\end{equation}
with $\tilde T(y)  =  (-i  + i \rme^{-i u})/u$.
The imaginary part of this function is 
\begin{equation}
    -\rmI\, \tilde T(u-i \hat \chi) = \frac{u - \rme^{- \hat \chi}(u \cos u + \hat \chi \sin u)}{u^2 + \hat \chi^2} \,.
\end{equation}

%%%%%%%%%%%%%%%%%%%%%%%%%%%%%%%%%%%%%%%%%%%%%%%%%%%%%%%%%%%%%
\subsection{Improved Opacity Expansion}
\label{app:finitez-ioe}
%%%%%%%%%%%%%%%%%%%%%%%%%%%%%%%%%%%%%%%%%%%%%%%%%%%%%%%%%%%%%
The IOE is similar to that we used in Sec.~\ref{sec:ioe}, one takes a perturbative expansion in $\mu|\bm x| \ll 1$ of Eq.~\eqref{eq:Def_v_potential} in Eq.~\eqref{eq:Def_vba_potential}. By including the color and $z$-dependence of the splitting, an effective jet quenching parameter can be defined as\footnote{Our definition includes sub-leading $\sim  z^2\ln z^2$ terms to the HO term and thus these terms get resummed. This should make the IOE expansion converge faster. The leading form without these terms would look like
\begin{equation}
    \label{eq:finitez-qhat}
    \hat q_{ba}(z,t)=\hat q_0(t)\left[\frac{C_{cba}}{2C_A}+\frac{C_{acb}}{2C_A}z^2+\frac{C_{bac}}{2C_A}(1-z)^2\right]\ln\frac{Q^2}{\mu_\ast^2}\,.\nonumber
\end{equation}
}
\begin{align}
    \hat q_{ba}(z,t)&\equiv\hat q_0(t)\left[\frac{C_{cba}}{2C_A}\ln\frac{Q^2}{\mu_\ast^2}+\frac{C_{acb}}{2C_A}z^2\ln\frac{Q^2}{z^2\mu_\ast^2}+\frac{C_{bac}}{2C_A}(1-z)^2\ln\frac{Q^2}{(1-z)^2\mu_\ast^2}\right]\,, \nn
    &= \qhat_0(t) \sum_p \Cc_p z_p^2 \ln \frac{Q^2}{z_p^2 \mu_\ast^2} \,,
\end{align}
The harmonic oscillator potential is then $v_{ba}^{\text{\tiny HO}}(\bm x,t) = \frac14\hat q_{ba}(z,t)\bm x^2$, which in the soft limit recovers the expression below Eq.~\eqref{eq:v_IOE}, $\hat q_{ba} \to \hat q$. 

\paragraph{Harmonic Oscillator:} 
The harmonic oscillator spectrum is given by
\begin{align}
    \label{eq:HO_spectrum_finitez}
    \frac{\dd I^{\text{\tiny HO}}_{ba}}{\dd z}&=\frac{\alpha_s }{\pi}P_{ba}(z)\ln\left[\frac12\left(\cos\left(\sqrt{\frac{2\omega_c}{\momega}}\right)+\cosh\left(\sqrt{\frac{2\omega_c}{\momega}}\right)\right)\right]\,,
\end{align}
where $\omega_c = \frac12\hat q_{ba}(z)L^2$. In the soft limit, $z\frac{\rmd I}{\rmd z}\approx2\bar\alpha\sqrt{\hat q/(2z^{3})}$, that reproduces the formula from Eq.~\eqref{eq:ho-limits}. The time-differential rate that appears in the evolution equation is
\begin{align}\label{eq:HO_rate_finitez}
    \frac{\dd I^{\rm (HO)}_{ba}}{\dd z\rmd L}&=\frac{\alpha_s }{\pi}P_{ba}(z)\frac1L\sqrt{\frac{2\omega_c}{\momega}}\frac{\sinh\left(\sqrt{\frac{2\omega_c}{\momega}}\right)-\sin\left(\sqrt{\frac{2\omega_c}{\momega}}\right)}{\cos\left(\sqrt{\frac{2\omega_c}{\momega}}\right)+\cosh\left(\sqrt{\frac{2\omega_c}{\momega}}\right)}\,.
\end{align}

\paragraph{Next-to Harmonic Oscillator:} 
The NHO spectrum is given by
\begin{align}
    \frac{\rmd I^{\rm NHO}_{ba}}{\rmd z}&=\frac{\alpha_s}{\momega^2}P_{ba}(z){\rm Re}\int_0^\infty\rmd t_2\int_0^{t_2}\rmd t_1\int_{\bm z}\int_0^L\rmd s\,\partial_{\bm x}\partial_{\bm y}\cK_{ba}^{\text{\tiny HO}}(\bm x,t_2;\bm z,s)\delta v_{ba}(z,s) \nn
    &\times\cK_{ba}^{\text{\tiny HO}}(\bm z,s;\bm y,t_1)|_{\bm x=\bm y=0}\\ 
    &=\frac{\alpha_s}{\pi^2}P_{ba}(z){\rm Re}\int^L_0\rmd s\int_{\bm u}\frac{1}{\bm u^2}\delta v_{ba}(\bm u,s){\rm e}^{-k^2(s)\bm u^2}\,,
\end{align}
where we have defined
\begin{align}
    k^2(s)&=i\frac{\momega\Omega}{2}\left[\cot(\Omega s)-\tan(\Omega(L-s))\right]\,,\\
    \delta v_{ba}(\bm x,t)&=\frac{\hat q_0}{4}\bm x^2C_{ba}(z)\ln\frac{1}{\bm x^2Q^2}\,,
\end{align}
and we have used $\Omega = \sqrt{\hat q_{ba}/(2i\momega)}$, $C_{ba}(z)=\frac{C_{cba}}{2C_A}+\frac{C_{acb}}{2C_A}z^2+\frac{C_{bac}}{2C_A}(1-z)^2$ and some sub-leading terms have already been included in $\hat q_{ba}$.
The integral over the transverse position can be done
\begin{align}
    \int_{\bm u}\frac{1}{\bm u^2}\delta v(\bm u,s)\rme^{-k^2(s)\bm u^2}=\frac{\pi}{4}\hat q_0\frac{1}{-k^2(s)}\left(\gamma_{\rm E}+\ln\frac{-k^2(s)}{Q^2}\right)\,.
\end{align}
Collecting all terms, we get
\begin{align}
    \frac{\rmd I^{\rm NHO}_{ba}}{\rmd z} &= \frac{\alpha_s}{2\pi}P_{ba}(z) \hat q_0C_{ba}(z) \rmR \int_0^L \frac{\rmd s}{-k^2(s)}\left(\gamma_{\rm E}+\ln\frac{-k^2(s)}{Q^2}\right) \\
    &\approx
    \begin{cases}
        \frac{\alpha_s}{\pi}P_{ba}(z)C_{ba}(z)\frac{\hat q_0}{\hat q_{ba}}\sqrt{\frac{\omega_c}{2\momega}}\left\{\sqrt{\frac{\momega}{2\omega_c}}\left(\frac{\pi^2}{12}\tanh\left(\sqrt{\frac{\omega_c}{2\momega}}\right)-2\ln2\right)\right. \nn
        \left.+1+\tanh\left(\sqrt{\frac{\omega_c}{2\momega}}\right)\left[\gamma_E-1+\frac\pi4+\ln\left(\frac{\sqrt{\momega\hat q_{ba}}}{\sqrt2 Q^2}\right)\right]\right\}, & \text{for } \momega\ll\omega_c\,, \\
        \frac{\alpha_s}{2}P_{ba}(z)C_{ba}(z)\frac{\hat q_0L^2}{2\momega}\left[1+\frac{2}{3\pi}\frac{\hat q_{ba}L^2}{2\momega}\left(2\gamma_E-\frac{7}{12}+\ln\frac{\momega}{2LQ^2}\right)\right], & \text{for } \momega\gg\omega_c\,.
    \end{cases}
\end{align}
The rate can be given explicitly in the soft and hard limit
\begin{align}\label{eq:NHO_rate_finitez}
    &\frac{\rmd I^{\rm NHO}_{ba}}{\rmd z\rmd L}=\frac{\partial}{\partial L}\frac{\rmd I^{\rm NHO}_{ba}}{\rmd z}\\
    &\approx
    \begin{cases}
        \frac{\alpha_s}{\pi}P_{ba}(z)C_{ba}(z)\frac{\hat q_0}{\hat q_{ba}}\frac{1}{2L}\sqrt{\frac{\omega_c}{2\momega}}{\rm sech}^2\left(\sqrt{\frac{\omega_c}{2\momega}}\right)
        \left\{1+\frac{\pi^2}{12}+\cosh\left(\sqrt{\frac{2\omega_c}{\momega}}\right) \right.\nn
        +\left.\left(\sqrt{\frac{2\omega_c}{\momega}}+\sinh\left(\sqrt{\frac{2\omega_c}{\momega}}\right)\right)\left(\gamma_E-1+\frac\pi4+\ln\left(\frac{\sqrt{\momega\hat q_{ba}}}{\sqrt2Q^2}\right)\right)\right\}
       , & \text{for } \momega\ll\omega_c\,, \\
        \frac{\alpha_s}{2}P_{ba}(z)C_{ba}(z)\frac{\hat q_0L}{\momega}
        \left[1+\frac{2}{3\pi}\frac{\hat q_{ba}L^2}{2\momega}\left(4\gamma_E-\frac{5}{3}+2\ln\left(\frac{\momega}{2LQ^2}\right)\right)\right], & \text{for } \momega\gg\omega_c\,.
    \end{cases}
\end{align}
The approximated formulas capture the exact formulas up to a few percent deviances and are therefore suitable for numeric implementation, as they do not contain any integrals.

\paragraph{Matching Scale:} We already introduced the $\omega$ dependence of $Q^2(\omega)$, that relied on the soft limit ($\omega \ll \omega_c$) of the spectrum (where the finite-$z$ correction disappears), and thus we use the same definition as in Eq.~\eqref{eq:Running_Q2}. In the numerical implementation, we set $\hat q = \max(\hat q_0,\hat q(z))$ and $Q^2 = \max(\sqrt{e}\mu_\ast^2,Q^2(z))$. This will only become relevant if $L<\lambda$ and $\omegacbar<\omega<\omegaBH$ which is a small corner of the phase space, where instead of the IOE one should use the OE. We showed numerically that using the IOE with the frozen matching scale or using the OE for $L<\lambda$ does not matter, however, the latter would need the introduction of a new smoothing between OE and IOE at $L=\lambda$ that complicates the implementation (see also App.~\ref{app:Numeric_impl}).

To summarize this section, the full emission phase space is covered by using a similar formula that was presented in Sec.~\ref{sec:summary_spectra}
\begin{equation}
    \frac{\rmd I}{\rmd z}=
    \begin{cases}
        \frac{\rmd I^{\rm ROE}}{\rmd z}\,, & \momega<\omega_{\rm tr}\,, \\
        \frac{\rmd I^{\rm IOE}}{\rmd z}\,, & \momega>\omega_{\rm tr}\,, \        
    \end{cases}
\end{equation}
where $\omega_{\rm tr} = \min(\omegaBH,\bar\omega_c)$ and $\momega=z(1-z)E$. The condition on $\momega$ comes from the limits calculated in this section. Importantly, the conditions are the same as the ones derived in Sec.~\ref{sec:spectrum}. 

We would like emphasize the $z,1-z$ symmetry presented in the $\momega$ condition. The gluon spectrum is trivially symmetric in $z,1-z$ as the emitted particles' kinematics is equivalent. The quark spectrum, on the other hand, is strongly asymmetric in $z$. One can still use the symmetric condition on $\momega$ as we saw in the limiting formulas in this section.

%%%%%%%%%%%%%%%%%%%%%%%%%%%%%%%%%%%%%%%%%%%%%%%%%%%%
\section{Numerical implementation of the evolution equation}
\label{app:Numeric_impl}
%%%%%%%%%%%%%%%%%%%%%%%%%%%%%%%%%%%%%%%%%%%%%%%%%%%%

\begin{figure}
    \centering
    \includegraphics[width=\textwidth]{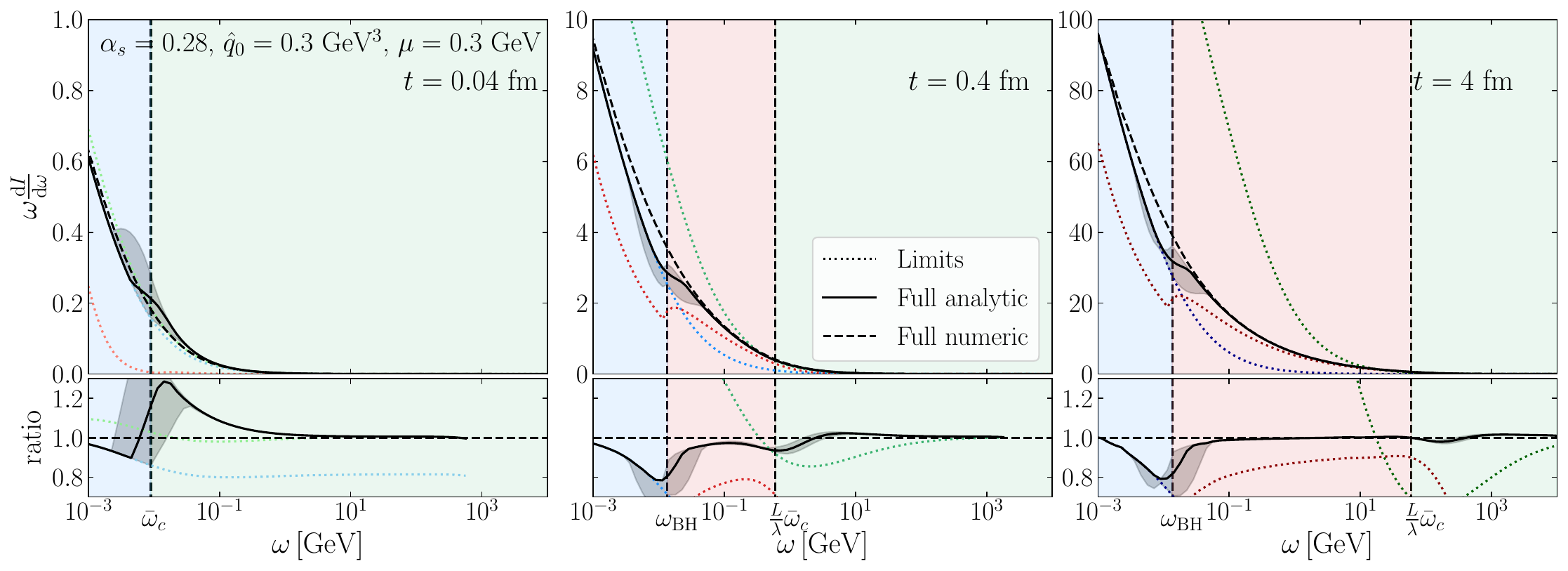}
    \caption{Same as Fig.~\ref{fig:spectra-numerics}, on a semilog scale.}
    \label{fig:spectra-numerics_semilog}
\end{figure}

The evolution equation is given in Eq.~\eqref{eq:Evolution_eq} and can be rewritten by introducing the variable $\xi = \frac{x}{z}$ ($\xi = xz$) in the gain (loss) term
 \begin{align}
     \partial_t D(x,t)&=\int_x^1\rmd\xi \,f(x,\xi,t)D(\xi,t)-D(x,t)\int_0^1\rmd\xi \,f(\xi,x,t)\,,\\
     f(x,\xi,t)&=\frac{x}{\xi^2}\left.\frac{\rmd^2 I}{\rmd z\rmd t}\right|_{E\mapsto\xi E}\left(z\mapsto\frac{x}{\xi}\right)\,.
 \end{align}
We implemented Eq.~\eqref{eq:N1_rate_finitez} (as $N = 1$), Eq.~\eqref{eq:Nr1_rate_finitez} (as $N_r = 1$), and Eqs.~\eqref{eq:HO_rate_finitez} and \eqref{eq:NHO_rate_finitez} (HO+NHO) in $f$ as
\begin{align}
    \left.\frac{\rmd^2 I^{\rm med}}{\rmd z\rmd t}\right|_E=
    (1-S)\left.\frac{\rmd^2 I^{N_r=1}}{\rmd z\rmd t}\right|_E
    +S\,\left.\frac{\rmd^2 I^{\rm IOE}}{\rmd z\rmd t}\right|_E\,, %& \momega>\omega_{\rm tr}
%    \begin{cases}
%        (1-S)\cdot\left.\frac{\rmd^2 I^{N_r=1}}{\rmd z\rmd t}\right|_E\,, & \momega<\omega_{\rm tr}\,,\\
%        S\cdot\left.\frac{\rmd^2 I^{\rm IOE}}{\rmd z\rmd t}\right|_E\,, & \momega>\omega_{\rm tr} \,,
%    \end{cases}
\end{align}
where $\omega_{\rm tr} = \min(\omegaBH,\bar\omega_c)$ and $\momega=z(1-z)E$. The upper formula is not smooth for any finite order of truncation in the transition between the IOE or ROE, and therefore in some cases, we used the switching function $S = \cos\left[\frac{\pi}{2}(1-\alpha)\right]$, with $\alpha = \frac{2\momega-\omega_{\rm tr}}{3\omega_{\rm rt}}$ if $\frac{1}{2}\omega_{\rm tr} < \momega < 2\omega_{\rm tr}$ to smoothing the transition. The uncertainty introduced by this procedure is smaller than the next higher-order contribution. One can study the matching uncertainty around the BH region by varying $\omega_{\rm tr}$ with a factor of 2. To study the matching condition of the IOE, $Q$ can also be varied by a factor of 2 in Eq.~\eqref{eq:Running_Q2} as it was done in Ref.~\cite{Barata:2021wuf}. In Fig.~\ref{fig:spectra-numerics_semilog} we show the deviation from the numeric solution including both of these variations. Other than the band, Fig.~\ref{fig:spectra-numerics_semilog} is equivalent to Fig.~\ref{fig:spectra-numerics}, we only use the soft limit ($z\ll1$). There is a further uncertainty coming from going to one higher order ($N_r=1\mapsto2$ and NHO $\mapsto$ NNHO), that we leave for future studies. We expect this uncertainty, however, to extend the error band in Fig.~\ref{fig:spectra-numerics_semilog} up to the numeric solution.

The integrals can then be divided into 
\begin{align}
    {\rm Gain}&=G_{\xi\to x}+G_{reg}+G_>=\left[\int_x^{x+\epsilon}+\int_{x+\epsilon}^{1-\delta}+\int_{1-\delta}^1\right]\rmd\xi \,f(x,\xi,t)D(\xi,t)\,,\\
    {\rm Loss}&=L_<+L_{reg}+L_{\xi\to x}=-\left[\int_0^\delta+\int_\delta^{x-\epsilon}+\int_{x-\epsilon}^x\right]\rmd\xi \,f(\xi,x,t)D(x,t)\,.
\end{align}
All divergences are present in the $\xi \to x$ terms, which cancel exactly and thus the trapezoid rule is used
\begin{equation}
    G_{\xi\to x}+L_{\xi\to x}\approx\frac{\epsilon}{2}\left[f(x,x+\epsilon)D(x+\epsilon,\tau)-f(x-\epsilon,x)D(x,\tau)\right]\,,
\end{equation}
which contributes to the regular part of the integrals. In our implementation $\epsilon = 10^{-6}$, and thus we have $x > \epsilon$. Similarly to vacuum physics, the $\epsilon$ cut was necessary to introduce because the soft divergence in the Bethe--Heitler region has to be regulated. The $G_{reg},L_<,L_{reg}$ are simple integrals and can be done numerically on a grid. So can $G_{>}$, however, we neglect this latter contribution by using the fact $\lim_{x\to1}D \to 0$ (the kernel is soft divergent and thus it moves quanta towards $x < 1$).

%%%%%%%%%%%%%%%%%%%%%%%%%%%%%%%%%%%%%%%%%%%%%%%%%%%%
%%%%%%%%%%%%%%%%%%%%%%%%%%%%%%%%%%%%%%%%%%%%%%%%%%%%
\section{HTL potential}
\label{app:htl}
%%%%%%%%%%%%%%%%%%%%%%%%%%%%%%%%%%%%%%%%%%%%%%%%%%%%
%%%%%%%%%%%%%%%%%%%%%%%%%%%%%%%%%%%%%%%%%%%%%%%%%%%%

In this appendix, we investigate what happens if employ the HTL potential in the multiple-scattering series. The potential now reads
\beq
\sigma(\q) = N_c n(t) \frac{\rmd^2 \sigma_{\rm el}}{\rmd ^2 \q} = \frac{4\pi\, \hat q_0}{\q^2 (\q^2 + m_D^2)} \,,
\eeq
where $m_D$ is the Debye screening mass in a thermal medium with temperature $T$ and $\hat q_0 = 4\pi \alpha_s m_D^2 T$ \cite{Aurenche:2002pd}.
In this case, we find that
\begin{equation}
    \label{eq:htl-sigma}
    \Sigma(\p^2) = \frac{\hat q_0}{m_D^2}  \ln\left(\frac{\p^2 + m_D^2}{\p^2} \right) \,,
\end{equation}
where we have a logarithmic divergence as $\p^2 \to 0$.
This implies that the mean free path has to be regulated by an IR cut-off so that $\mfp \sim \Sigma^{-1}(p^2_{\rm min})$. 

In this case, the OE remains unmodified because the divergence at $p^2_{\rm min} \to 0$ cancels order by order between the real and virtual contributions. The ROE, however, has to be treated with care when truncated at a finite order since we have to introduce an explicit IR cut-off in order to define the elastic Sudakov factor. Hence, the modified Sudakov reads
\begin{equation}
    \Delta(t,t_0) = \rme^{- \Sigma_\reg(t-t_0)} \,,
\end{equation}
in a static medium, where $\Sigma_\reg \equiv \Sigma(p^2_{\rm min})$ with $p^2_{\rm min}$ and unknown IR regulator. Clearly, an all-order resummation of the ROE series would remove the spurious IR dependence.

The role of the medium potential in the IOE was clearly elucidated in \cite{Barata:2020sav}, where the information about the scattering potential is fully contained in the definition of $\mu_\ast^2$. We refer the reader to this paper for an exhaustive discussion.

In order to clarify what modifications arise in the OE and ROE, we compute the respective first-order terms of the expansions, i.e. $N=1$ and $N_r=1$, explicitly here. In the OE, we find
\begin{equation}
    \label{eq:htl-dilute}
    \omega \frac{\rmd I^{N=1}}{\rmd \omega} = 2 \abar \,\frac{\hat q_0 L}{m_D^2}\, \int_0^\infty \rmd u\,  \ln \left( \frac{u+y}{u}\right)\, [-\rmI \, T(u)] \,,
\end{equation}
where now we defined $y = \bar \omega_c/\omega$ with $\omegacbar = m_D^2 L/2$. The prefactor contains $\hat q_0/m_D^2$, which is similar to the inverse mean free path (in fact, in the GW model it would be exactly equal to $\mfp^{-1}$), with a missing logarithmic factor. Using \eqref{eq:tfunc-small}, we can immediately extract the limiting behaviors, which yield
\begin{equation}
    \label{eq:htl-dilute-limits}
    \omega \frac{\rmd I^{N=1}}{\rmd \omega} = 2 \abar \, \frac{\hat q_0 L}{m_D^2} \begin{cases}
         \ln(\frac{\omegacbar}{\omega}) \left[ -1 + \gamma_E +  \ln \left( \frac{\omegacbar}{\omega}\right)\right] & \text{ for } \omega \ll \omegacbar \\
        \frac{\pi}{4} \frac{\omegacbar}{\omega} & \text{ for } \omega \gg \omegacbar \,. 
    \end{cases}
\end{equation}
As expected, in this limit there is no sensitivity to the mean free path {\it per se}. However, compared to the $N=1$ result in \eqref{eq:n1-limits}, there is an additional logarithmic enhancement $\sim  \ln \omegacbar/\omega$ in the soft limit. Comparing to Eq.~\eqref{eq:htl-sigma}, we see that this logarithm can be absorbed into the prefactor $\sim \hat q_0/m_D^2$ to recreate an effective, regularized mean free path in the HTL theory, i.e.
\begin{equation}
    \left.\mfp_\reg^{-1} \right|_{L\ll \mfp} = \frac{\hat q_0}{m_D^2}  \ln\left(\frac{\omegacbar}{\omega}\right) \,.
\end{equation}
This regularization follows also from the discussion in Refs.~\cite{Wiedemann:2000za,Andres:2020kfg}. With this modification, the soft limit for low medium opacity is equivalent in the GW and HTL theory. 

In the hard limit, $\omega \gg \omegacbar$, there is no additional logarithmic enhancement in the HTL compared to the GW theory, see \eqref{eq:n1-limits}, and the mean free path is simply $\mfp_0^{-1} = \hat q_0/m_D^2$. Apart from this subtlety, the expressions are again equivalent.

Let us now turn to the ROE resummation which is valid in dilute (small) media or in the soft limit for dense or large media. At first order $N_r=1$, we find now
\begin{equation}
    \label{eq:htl-dense}
    \omega \frac{\rmd I^{N_r=1}}{\rmd \omega} = 2 \abar \,\frac{\hat q_0 L}{m_D^2}\, \int_0^\infty \rmd u\,  \ln \left( \frac{u+y}{u}\right)\, [-\rmI \, T(u - i \chi_\reg)] \,,
\end{equation}
where the opacity is $\chi_\reg = \Sigma_\reg L$ and $\Sigma_\reg = \int_\q \sigma(\q)\Theta(\q^2 - q^2_\min)$ is the regularized inverse mean free path.\footnote{For consistency, compared to the ``regularized expansion in Eq.~\eqref{eq:eoe-mom}, one should also include an IR regulator $\sim q^2_\min$ in the lower limit of the integral. We will neglect this subtlety for now.} As discussed at length in Sec.~\ref{sec:roe}, at small opacities the ROE is equivalent order by order to the OE. This was discussed in the paragraphs above. For large opacities, $\chi_\reg \gg 1$, we use \eqref{eq:tfunc-large} to solve the integral analytically. As discussed below Eq.~\eqref{eq:tfunc-large}, the expressions permits a transmutation of the relevant scale from $\omegacbar$ to $\omegaBH$, where now $\omegaBH=m_D^2 / (2 \Sigma_\reg)$. It is then straightforward to extract the following limiting behavior,
\begin{equation}
    \label{eq:htl-dense-limits}
    \omega \frac{\rmd I^{N_r=1}}{\rmd \omega} = 2 \abar \, \frac{\hat q_0 L}{m_D^2}\, \begin{cases}
        \frac{1}{24} \left[5 \pi^2 + 12  \ln^2\left(\frac{\omegaBH}{\omega} \right) \right] & \text{ for } \omega \ll \omegaBH \\
        \frac{\pi}{2} \frac{\omegaBH}{\omega} & \text{ for } \omega \gg \omegaBH \,.
    \end{cases}
\end{equation}
The soft and hard limits have again subtly different characteristics. In the former case, we again observe a double-logarithmic enhancement, similar to the soft limit in dilute media \eqref{eq:htl-dilute-limits} and stronger than the single-logarithmic behavior in the GW model, see Eq.~\eqref{eq:nr1-limit-highop}. We could again absorb one of these factors in an effective mean free path, by defining
\begin{equation}
    \left.\mfp_\reg^{-1} \right|_{L\gg \mfp} = \frac{\hat q_0}{m_D^2}  \ln\left(\frac{\omegaBH}{\omega}\right) \,,
\end{equation}
which demonstrates once more the transmutation of relevant scales. Note that the spurious IR regulator $q^2_\min$ appears now on the level of $\sim  \ln  \ln(q^2_\min)$. In the hard limit, the result is again equivalent to the GW model, see \eqref{eq:nr1-limit-highop}, by identifying the ``bare'' mean free path and rescaling the Bethe--Heitler energy $\omegaBH$. We recall that in this limit, the ROE opacity breaks down and should be replaced by the IOE resummation.

%%%%%%%%%%%%%%%%%%%%%%%%%%%%%%%%%%%%%%%%%%%%%%%%%%%%
%%%%%%%%  REFERENCES
%%%%%%%%%%%%%%%%%%%%%%%%%%%%%%%%%%%%%%%%%%%%%%%%%%%%
\bibliography{ref}

\end{document}